\documentclass[12pt,halfline,a4paper]{pyrrha}

\begin{document}

\title{Understanding Smart Contracts: \\ Hype or Hope?}

\author{
\name\Large{{Elizaveta Zinovyeva \hspace{1cm} Raphael C. G. Reule}}
\address\footnotesize{{Blockchain Research Center, Humboldt-Universit\"at zu Berlin, Germany.\\
International Research Training Group 1792, Humboldt-Universit\"at zu Berlin, Germany.}\\
\email{elizaveta.zinovyeva[at]wiwi.hu-berlin.de \hspace{0.3cm} irtg1792.wiwi[at]wiwi.hu-berlin.de}}
\and
\name{\large{Wolfgang Karl H\"ardle}}
\footnotesize{\address{Blockchain Research Center, Humboldt-Universit\"at zu Berlin, Germany. \\ Wang Yanan Institute for Studies in Economics, Xiamen University, China. \\ Sim Kee Boon Institute for Financial Economics, Singapore Management University, Singapore. \\ Faculty of Mathematics and Physics, Charles University, Czech Republic.\\ National Chiao Tung University, Taiwan.}
\email{haerdle[at]wiwi.hu-berlin.de}}
}

\abstract{Smart Contracts are commonly considered to be an important component or even a key to many business solutions in an immense variety of sectors and promises to securely increase their individual efficiency in an ever more digitized environment. Introduced in the early 1990's, the technology has gained a lot of attention with its application to blockchain technology to an extent, that can be considered a veritable hype. Reflecting the growing institutional interest, this intertwined exploratory study between statistics, information technology, and law contrasts these idealistic stories with the data reality and provides a mandatory step of understanding the matter, before any further relevant applications are discussed as being ``factually" able to replace traditional constructions. Besides fundamental flaws and application difficulties of currently employed Smart Contracts, the technological drive and enthusiasm backing it may however serve as a jump-off board for future developments thrusting well in the presently unshakeable traditional structures.\\

\footnotesize{
\vspace{0.5cm}
}

}\vspace{1cm}

\date{\today}

\maketitle

\newpage

%%%%%%%%%%%%  Start  %%%%%%%%%%%%
%%%%%%%%%%%%  Start  %%%%%%%%%%%%
%%%%%%%%%%%%  Start  %%%%%%%%%%%%

\normalsize

\section{Introduction}
\label{sec1}

\blfootnote{
\textit{Financial support of the European Union's Horizon 2020 research and innovation program ``FIN-TECH: A Financial supervision and Technology compliance training programme" under the grant agreement No 825215 (Topic: ICT-35-2018, Type of action: CSA), the European Cooperation in Science \& Technology COST Action grant CA19130 - Fintech and Artificial Intelligence in Finance - Towards a transparent financial industry, the Deutsche Forschungsgemeinschaft's IRTG 1792 grant, the Yushan Scholar Program of Taiwan, the Czech Science Foundation's grant no. 19-28231X / CAS: XDA 23020303, as well as support by Ansar Aynetdinov (ansar.aynetdinov[at]hu-berlin.de) are greatly acknowledged.}}

\blockquote{All great truths begin as blasphemies.}
\vspace{-0.4cm}
\begin{displayquote}
\small{\citet{Shaw:1919}}\\
\end{displayquote}

\vspace{-0.6cm}

$Smart$ $Contracts$: a key buzzword of our modern, fully digitized, and increasingly globalized media and scientific world amongst ``crowdfunding", ``decentralization", ``resource management", ``prosumer", ``blockchain", ``cryptocurrency", ``internet-of-things", ``digital asset regulation", and many more. The Smart Contract (SC) hype has reached the non-tech-mainstream, driven by an ever-increasing number of FinTech, RegTech, SupTech, LegalTech, and OtherTech startups, as well as governmental initiatives -- yet most often disappear into the ether after some hype.\\

%A subject, which seems to render like an archaic branch of alchemy in a new framework given the plethora of descriptions of this entity. At times, when reading certain outlets - may they be for popular or academic readers - one might think that the $Philosopher's$ $stone$ was finally discovered - a legendary substance capable of doing practically anything, from curing diseases, to providing nourishment for those in need, as well as preventing natural catastrophes. Yet, can smart contracts really revolutionize healthcare, provide aid for the third world, or manage reforestation incentives?\\
 
The basal common ``understanding" is that $blockchain$ (BC)  $applications$ of so-called SCs are $self$-$executing$ pieces of software/algorithms/codes/processes with a predetermined set of rules to be followed in calculations and other problem-solving operations. SCs are supposed to decentralize the authority over agreements and their enforcement by being purposely $self$-$enforcing$, with people across the whole world starting binding relationships without ever trusting -- leave alone knowing -- each other. This ideal in a new appearance represents a fundamentally different approach on organically grown contractual systems. A technology from the 1970s, the so-called $Electronic$ $Data$ $Interchange$ (EDI), was driven by a similar ideal and equally hyped but eventually failed to deliver on replacing traditional contracting. While it was able to reduce some costs, it effectively enshrined the human inefficiency in decision making in the process \citep{Sklaroff:2017}.  Translated to today's even more information-driven times, announcing to work on some BC application attracts a lot of attention and investments to companies -- even though the rate of delivery of BC-driven outcomes is negligible with such announcements incorporating respective buzzwords are observable as an attempt to take advantage of artificial premia \citep{Aky:2021}. Therefore research on a possible gap between an actual status of SCs (given by data) and their target status is mandatory.\\

From crops insurance, over supply chain management, to development assistance, manifold imaginary fields of SC application turn out to be either philosophical ``How nice the world could be"-papers or ``This is what we used, use, and what we will use"-outlets of very specialized fields to forcefully foster the idea of SCs. They all have one thing in common: they do not access the broad view on (1) what the technology can really deliver and/or (2) where it can be applied to real-world situations. Evolutionary, humans used clay, wood, or paper to enshrine their agreements to trade grain, Schnapps, or other values within a system supervised by some hierarchical structures. BC SCs promise to overcome this homeostasis. Seeing that many forwarded SC-oaths are surreal to be accomplished even through traditional means, we critically assess and identify these areas to find out, if this is all just hype or if there is hope. Through the use of datasets depicting in depth the activity and information flow of SCs on the Ethereum (ETH) network, we are able to research the facts and what SCs are used for contrary to the common belief:

\begin{itemize}[leftmargin=*]
	\item[$\boxdot$] Can SCs increase technological literacy, assure increased inclusion, and participation rights?
\vspace{-0.3cm}
	\item[$\boxdot$] Does it make sense for every economic endeavour to adapt to such systems?
 \vspace{-0.3cm}
	\item[$\boxdot$] Are they more than just a financial vehicle and really able to, for example, fight hunger in underdeveloped countries? 
 \vspace{-0.3cm}
	\item[$\boxdot$] Do they really increase efficiency and security whilst reducing associated costs and procedural risks?
 \vspace{-0.3cm}
	\item[$\boxdot$] Are SCs, that follow strict structures in order for their coding to work properly, really more flexible than traditional natural language agreements?
\end{itemize}

With many teething issues to be overcome by such applications beyond being simple transaction monitoring software, the result to date is a realm of uncertainty which the construction in itself tries to prevent ironically. Especially terms like $smart$, $contract$, $authority$, and self-enforcing are highly ambiguous and not as safely defined as in traditional schemes. The most obvious issue, and one of the common misconceptions, is the colported independent self-executability of SCs - which is wrong, as an external input is always needed to order an SC to execute its code. The code is then indeed self-enforcing, possibly acting as a vigilante, leading to a superfluity of problems. Our work is inspired by the work of \citep{oliva2020exploratory} with the aim to extend it with a legal perspective and a proposal of approaches to gain opportunities for further exploration. In this sense, we will crystallize, that as a fact, SCs are just another -- yet ambitious -- ``pathetic dot" regulated by the forces of the market, technical infrastructure, law, and social conception \citep{Lessig:99}.\\

%, hence by solving Thomas Reid's famous dilemma with the Scottish School of Common Sense, that there is no greater impediment to the advancement of knowledge than the ambiguity of words, is what smart contract developers are aiming at. 

 %We engrave a fundamental understanding of this blockchain application by a brief overview of the redundance of potential use cases, and point out the respective technical and real world limits, while providing a historical overview of what smart contracts have really been used for.\\

In this research, we first identify the sources of the hype on SCs; secondly, we present algorithms of SCs to understand what SCs are actually doing. This responds to many inherent questions by creating a taxonomy and critical understanding of SCs. By outlining their framework and providing an extensive mapping of SCs based on the ETH network. Visually, we de-frame the Storytelling by outing their real fields of application and the costs that come with this fractional delivery on the promised list of miracles. Based on the research insight we got from the analysis of existing SCs, we propose an algorithm to identify a proper answer and solution to the question, if it makes sense for a given endeavor to invest into creating BC/SC frameworks.\\

The paper is structured by giving an introduction to the existing research literature and subject in sections \ref{Literature_Review} and \ref{Sec:FrameworkInfo}, followed by a dataset disclosure in section \ref{DatasetSection}, a taxonomy of SCs in sections \ref{clustering} (Clustering) and \ref{sec:Classification} (Classification), eventually culminating in a critical status analysis and improvement proposal in section \ref{Field}, and ending with closing remarks and an appendix in sections \ref{Sec:Closing} and \ref{Appendix}. All presented graphical and numerical examples shown are reproducible and can be found via \includegraphics[keepaspectratio,width=0.4cm]{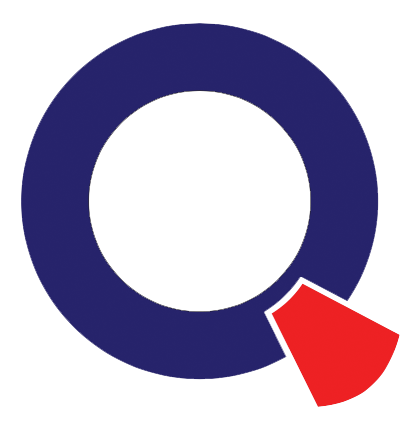} \url{www.quantlet.com} \citep{BH:2018} and are indicated as \href{https://github.com/QuantLet/USC}{\includegraphics[keepaspectratio,width=0.4cm]{media/qletlogo_tr.png}}\href{https://github.com/QuantLet/USC}{ USC}.  Research data is obtainable through the
\href{https://blockchain-research-center.de}{\includegraphics[keepaspectratio,width = 0.4cm]{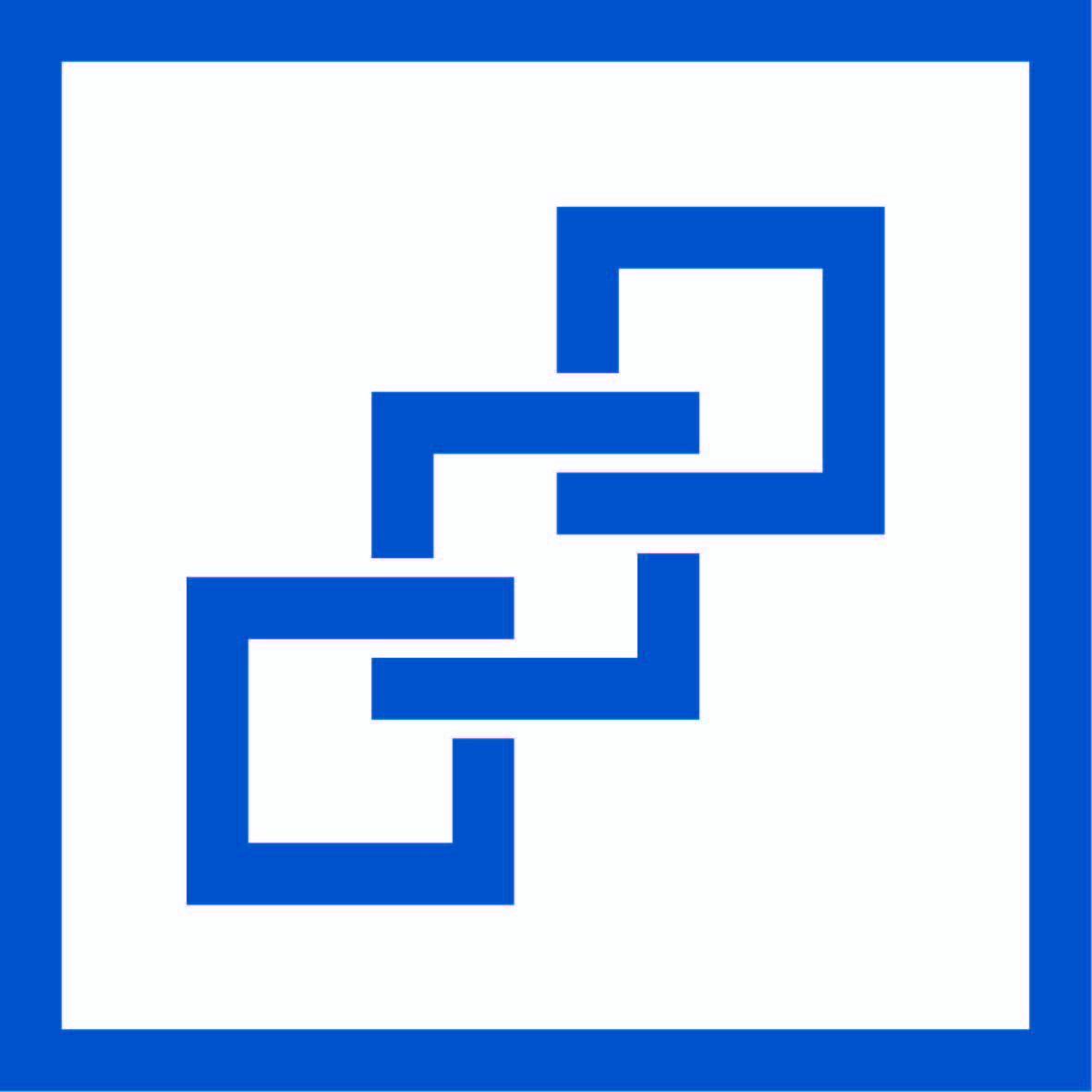} Blockchain-Research-Center.de}.

 %\tableofcontents

%Publicly most noted, Facebook's cryptocurrency coined ``Libra" is designed to be a globally accessible high performance blockchain system construction, envisioned by heavily lending on especially the ideas of $Ethereum$ with programmable features and a share oriented consensus mechanism. Some of the inspiring features are that anyone can create pegged subset replicas of the original Libra blockchain, or run commands associated with objects such as smart contracts or a set of $wallets$. Furthermore, as with Ethereums smart contracts, running actual code in the field comes at a cost and all operations require a payment of Libra as gas in order to run.\\

%%%%%%%%%%%%NEXT%%%%%%%%%%%%%%

\section{Outlet Review}
\label{Literature_Review}

\subsection{Recent Research Review}

To outline the dissonance between the mass of whitepapers/prosaic outlets in contrast to the actual data available for SCs, we conduct a data-driven literature overview gathering a rather diffuse impression. Researchers, for example, focus on a variety of aspects, among others, ranging from applications of SCs in healthcare and quality monitoring \citep{Celesti2020, Khatoon2020, Yu202012479}, technical implementations and vulnerabilities \citep{Ajienka2020, Pierro20201, Li2020438} to games \citep{scholten2019ethereum}.\\

One of the first research works dedicated to the empirical analysis of SCs is the work by \cite{bartoletti2017empirical}. The authors gathered a dataset of 834 SC source codes, which were then manually classified into different categories using the proposed taxonomy. The identified categories contained types like Notary, Financial, Game, Wallet, Library. Moreover, the authors were able to identify the design patterns found in the source code. \cite{wohrer2018design} also dedicate their work to design patterns that are aimed to improve and facilitate the work of software developers and engineers working on applications for the BC, as similar problems may require similar solutions. One of the later empirical studies created by \cite{oliva2020exploratory} propose the approach of cross-linking the data from various sources in order to scrutinize better their behavior: their activity level, category of what these contracts are, and source code metrics, like complexity.\\

An important research direction in this context is automatic source code classification. Although the idea of using machine learning for code classification is not very new, it is not that present in the ETH realm, as we will see later. \cite{norvill2017automated} use unsupervised techniques and run clustering on bytecodes to group contracts with a similar purpose. \cite{chen2018detecting} use the machine learning method XGBoost on the opcodes and account data to identify Ponzi schemes. \cite{wang2020contractward} use different machine learning methods to identify security vulnerabilities in SCs. Many codes on the ETH BC are very similar and are quite often copied. \cite{kondo2020code} investigate code cloning on ETH and identify clusters of the code clones and inspect their behavior.

\begin{figure}[H]
\centering
\includegraphics[width=15.5cm]{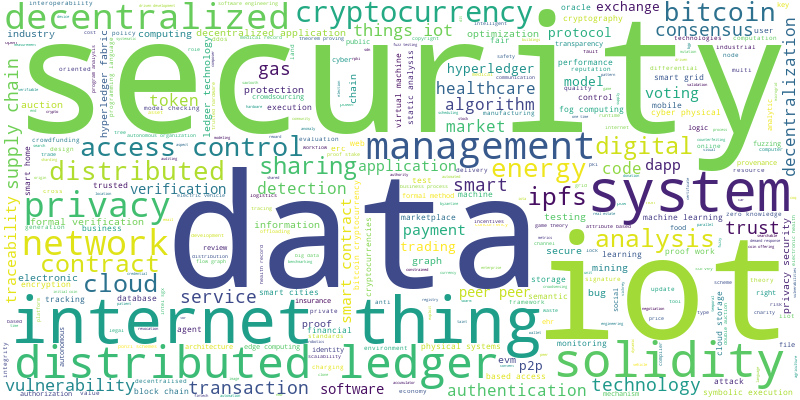}
\caption{Wordcloud of existing research keywords \href{https://github.com/QuantLet/USC/tree/master/SC-literature-research}{\includegraphics[keepaspectratio,width=0.4cm]{media/qletlogo_tr.png}}}\label{fig:research}
\end{figure}

\subsection{Preliminary Data Basis}

In order to scan a wide-angle of exhaustive SC outlets, we utilized the \href{https://www.scopus.com/home.uri}{Scopus citation and abstract database} that gathers peer-reviewed research from around 11.7K publishers from different directions in science. The query retrieves the scholarly work from conferences and journals in the English language containing \textit{ethereum} or \textit{smart contracts} in the title. Furthermore, the query is limited to work conducted in the areas like Business, Computer and Decision Science, and Mathematics. The exact query can be found in the respective Quantlet \href{https://github.com/QuantLet/USC/tree/master/SC-literature-research}{\includegraphics[keepaspectratio,width=0.4cm]{media/qletlogo_tr.png}} on Github. In total, the data pool lists 839 research outlets. Research publishing on ETH SCs began in 2016 and achieved a peak in 2019. At the date of retrieving the query, the year 2020 was ongoing, hence, we could not fully measure whether the amount of research in 2020 is higher or lower than in 2019. Nonetheless, we can see that the SC area's interest is not fading and is staying on a relatively high level. The topics of the examined research outputs were distilled from their abstracts and keywords. We present the most frequent keywords as a wordcloud in Figure \ref{fig:research}. The keywords \textit{blockchain}, \textit{smart contract}, and \textit{ethereum} were filtered out since they were used to limit the query and therefore do not deliver any additional semantic information to differentiate between the outlets. Technical keywords such as \textit{security}, \textit{data}, \textit{iot}, \textit{solidity}, \textit{distributed ledger}, and \textit{decentralized} are the most frequently used. Unfortunately, these do not provide us with sufficient additional information to identify clear tendencies in the research. Each of these keywords could belong to any given paper in the field of SC research. Hence, bibliometric research on SC outlets needs to be conducted on a more granular level.

\vspace{-0.5cm}

\begin{figure}[H]
\centering
\includegraphics[width=15cm]{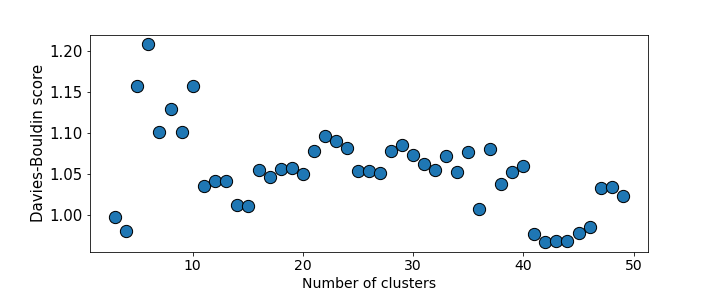}
\caption{Identification of optimal number of groups using elbow-curve method with Davies-Bouldin score \href{https://github.com/QuantLet/USC/tree/master/SC-literature-research}{\includegraphics[keepaspectratio,width=0.4cm]{media/qletlogo_tr.png}}}
\label{fig:Elbow_curves_literature}
\end{figure}

The gathered research papers were categorized into their dedicated topics using dimensionality reduction, clustering, and natural language processing techniques (Bidirectional Encoder Representations from Transformers, BERT) to subsequently inspect the most important words used therewithin. This analysis was performed on the abstract to achieve higher specification. This step's technical implementation is described in greater detail in section \ref{clustering}. The optimal amount of groups/clusters for this step was determined by the elbow (curve) method: a heuristic to identify the number of clusters in a dataset \citet{thorndike1953belongs}. This method measures the identified performance metric for different numbers of groups, plots them against each other, and identifies points that are knee or elbow of a curve. The possible metrics for clustering could be an intra- or intercluster variance, the Davies-Bouldin, and Silhouette scores. Since these metrics might produce contradictory results, we decided to utilize only one of them -- The Davies-Bouldin score. This score \citep{davies1979cluster} shows the average similarity measure of each cluster with its most similar cluster. Figure \ref{fig:Elbow_curves_literature} displays the Davies-Bouldin score, where similarity is the ratio of within-cluster distances to between-cluster distances. It evaluates intra-cluster similarity and inter-cluster differences. Applying the elbow criterion to these diagnostic plots leads us to a value around 14 -- one of the first \textit{elbows} of the plot, as we prefer rather lower values of the Davies-Bouldin score and the lower amounts of topics an-ease the interpretability. Thus, the following Figure \ref{fig:Literature_scatter} presents the scatter plot depicting our clustering into 14 various topics of the existing research.\\

\begin{figure}[H]
\hfill
\subfigure[Topic 1]{\includegraphics[width=4.8cm]{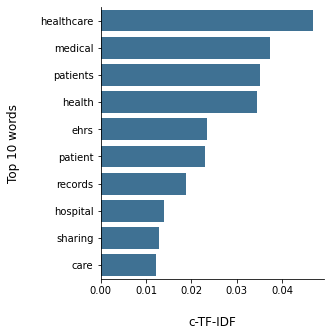}}
\hfill
\subfigure[Topic 2]{\includegraphics[width=4.8cm]{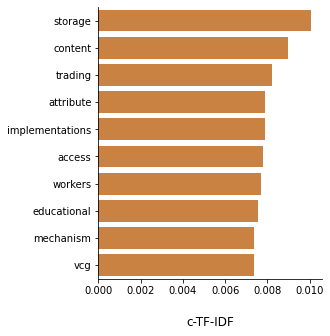}}
\hfill
\subfigure[Topic 3]{\includegraphics[width=4.7cm]{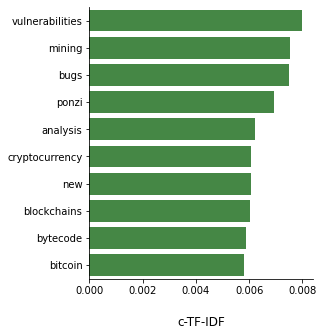}}
\hfill
\caption{Top 10 the most important words per topic identified in the existing SC research \href{https://github.com/QuantLet/USC/tree/master/SC-literature-research}{\includegraphics[keepaspectratio,width=0.4cm]{media/qletlogo_tr.png}}}
\label{fig:topics_research_example}
\end{figure}

 The manual assessment of the top ten identified most \textit{important} words pointed to the following possible topics:  \textcolor[HTML]{3174a1}{Healthcare/medicine},  \textcolor[HTML]{e1812c}{Implementation/Storage},  \textcolor[HTML]{3a923a}{Vulnerabilities/Ponzi/Scam/Bugs},  \textcolor[HTML]{c03d3d}{Voting/Trust}, \textcolor[HTML]{9372b2}{Research/Categories of Research},  \textcolor[HTML]{845b53}{Programming language/Solidity/Software Development},  \textcolor[HTML]{d684bd}{Crowdsoursing/Sharing}, \textcolor[HTML]{7f7f7f}{Applications Development/Protocols}, \textcolor[HTML]{a9aa35}{IOT/Smart devices}, 
\textcolor[HTML]{2eaab8}{Verification/Deployment/ Execution of SC}, \textcolor[HTML]{465E81}{Food/Agriculture}, \textcolor[HTML]{807E07}{BTC/Finance/ICO}, \textcolor[HTML]{E8BD1B}{Energy/Electricity/Gas}, and \textcolor[HTML]{E81B53}{Payments/Audit}. In Figure \ref{fig:topics_research_example}, you can find the example of three topics, where each bar represents the most \textit{important} word and its length -- its \textit{importance} in terms of c-TF-IDF, which will be described in Section \ref{clustering}. All 14 topics are available in the Appendix \ref{Appendix:Topics_literature}. The colors of already mentioned categories correspond to the colors in the Appendix section \ref{Appendix:Topics_literature}. We observe in Figure \ref{fig:Literature_scatter} that the topics are very diverse, yet inter-related, and express different distances to each other. Whilst there are no dominating topics present, we observe that some of the clusters are relatively similar in the size: e.g., \textcolor[HTML]{E81B53}{Payments/Audit}, \textcolor[HTML]{3a923a}{Vulnerabilities/Ponzi/Scam/Bugs}, \textcolor[HTML]{807E07}{BTC/Finance/ICO}, and \textcolor[HTML]{845b53}{Programming language/Solidity/Software Development}. The \textcolor[HTML]{3174a1}{Healthcare/Medicine} cluster is very isolated, which points out that these papers are rather unsimilar to all the others. As well as \textcolor[HTML]{E8BD1B}{Energy/Electricity/Gas} cluster seems to have its very own distinct features, as there are space gaps, which are meaningful in UMAP dimensionality reduction (see section \ref{clustering} and \ref{Appendix:umap}), between this cluster and the others. Interestingly, there are no clusters containing distinctly related machine learning key-words. Moreover, there are no clusters that would obviously outnumber all the others. This will be later of importance where we show that the existing applications are mostly dominated by the \textcolor[HTML]{807E07}{Finance/ICO}.\\

Hence, after this first step into understanding SCs and structuring outlets on SCs, we see that there are no dominating topics and that they are diverse. This goes hand in hand with the common understanding that SCs can be readily applied to many diverse use cases. Further evidence to that can be readily collected to a basal Google search query or by respective Google Search Query Trends. Therefore, one of the aims of our work is to show whether the use-cases promised in the media or research can be realized within the SC concept or otherwise that the SC environment has significant technical and legal constraints and, in the end, is used primarily for one purpose -- as a financial instrument. This fact would make BCs using the SC concept not significantly different enough from other CCs.

\begin{figure}[H]
\centering
\includegraphics[width=10cm]{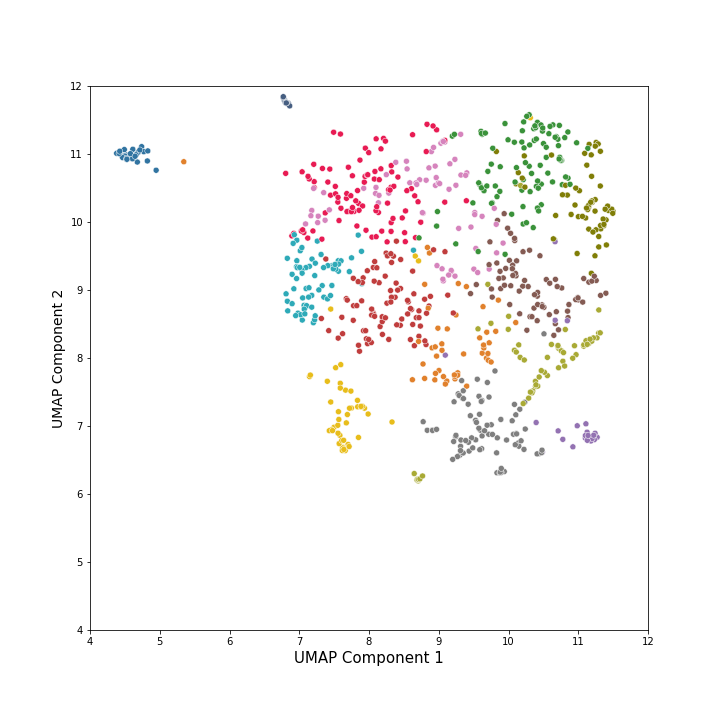}
\caption{Grouping the existing research into 14 topics via UMAP \href{https://github.com/QuantLet/USC/tree/master/SC-literature-research}{\includegraphics[keepaspectratio,width=0.4cm]{media/qletlogo_tr.png}}}\label{fig:Literature_scatter}
\end{figure}

%%%%%%%%%%%%NEXT%%%%%%%%%%%%%%

\section{Labore et Scientia}
\label{Sec:FrameworkInfo}

%The intermarriage of human nature with scientific or science-fictious ideas - may it be Wilhelm Schickard's $Rechenuhr$, Fritz Lang's $Metropolis$, or Elizabeth Holmes' $Theranos$ - has always fascinated the public and left the gates open for many venues of evolution, with the ultimate goal always being to eliminate the inherent risks that the human species poses to anything around and among itself.\\

Governments, companies, clients, humans, in general, are making manifold kinds of transactions and agreements each day. Most are dependent on government-backed third-party entities to create \textit{trust} in the given system. These highly centralized systems inject costs and risks into otherwise simple processes by being dependent on a stable workforce and intermediaries in order to keep this organically grown and comparably slow traditional \textit{trust machine} running. This machinery is confronted by a rapidly increasing load of information and a plethora of interdependent data streams with increasing digitization. Smart phones and smart watches, for example, provide us with instant access to the whole of available knowledge of humankind, may it be location tracking or health data -- yet, whilst the entry to knowledge may be easy, the real great leap forward is to understand and correctly employ this knowledge. Putting ``smart" into the name of any given product does not make it \textit{smart} or even an \textit{intelligent} idea to begin with. However, the idea of SCs as a tool to increase efficacy and objective decentralized controlling mechanisms is an essential approach on minimizing the possibilities of uncertainty and risks originating from human interference and ``allow a quarrelsome species ill-suited to organizations larger than small tribes to work together on vast projects like manufacturing jumbo jets and running hospitals" \citep{Szabo:1997}.

%- like $SMALT$ (Smart Salt Shaker - to not e.g. overdo on salt ingestion) or $Digitsole$ $Smartshoe$ (Rechargable Intelligent Sneakers - to e.g. keep you even more healthy).

\blockquote{``A smart contract is a computerized transaction protocol that executes the terms of a contract. The general objectives of smart contract design are to satisfy common contractual conditions (such as payment terms, liens, confidentiality, and even enforcement), minimize exceptions both malicious and accidental, and minimize the need for trusted intermediaries. Related economic goals include lowering fraud loss, arbitration and enforcement costs, and other transaction costs."}
\vspace{-0.7cm}
\begin{displayquote}
\citet{Szabo:1994}
\end{displayquote}

Nick \cite{Szabo:1994} envisioned the now-famous concept of SCs to increase contractual efficiency and published it around fourteen years before the popularization of the BC technology through Bitcoin (BTC). Unfortunately, technical constraints forced this idea to become a more theoretical and abstract construction for even more complex use cases like the transfer of real estate ownership, shares in a company, or intellectual property. Szabo explained that a vending machine is the simplest form of an abstract SC: it is a technical device accessible to anybody who is eligible, designed to transfer ownership of a good when provided with a predefined input. Given the nature of a vending machine, it therefore also \textit{controls} and \textit{oversees} the actual value for the involved parties: the seller, by storing it, and the buyer, by granting access to it after corresponding valid input. It can hence also enforce the terms of the given agreement by either granting access to the requested value or asking for a fitting input -- otherwise, the request gets rejected. Envisioned input for SCs is anything from payments, votes, or any other condition that can be expressed by code \citep{Szabo:1997a}.

\subsection{Basic Understanding}
\label{Sec:UnderstandBC}

The previously abstract ideas on SCs were ready to be transformed into a basic proof-of-concept status with BC technology's emergence. BCs are technically speaking a sequential and shared database managed and controlled by consensus algorithms for internal consistency. They can be designed as \textit{permissionless/public} or \textit{permissioned/private} system constructions, i.e., are defined by their read/write  ``rights" to access the network \citep[see further][]{ucc:2020,MBL:2020}. ETH is a community-driven, open-source software platform that was the first of such BC systems to support SCs. Its Whitepaper was originally published in 2013 by Vitalik Buterin, before its public launch in 2015 \citep{ETH:WP}. The main elements of interest for ETH BCs, i.e., a chronological ledger of human-induced technological happenings, are peer-to-peer networks, consensus mechanisms, hash values, Merkle trees, and asymmetric key encryption. These are also some of the buzzwords employed by marketeers promoting endeavors based on ETH. They are repeated over and over again without explanation to confuse and subsequently impress possible investors -- we will touch on these in sections \ref{Sec:UnderstandETH} and \ref{Sec:UnderstandSC}, as well as appendix \ref{ECDSA}.\\

\begin{figure}[H]
	\centering
	\includegraphics[keepaspectratio,width=16cm]{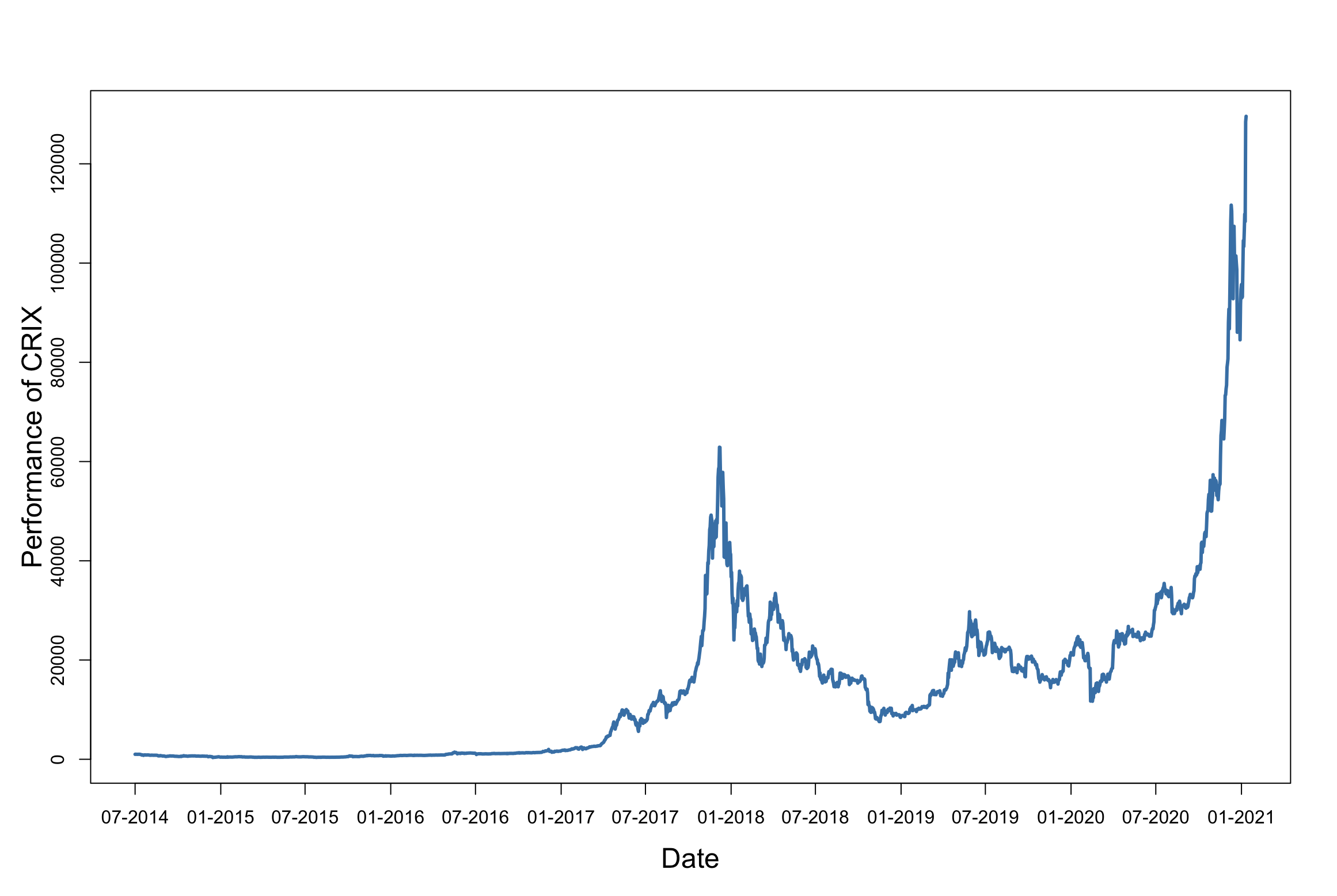}
        \caption{\href{https://thecrix.de}{CRyptocurrency IndeX CRIX \includegraphics[keepaspectratio,width=0.7cm]{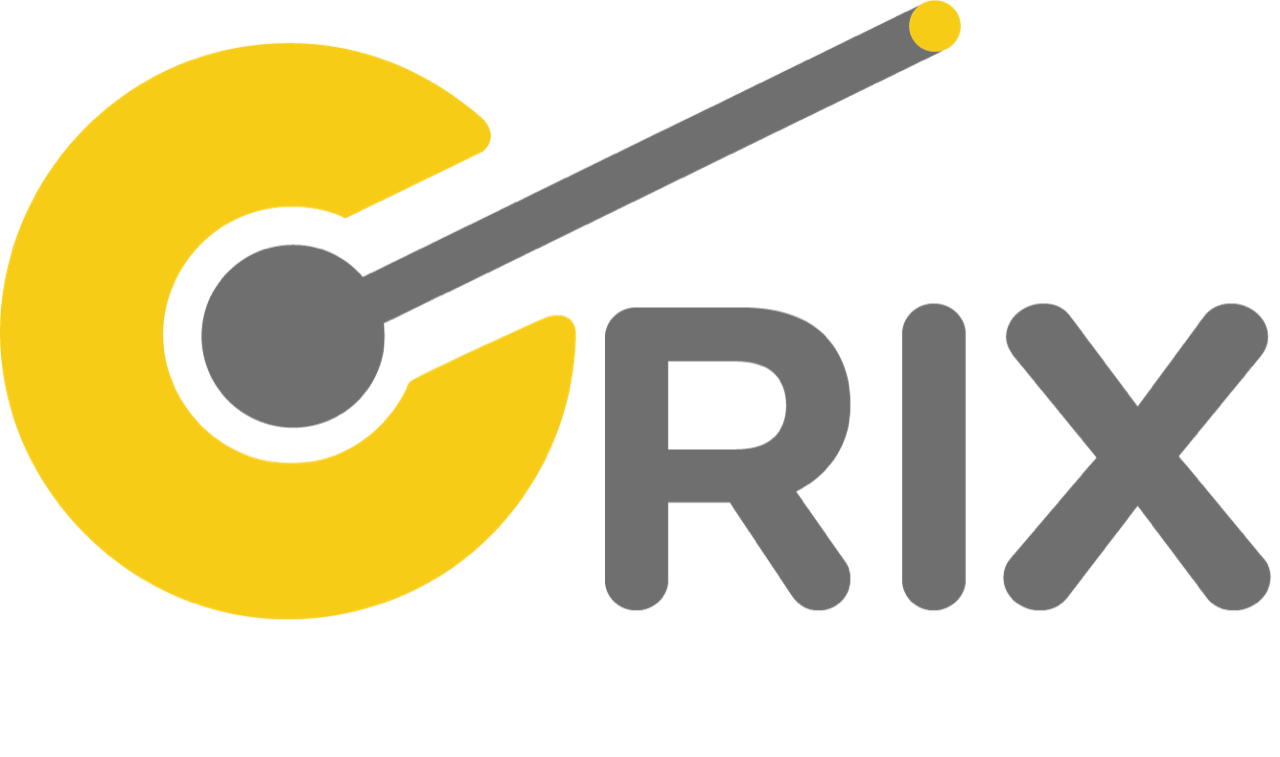}}, 20140731-20210211  \href{https://github.com/QuantLet/USC}{\includegraphics[keepaspectratio,width=0.4cm]{media/qletlogo_tr.png}}}
        \label{Fig:CRIX}
\end{figure}

\begin{figure}[H]
	\centering
	\includegraphics[keepaspectratio,width=16cm]{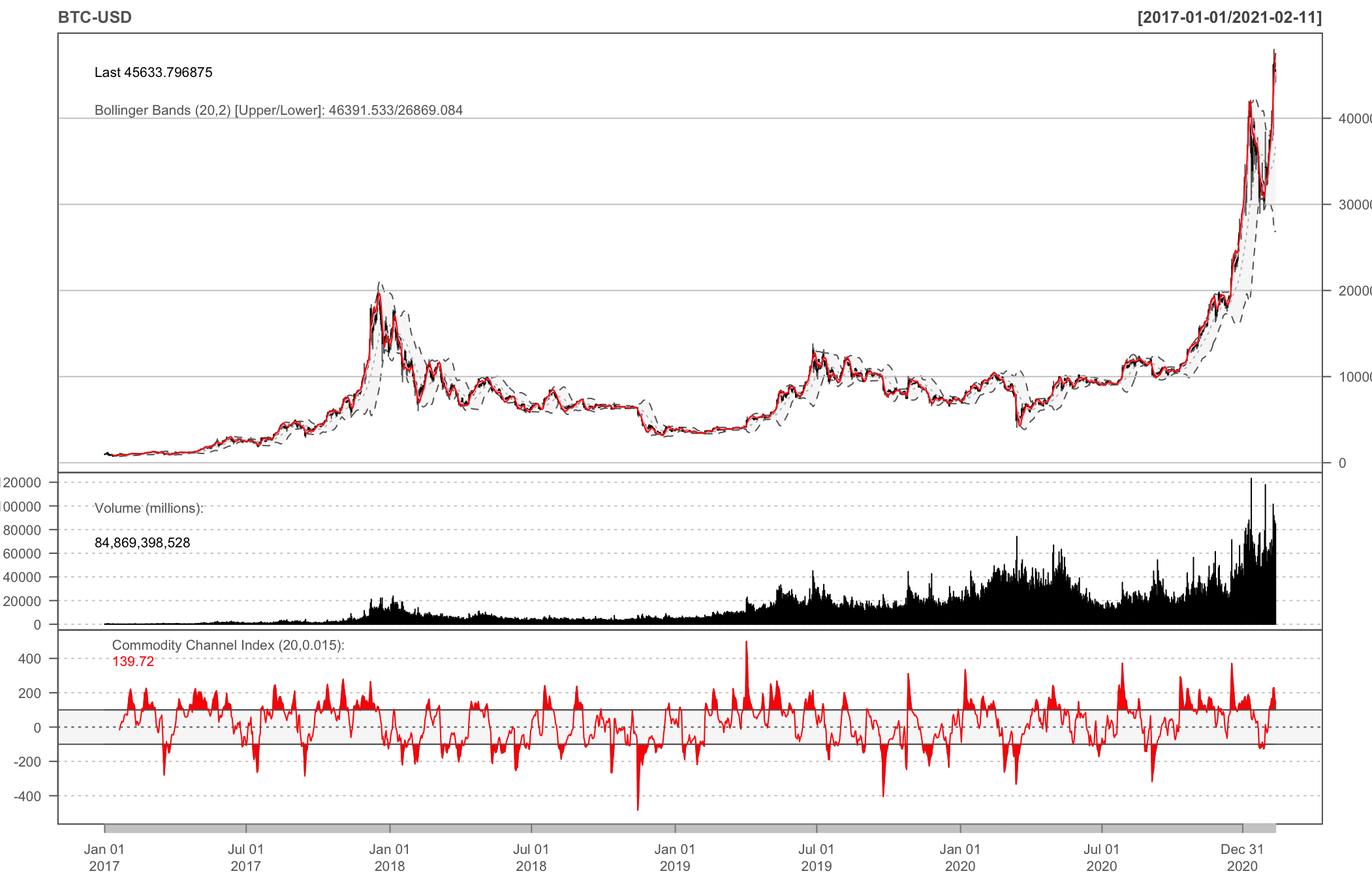}
        \caption{BTC-USD, 20170101 - 20210211 \href{https://github.com/QuantLet/USC}{\includegraphics[keepaspectratio,width=0.4cm]{media/qletlogo_tr.png}}}
        \label{Fig:BTC}
\end{figure}

\begin{figure}[H]
	\centering
	\includegraphics[keepaspectratio,width=16cm]{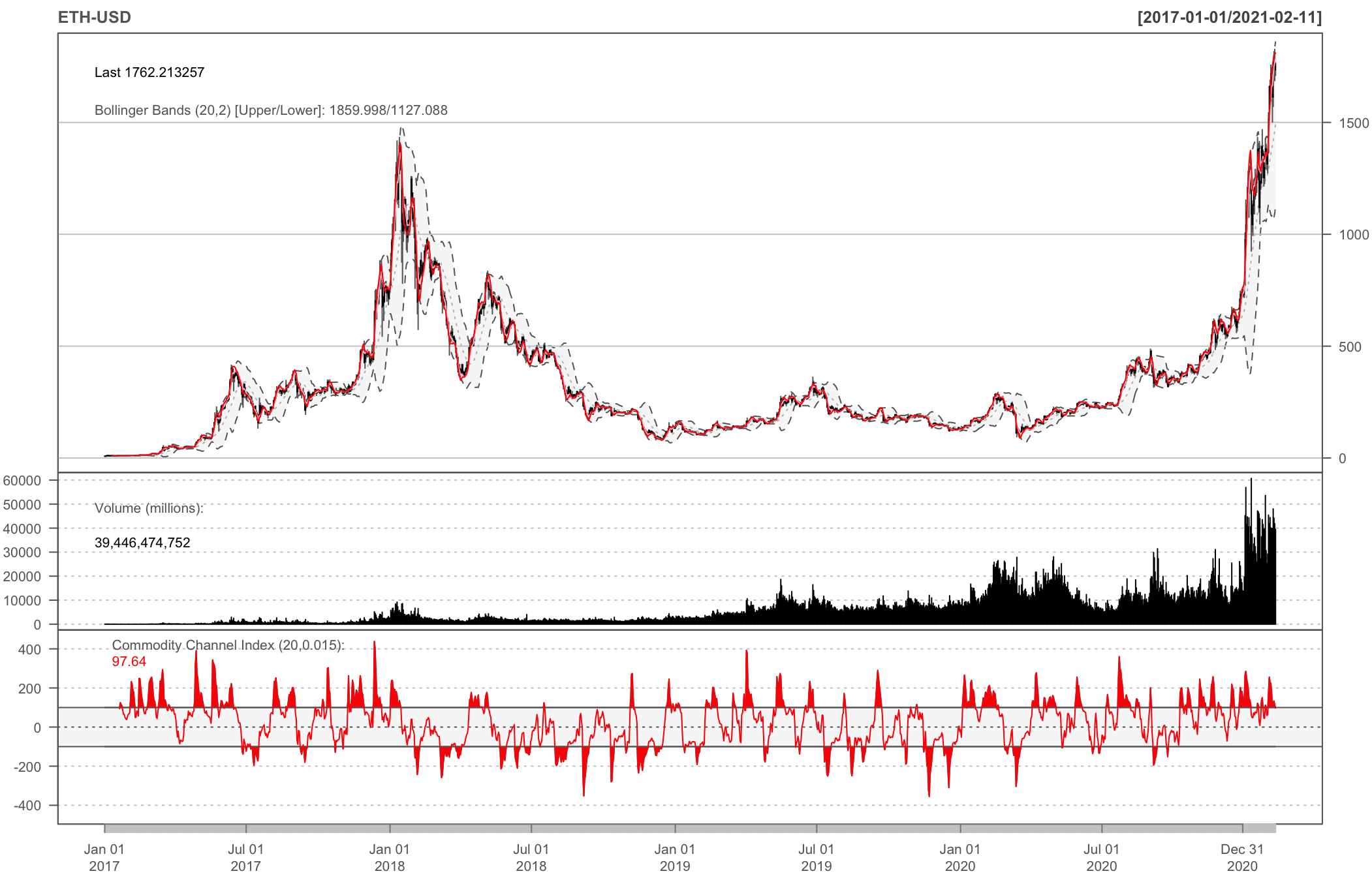}
        \caption{ETH-USD, 20170101 - 20210211 \href{https://github.com/QuantLet/USC}{\includegraphics[keepaspectratio,width=0.4cm]{media/qletlogo_tr.png}}}
        \label{Fig:ETH}
\end{figure}

One of the key-points of ETH BC is to be not only just a \textit{cryptocurrency} (CC) but also to allow a variety of opportunities through SC decentralized applications. But is it similar to other CCs? BTC is the de facto CC market driver (see also Figure \ref{Fig:CRIX}), we observe in Figures \ref{Fig:BTC} and \ref{Fig:ETH}, that BTC can be seen as a speculative and very volatile financial vehicle, while ETH seems to be historically more of a technological driver which not submitting to this BTC regime as extremely -- it is of course heavily influenced nevertheless \citep{PetTriHaeEle:2019,GK:2020,Rise:2020}.

% We contrast the \href{https://www.bitfinex.com/}{Bitfinex} USD exchange value of BTC to that of ETH in the following tables \ref{Tab:Num1}, \ref{Tab:Num2}, \ref{Tab:Num3}, and \ref{Tab:Num4} to visualize their trading behaviour after having been listed .

\vspace{-0.5cm}

%\vspace{-0.5cm}

%\vspace{-0.5cm}

\newpage

%In contrast to BTC, which is just a CC, \textit{Ether} is ETH's own CC yet can also be considered to be a \textit{token}. The difference between a token and a CC is that a CC runs independently by being mounted on its own construction like say BTC and can be considered to be a surrogate ``currency", whereas a token is an asset like a CC mounted on another pre-existing BC like for example so-called ``colored coins" that can also represent real-world assets.\\

Some media outlets and academic sources differentiate between two types of crypto assets: coins and tokens. One of the possible definitions is the following: a \textit{coin} is a CC that runs on its own BC, whereas a \textit{token} does not have their own BC and are run on another BC. BTC and \textit{Ether} -- the native coin of SCs and ETH BC are considered to be coins, as they run on their own BCs. Dai \citep{dai_coin} and  \citep{golem_coin} living on ETH are usually perceived as tokens. This difference may be important when different crypto assets are traded. While one can directly exchange tokens via internal applications, the coins, due to their non-standardized code protocols, can be exchanged only through external cryptocurrency exchange platforms \citep{token_coin}. The difference is less visible in all other cases, and quite often, these terms are used interchangeably, or the tokens are called coins. In the following, we will differentiate between coins and tokens. \\

One important type of crypto asset running on ETH BC is a \textit{stablecoin} -- a type of token, which price is designed to stay fix or, as the name says -- stable with respect to another external asset, like gold or U.S. dollar. It is designed this way to be robust against the volatility of the \textit{Ether's} price. One example of the \textit{stablecoin} is the \href{https://etherscan.io/token/0xa0b86991c6218b36c1d19d4a2e9eb0ce3606eb48}{USDC} \citep{USDC_coin}. The already mentioned Dai token \citep{dai_coin} is also a \textit{stablecoin}. \\

%Many tokens can therefore also commonly said to be CCs, yet not all CCs are tokens - in other words: BTC is a CC, Ether is a CC and token, while \href{https://etherscan.io/token/0xa0b86991c6218b36c1d19d4a2e9eb0ce3606eb48}{USDC} is a \textit{stablecoin} (i.e., it has a stable price through being ``pegged" to a fiat currency, i.e., this one is purported to maintain full value reserves of the equivalent fiat currency) on the ETH BC, while \href{https://www.bitfinex.com/wp-2019-05.pdf}{LEO} is commonly called a CC (it is actually \href{https://etherscan.io/token/0x2af5d2ad76741191d15dfe7bf6ac92d4bd912ca3}{just a token} on the ETH BC), amongst \href{https://coin360.com/?group=Ethereum}{other facettes} of that same play (see further section \ref{DatasetSection} and appendix \ref{Appendix:ListCC}). 

Within the ETH realm, the tokens must follow certain standards to comply with its coding codex for their inception via an SC. These standards are proposed as \href{https://eips.ethereum.org/}{\textit{Ethereum Improvement Proposals} (EIP's)} and \textit{Ethereum Request for Comments} (ERC) that need to be \href{https://github.com/ethereum/EIPs/issues}{peer-reviewed and accepted by the community}. ERCs are application-level standards for token standards, name registries, library/package formats, et cetera. Respectively for SCs, the \href{https://eips.ethereum.org/EIPS/eip-20}{ERC20 token standard} is a set of functions that have to be followed as a mandatory ``styleguide".\\

The majority of ETH tokens are ERC20 tokens and have become famous through their \href{https://cryptoradar.org/ico-calendar/?platform=ethereum
}{\textit{Initial Coin Offering}} (ICO, remotely similar to Initial Public Offering IPOs in the traditional field). Besides ICOs, ETH's SCs also enable and define constructions like \href{https://ethereum.org/en/DApps/
}{Distributed Applications} (DApps), yet they most often just plainly represent transaction flows (see sections \ref{sec:Classification} and \ref{clustering}). As said, Tokens are commonly emitted via ICOs, which can be seen as a crowd funding scheme -- in other words: a lot of the created CCs are just tokens based on the ETH network rather than having their own BC network construction. Tokens can be \textit{Usage Tokens} or \textit{Work Tokens}. Usage tokens act like Ether -- just within their respective DApp. For example, if one wants to use services of a \href{http://quantlet.com}{Quantlet App \includegraphics[keepaspectratio,width=0.3cm]{media/qletlogo_tr.png}}, then one needs to pay with the corresponding Quantlet coin (``coin", as it is indeed just a token in the ETH system as outlined). Work tokens can be seen as shareholder identificators or membership cards, which means that one can interact within the DApp through this token, for example, to vote on who becomes the speaker of a given group.

%ETH provides several standards to standardize tokens behaviors and their implementation in pegged applications like $wallets$ or $exchanges$.  The \href{https://ethereum.org/en/developers/docs/standards/tokens/erc-20/}{ERC-20 standard} is the most used token standard on ETH and defines standard interfaces and standard events that have to be followed for respective bug free processes.

%the standard method transfer is declared as “function transfer (address to, uint256 value) public returns (bool success), which is used to transfer a number of tokens to address to. The function should fire the TRANSFER event to inform whether the tokens are transferred successfully. The function also should throw an exception if the message callers account balance does not have enough tokens to spend

\vspace{-0.5cm}

\begin{figure}[H]
	\hspace{-0.6cm}
	\includegraphics[keepaspectratio,width=17cm]{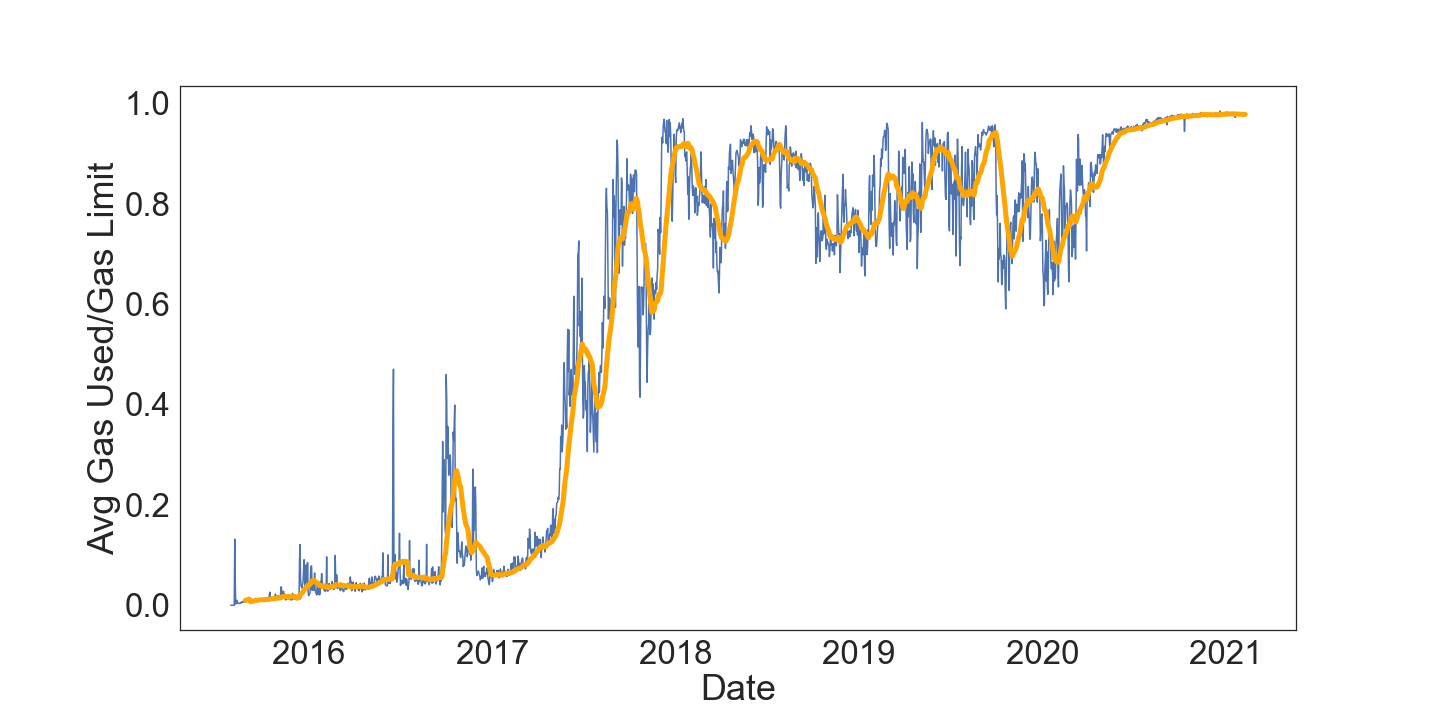}
        \caption{ETH Network Utilization (\textbf{\textcolor{orange}{Mean}}), 20150630-20210211
		\href{https://github.com/QuantLet/USC}{\includegraphics[keepaspectratio,width=0.4cm]{media/qletlogo_tr.png}}}
        \label{Fig:NetUtiliz}
\end{figure}

For purposes of understanding the term ``Smart Contract" -- as Hanna \cite{BoC18} with the Bank of Canada very critically outlines -- we agree that the concept and actual modus operandi for the absolute majority of BC-based SCs is not new. Automated recurring payments between entities are an example of how SCs effectively define themselves nowadays. However, the elemental misunderstanding is with the idea of BCs. BCs are indeed not needed to gain the benefits from SCs when leaving the crucial aspect of immutability/non-modifiability and the utilization of BC technology out, i.e., by strictly following \cite{Szabo:1994}'s initial definition of \textit{unbiased controlling}. In advance of section \ref{DefiningSC}, whereas \cite{Szabo:1994} defines SCs as a ``computerized transaction protocol that executes the terms of a contract" -- hence every self-executing agreement like, for example, related to a vending machine, an ATM, or an AppStore process -- \cite{ETH:WP} eventually defines them as any sort of script running on the ETH BC that follow certain defined standards for specified SC fields of employment \citep{ERC:20}. Of course, there are now plenty of other system constructions also supporting their individual kind of technological application for their individual purposes, yet we will only specifically look at ETH SCs as it has proven to be a stable system that has gained great market acceptance and steady network utilization (see further section \ref{DatasetSection}, as well as Figures \ref{Fig:NetUtiliz}, \ref{Fig:SC_New_Total_Time}, and \ref{Fig:Mining}). ETH SCs are therefore realizing Szabo's idea of \textit{control} by constructing BC-based, downtime-resistant, fraud-proof, and network audited systems. These operate without interference from a third party, by storing, verifying, and enforcing agreements automatically, therefore making human interaction redundant, hence aim to reduce process time and costs, while also improving transparency and trust.

%\textcolor{red}{""""BOC"""" A more careful look into the technology reveals that most of the proposed benefits of so-called blockchain technologies do not actually come from blockchain. What gets bundled up as blockchain technologies—smart contracts, encryption and a distributed ledger—are separate concepts. The three may be implemented together, but they do not need to be. We analyze them separately and argue that most of the proposed benefits come from encryption and smart contracts. But encryption and smart contracts do not need blockchain.}\\

%So, can this product called $Smart$ Contracts even be considered to be $smart$, or is it just another $snake$ $oil$ product?\\

%%%%%%%%%%%%NEXT%%%%%%%%%%%%%%

%%%%%%%%%%%%NEXT%%%%%%%%%%%%%%

%%%%%%%%%%%%NEXT%%%%%%%%%%%%%%

%%%%%%%%%%%%NEXT%%%%%%%%%%%%%%

%%% NEXT SUB %%%

\subsection{Understanding ETH}
\label{Sec:UnderstandETH}

ETH runs a so-called \textit{account-based} BC, which means that these are the keypoints and central issue we need to look at after understanding the basic functional framework. An ETH BC block header contains, besides the hash of the previous block and \href{https://ethereum-classic-guide.readthedocs.io/en/latest/docs/world_database/blocks.html}{other relevant information}, three so-called \textit{Modified Merkle Patricia tree}'s (tree also subsidized as \textit{trie}) for its \textit{state} system: state root (system state: account balances, contract storages, contract codes, and account nonces), \textit{transactions} (TXs), and receipts (the outcome of individual TXs). These touch the overall state, i.e., information linked to blocks being added sequentially to the BC, as well as the individual state of certain entities, like some balance of value stored on an entity such as a given account.\\

\begin{figure}[H]
	\centering
	\includegraphics[keepaspectratio,width=15cm]{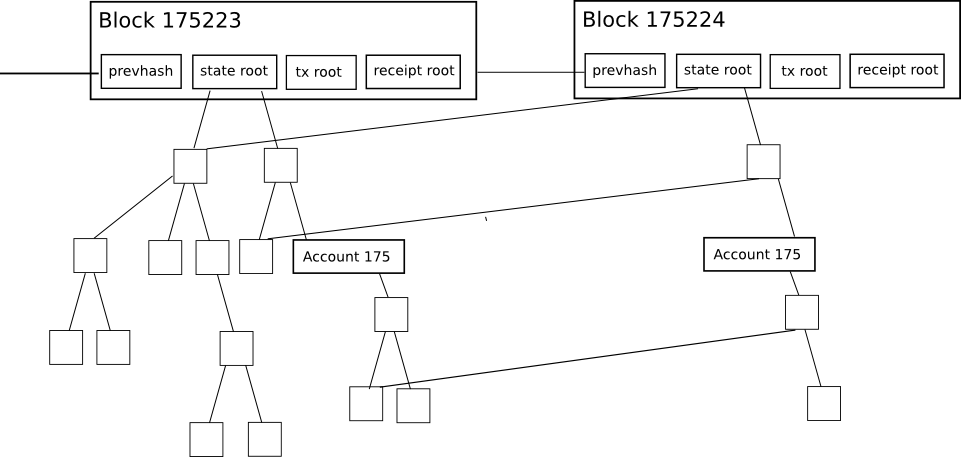}
        \caption{ETH BC Visualization \citep[adapted from][]{MerklingETH}}
        \label{Fig:ButerinETH_BC}
\end{figure}

In a Merkle tree parent/non-leaf nodes contain the hash of their children/leaf nodes, and the child nodes contain the hash of a block of data. Therefore, any change to the underlying data, i.e., the state, causes also the hash of the respective reference node to change. Since each parent node hash depends on the data of its children, any change to the data of a child node causes the parent hash to change. This would of course trigger a reaction up to the root node hash, but only the changed hashes will be propagated to a later block with respective parents of the later block - previous information contained in a already created block is unchanged. This means, that one does not need to compare all data across the network of leaf nodes for verification, as the root node hash provides a quick answer to that when looking up individual \textit{addresses}. In return, confirming that some specific small amount of information belongs to the whole tree in question, is done easily via a so-called \textit{Merkle Proof}. Only the minimum amount of nodes are modified to recalculate the root hash for a single operation as visualized in Figure \ref{Fig:ButerinETH_BC}.

This creates systemic benefits as it is possible to store only a hash of the root node to represent the data at that point in time whilst keeping the data immutable on a previous block - see further \citep[Appendix D in][]{YellowPaper}, the \href{https://eth.wiki/en/fundamentals/patricia-tree}{Ethereum Wiki}, and  \citet{MerklingETH}. As this is mandatory understanding for the following sections: the state contains every account information present in the BC, however, the state itself is not stored in each block of the BC. It is, as outlined in \citet{YellowPaper}, generated by processing each block since the genesis block, that is the initial/inception block of the chain of blocks, whereas each individual block will only modify parts of the state, but not the whole state. State trie information therefore exists independently of a referenced state root.\\

Therefore Figure \ref{Fig:ButerinETH_BC} presents that the state root in block 175223 (left) refers to two child nodes, and that the state root block 175224 (right) refers to one of the same child nodes as block 175223 with one new child node not referenced by the prior state root. Given that 175224 is the last block, the previous state trie information of block 175223 becomes ``irrelevant", when referenced in the state root of block 175224. This means, that only altered information is connected to the state root of a respective block containing this information and linking it to previous blocks. As each hash is unique, this means, that an imaginary block 175230 can also refer to information stored in the state root of block 175223, given that this old information is updated with new input - i.e., the overall size of data transferred can be kept low without loosing information.\\

Two aspects are important for understanding this system: Firstly, adding subsequential blocks to the ETH BC that do not contain previous information, but only new or updated information, does not mean, that previous entries are forgotten/deleted/redundant. In contrary, each unique information has a corresponding unique hash, that forms a \textit{hashmap}. The state root of block 175224 still references to previous entries present ad infinitum within the BC, hence allowing to reference values existant in the state trie of block 175223. Secondly, and this is important for the upcoming exegesis on accounts and TX, this \textit{hashmap} creates a reference point for interaction on the BC. However, this referencing might be missing previous points contained in block 175223 or earlier, if one only loads data starting at block 175224, which might be referencing to information of block 175223. Hence that hashmap might miss information referenced in the state root of block 175223, for example, if an account has sufficient funds. Remember, that the whole state is not stored, but only the root hash. Conveniently, the ETH network protocol solves this by requesting missing information from peers - recall the structure and modus operandi of a BC. Even with only recent information at hand, i.e., missing the non-altered information of previous blocks, peers commit missing values needed.\\

%Lets say 175224 is having the transaction (A->B 50 eth). So it means if I refer to block 175224 then based on state root hash I can find Account A's state is 50 eth and B's account state is 150 eth (as B's initial state was 100). 4. If I refer to 175223 having a transaction where A's total is becoming 100 eth then based on state root hash I can find Account A's state is 100 eth

%https://ethereum.stackexchange.com/questions/15288/ethereum-merkle-tree-explanation/15294

%Only the state tree is required. The transaction and receipts trees are to build from the block data. It is only used to validate the block. It is necesary to store them, because they will not be used again. 2. Yes, if your node store the whole state history you could retrieve the state in any point in the history. But some nodes can prune old history. 4) y 5) The state will only show the final balance. For block 175223 A: 100, B: 100 and for block 175224 A: 50, B: 150.

\newpage

With a basic understanding of the ETH BC at hand, we can now proceed to dissect which actors with which corresponding acts can be observed in this system. A TX can be described as simply sending some value, like say 1 Ether, from Alice to Bob, i.e., from one \textit{address} to another, usually by using a \textit{wallet}. That said value, however, is not in a ``wallet", but on the ETH BC linked to an \textit{account}. TXs themselves do not need to be financial operations, but can also initiate systemic changes like creating an account. Therefore we first need to understand what accounts are, how they act and can be interacted with, and how these are uniquely identifiable. There a \href{https://ethdocs.org/en/latest/contracts-and-transactions/account-types-gas-and-transactions.html#eoa-vs-contract-accounts}{two distinct classes of accounts} in ETH: \textit{Externally} \textit{Owned Accounts} (EOA) and \textit{Contract Accounts} (CA, often just called \textit{contracts}, hence unanimously used in relation to SCs).\\

An EOA is an entity that can be accessed and can validate TX through the individual secret \textit{private key} (see further appendix \ref{ECDSA}). It can consequently be identified by the extracted \textit{public key} and the complementary \textit{public address}. The public address can also be derived from respective in-/outgoing TXs connected to, for example, a given EOA. Moreover, it serves to store value like Ether and controls applications like wallets. One private key can generate multiple public keys and multiple other ``private keys" for certain applications (these are in fact not the real private key, but just private keys for an individual application, hence public keys that are kept secret within that system which then create further public keys in a mimikry fashion).\\

A wallet is a software client that allows to manage EOAs. One or multiple EOAs can be accessed via different wallets and the majority of \href{https://cointelegraph.com/ethereum-for-beginners/ethereum-wallets}{wallet offerings} can generate EOAs upon request. Wallets can be linked to further services like \href{https://coinmarketcap.com/rankings/exchanges/}{CC exchanges}. Without going into further details, there are different kinds of wallets, like desktop, web, physical, and mobile app based ones, which allow features such as TX logging, multisignature, withdrawal limits, and many other functions. Wallets are therefore user interfaces, which are controlled via code and authenticated by an EOA's private key.\\

CAs, on the contrary to EOAs, do not possess a corresponding private key and are controlled by the inhibited information stored within: an SC with respectively coded functionalities. Individual CAs are generated, when an SC is \textit{deployed} with a TX -- hence these two terms, CA \& SC, are often used synonymously -- and which is linked to the deploying EOA through means of \textit{public key cryptography/asymmetric cryptography} and its \textit{nonce} -- and hold respective code and/or data (see above and section \ref{Sec:UnderstandSC}). The state refers to information about individual accounts and consists of the accounts' balance (in Wei, see further \ref{Appendix:Wei}), the nonce (the sequential number of TX sent from an EOA, which includes the number of deployed SCs/created CAs made by an EOA/CA, as they are TX as well), the storageroot (the root node hash of the account storage trie), and the codeHash (hash of the compiled SC, the  \textit{Ethereum  Virtual Machine} (EVM) \textit{bytecode}).  Note: the nonce -- differently to the \textit{Proof-of-Work} nonce, which is a random number -- is simply the sending accounts' sequential TX count and prevents \textit{double spending}. CAs have a balance, a nonce, bytecode, and the root hash of a storage trees (the beforehand explained Merkle trees for TXs, receipts, and state). In contrast, EOAs do not host an SC, hence the bytecode and storage hash are empty. For CAs, the bytecode is the SCs code and the storage hash is the merkle root hash of all the key-value pairs in the CAs state. Only the codeHash represents an immutable field, as it contains the bytecode of the SC deployed to the BC -- hence, an error in the code will exist persistently. The other fields are mutable, like the nonce, which increments when a TX created by this account.\\

%Creating the keys (for our wallet) we generate a 256-bit private key, and then the public key is a point on the secp256k1 ECDSA curve (x,y point). This key is then hashed using Keccak-256 (aka SHA-3), and the lower 160 bits becomes the public Ethereum address.

There are two different kinds of interactions in ETH: TXs and \textit{message call}s (MGSs). TXs are information packages, like debit or credit instructions of a given value, that are authorized by the interactor/sender through the \textit{individual} TX \textit{signature}. The authorizing signature is mathematically generated through hashing the respective information, like the above example of sending value from one address to another together with the private key by using the Keccak-256 algorithm -- the result is the signed and hence authorized information. A MSG is equivalent to a TX, being only a local invocation of an SC function that does not broadcast or publish anything on the BC. They contain the source and target addresses, some data payload, some value like Ether, Gas and return data.  A simple TX between two EOAs or an EOA to a CA is therefore straightforward with only three main parameters required: the interactor/sender account address, recipient account address, and the TX value. Parameters like the nonce, the \textit{gasLimit}, and the \textit{gasPrice} will be input automatically, for example, by using the \href{https://geth.ethereum.org/docs/interface/javascript-console}{Geth console} (gasLimit and gasPrice will be touched on later in section \ref{Sec:UnderstandSC}). Further parameters are related to ECDSA and accordingly used to sign a TX. The parameters \textit{data} and \textit{init} refer to SCs. Data is relevant for value transfer and sending a MSG to an SC, like altering a specified value. The init parameter is relevant for, as implied, CA creation, i.e., the bytecode utilized for initialising one.\\

MSGs are created, for example, when a CA executes certain SC \textit{opcodes}, or when interacting through an EOA with an SC to check on a certain variable saved in the CA's state. CAs can communicate with other CAs through this functionality given respective coding (see further section \ref{Sec:UnderstandSC}). MSGs are also called \textit{internal transactions} -- a very perplexing misnomer that may not be confused with TX, as they are not recorded on the BC, but can be derived from other ETH system functionalities and are actually reflected in the overall balance of the individual account over time, as they also consume \textit{Gas} (which is \href{https://etherscan.io/gastracker}{a fraction of Ether} -- more on that later and in section \ref{Appendix:Wei}). MSGs are, for example, the result of an SC initiating a value transfer to an EOA -- like sending a Token after a valid and accepted TX has been received from an EOA for an ICO -- or of calling another CA/SC to, for example, check their funds. Hence, CAs can only MSG an EOA to validate/execute a then \textit{recorded} TX (that means publicly displayed on the BC). A MSG doesn't need to be signed, whereas the called account automatically has access to the identity of the calling account, which is most basically refered to initially in a variable called msg.sender. This variable information is stored in the constructor, which is a non-callable/MSGable function that is executed during the creation of a CA/deployment of an SC and permanently stores the address of the entity creating the CA/SC in the beginning. Note, that the actual ownership of a CA can be altered afterwards, given the ownership transfer function is coded in the SC. In other words: In the ICO example, one first needs to send some value, like Ether, to the CA that hosts the ICO SC in order to receive a defined value like a respective token in return through a MSG from the CA, or through the CA calling a EOA, which then transfers value to the initially interacting account through a BC-recorded TX or non-BC-recorded MSG after release.\\

Therefore the term ``self-execution" of SCs, as mentioned in the beginning, is wrong. Every chain of motion is triggered originally by an non-CA input. Whilst CAs are only able to send MSGs (which EOAs can as well) and transfer value through these MSG's respective to coded functions therewithin to other accounts.  Importantly, TX can not be created by SCs (only ``internal TX", that are MSG). TX cause state changes by either a simple transfer of value between accounts (which changes their respective balance), similarly by sending a MSG to a CA (for example to alter an SC value, which can be thought of like altering the balance of an account), or by deploying an SC and herewith creating a CA.\\

To summarize: EOAs are controllable via private keys, while CAs are controlled via the individual SCs bytecode (see further section \ref{Sec:UnderstandSC}). EOAs have private keys and are controlled by, for example, wetware through using software like wallets. Private keys can be used to create, through the trapdoor function ECDSA, multiple public keys that uniquely form derived addresses. Therefore, one EOA on the ETH BC can have multiple CAs linked through deploying SCs. Only EOAs can sign and hence send TXs, which are recorded on the BC, whereas a MSG are not recorded on the BC. TX/MSG's can be sent to a CA in which the respective SCs functions must run accordingly. Finally, only the most crucial information is stored on the BC like changes in account balance but not, for example, MSG's to check on a certain accounts' balance (these information can however be obtained through different channels).

%Once again, CA's are completely dependent on external EOA input to create TX, which are recorded on the BC, due to their respective lack of a private key. 

%SC reacts to TX, cannot TX itself. MSG = internal TX // send ETH to SC receive token https://etherscan.io/address/0xa33e729bf4fdeb868b534e1f20523463d9c46bee\\

%https://etherscan.io/address/0x7a41e0517a5eca4fdbc7fbeba4d4c47b9ff6dc63#code

%%%%%%%%%%%%NEXT%%%%%%%%%%%%%%

%%%%%%%%%%%%NEXT%%%%%%%%%%%%%%

\subsection{Understanding SCs}
\label{Sec:UnderstandSC}

Occasionally, SCs are seen as so-called \textit{programmatically executed transactions} (PETs),  i.e., technology for enforcing agreements. Simply put we can state that the execution of an SC just replaces third party trust with mathematical proof, that some conditions were fulfilled -- nothing more, nothing less (given we are not dealing with nefarious external input). SCs can accordingly be seen as an enhanced \textit{zero-knowledge protocol} or \textit{proof} (see further appendix \ref{Appendix:ZeroKnowledgeProof}). It only returns, that the predetermined conditions are fulfilled or non-fulfilled (\textit{completeness}), that the conditions are clearly defined if conditions are not-fulfilled (\textit{soundness}), and that nothing else was disclosed but the fact that the conditions are met by successfully terminating the process (\textit{zero-knowledge}), and eventually it terminates the process with a given outcome like granting access to a motel room.\\

% \textcolor{red}{Rental raus oder kritisch\\
% https://www.schoenherr.eu/publications/publication-detail/ricardian-contracts-a-smarter-way-to-do-smart-contracts}\\

Creating agreements for a variety of codeable use cases as SCs, besides the above mentioned financial vehicles, has therefore several benefits:

\begin{itemize}[leftmargin=*]
	\item[$\boxdot$] \textbf{Standardization}: Code, following systemic standards, modeled after traditional contractual counterparts, and proven by best practices, combined with a deterministic execution dependent on clearly pre-defined parameters, reduces risks and costs, while saving valuable time.
\vspace{-0.3cm}
	\item[$\boxdot$] \textbf{Automation}: Reduced procedural friction, by removing intermediaries, and increased transaction flow based on a BC or associated consensus system.
 \vspace{-0.3cm}
	\item[$\boxdot$] \textbf{Certainty}: Code executes automatically as pre-defined, hence reduces risk (e.g. counterparty or settlement risk), and therefore creates trust in these deterministic digital processes.
 \vspace{-0.3cm}
	\item[$\boxdot$] \textbf{Security}: Transactions are encrypted and uniquely identifiable by their hash and stored on a BC or associated construction.
\vspace{-0.3cm}
	\item[$\boxdot$] \textbf{Innovation}: Modular and adaptable code that automatises the flow of digital assets and payments to incubate new business models. Coded regulatory compliance, for example data reporting, and specialized interdisciplinary regulatory BC-based nodes to comply with digital advancements.

\end{itemize}

To create a CA is therefore similar to how a self-extracting software installer works. During its execution, it configures the system environment and handles the information according to the given algorithm. In technical terms, SCs are account holding objects on the ETH BC that contain code with functions to store information or to trigger others. SC code is immutable once stored/deployed within the ETH BC and thus developing SCs is closer to hardware programming than say web development. CAs consist each of an individual SC code in a specific binary format called \textit{Ethereum Virtual Machine} (EVM) bytecode representing opcodes and further respective data related to its state (see further section \ref{Sec:UnderstandETH}). To deploy an SC, a CA needs to be created by sending a TX from an EOA to an empty address with the SC bytecode as data. CAs can also create other CAs, if the contained SC is respectively coded and the oricess in initiated by an authorized EOA -- usually these CA/SCs are called \textit{factory contracts}. This process can be done, for example, through the \href{https://remix.ethereum.org/}{Remix integrated development environment (IDE)}, which is the ETH-native software for building applications combining common developer tools and making it easily accessible through a comprehensive graphical user interface (GUI). The same process can be repeated ad infinitum resulting in CA ``clones", which just happen to have unique addresses as each EOA TX is created using a unique nonce (see section \ref{Sec:UnderstandETH}). Modifying data in any of these does not have an impact on the other clones that have been created unless coded to do so).\\

Most SCs are typically written in \textit{Solidity} before being compiled into bytecode, which expresses similarities found in object-oriented languages that include variables, arithmetic operations, functions, classes, string manipulations, and further programming elements. Access to this is therefore easy with some knowledge in Python, Perl, JavaScript, or C++. Other languages like \href{https://github.com/Betterpath/Bamboo}{Vyper} or \href{https://github.com/vyperlang/vyper}{Bamboo} can be used as well, but we will not touch further on these. The human-readable code in Solidity can not be executed directly but needs to be compiled to machine-readable opcodes/bytecode that is executable. These opcodes are allocated a byte each and create an efficient way of process control with, for example, ``stop" being plainly called with \verb=0x00=. Accordingly, these define the management of contextual procedures, like how to handle certain functions, variables, and parameters. The SC runtime environment -- that is embedded within ETH network nodes and responsible for executing contract bytecode -- is called the EVM (see section \ref{Sec:UnderstandETH}). The term Virtual Machine (VM) has a strong overtone, and to clarify: the EVM is not a VM -- it is an \textit{interpreter} for the \href{https://solidity.readthedocs.io/en/v0.4.24/assembly.html}{\textit{assembly language} defined through Solidity} (the beforehand mentioned bytecode), with every participating node running that interpreter. The actual hash of a given TX is not relevant for this operation until being recorded into the BCs block Merkle tree (see above section \ref{Sec:UnderstandETH}), but only the data, the amount of Gas available and used and potentially also the SC code that is being called if it is such an operation. The EVM handles the state transition function which is composed of updates to accounts (balances \& nonces), Gas (non-/used), bytecode execution, and mining (block creation rewards and block handling) -- this is explained in extenso in the Yellow Paper \citep{YellowPaper}. 
A buzzword often used to describe the EVM is \textit{Turing completeness}, which simply refers to a system capable of solving any computational problem given enough resources -- which in turn does practically not mean, that we can calculate the answer to everything (which is of course \href{https://www.scientificamerican.com/article/for-math-fans-a-hitchhikers-guide-to-the-number-42/}{``42"}). To make this more tangible: different from just the hosting of software -- like the beforehand used example of an AppStore -- ETH nodes not only process and manage but also execute these codes through their individual instance of the EVM, before they or their respective outcome are being propagated. As network participants predominantly care about gaining wealth, every node has to verify the results of individual computation thus benefiting fractionally from the respective fees for individual opcodes. Every function of SCs -- in general every interaction, successful or not successfully, like transferring Ether -- consumes computational power, which is monitored by the EVM and needs to be compensated through Gas to the participating nodes, i.e., the \textit{miners} \cite [see further][]{ucc:2020}.\\

\vspace{-0.5cm}

\begin{figure*}[!h]
\begin{multicols}{2}
    \includegraphics[width=\linewidth]{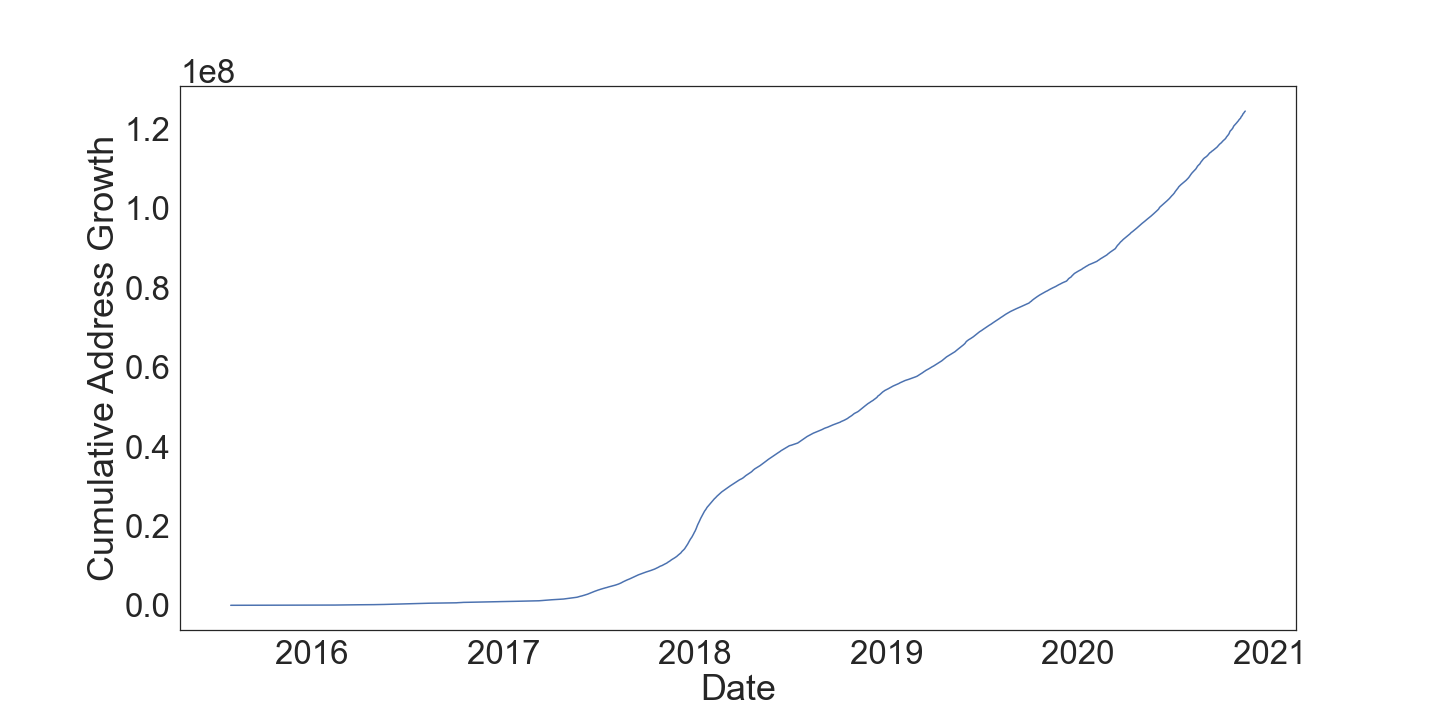}\par 
    \includegraphics[width=\linewidth]{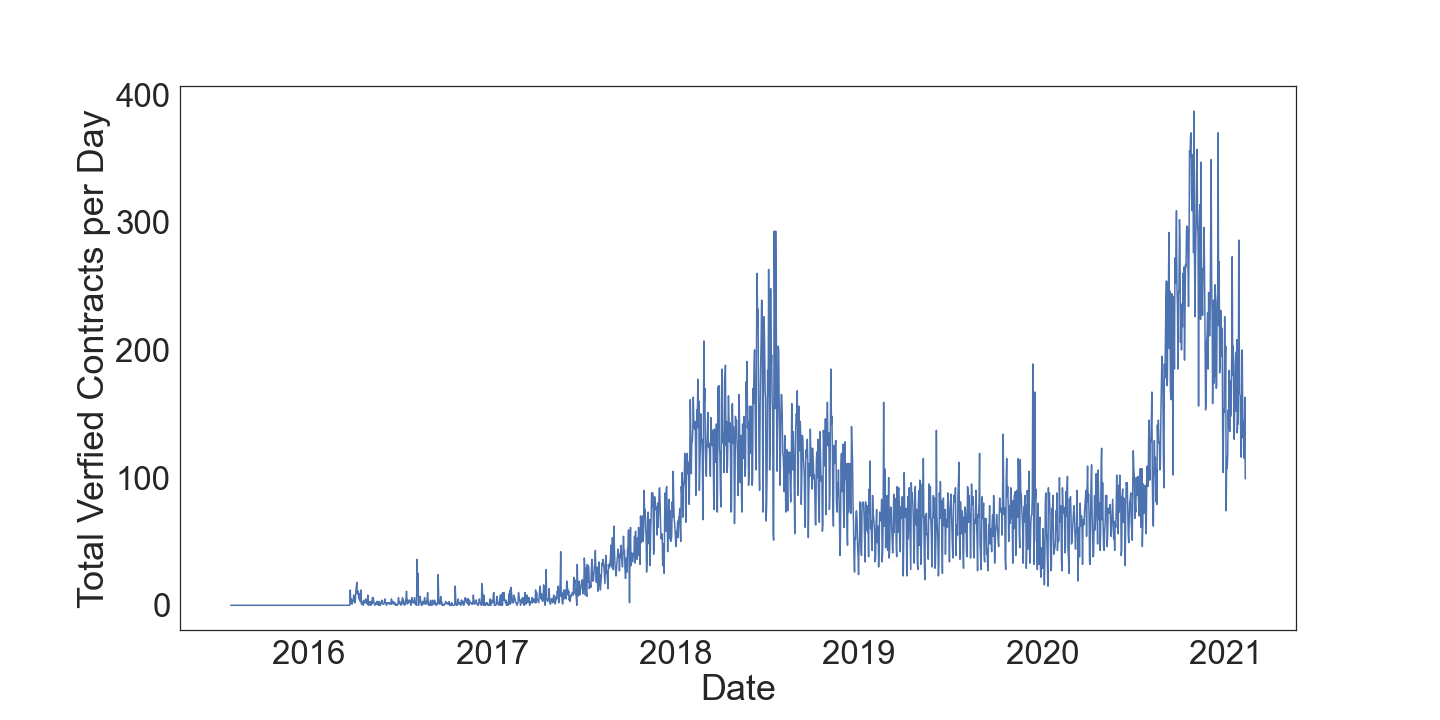}\par 
\end{multicols}
\begin{multicols}{2}
    \includegraphics[width=\linewidth]{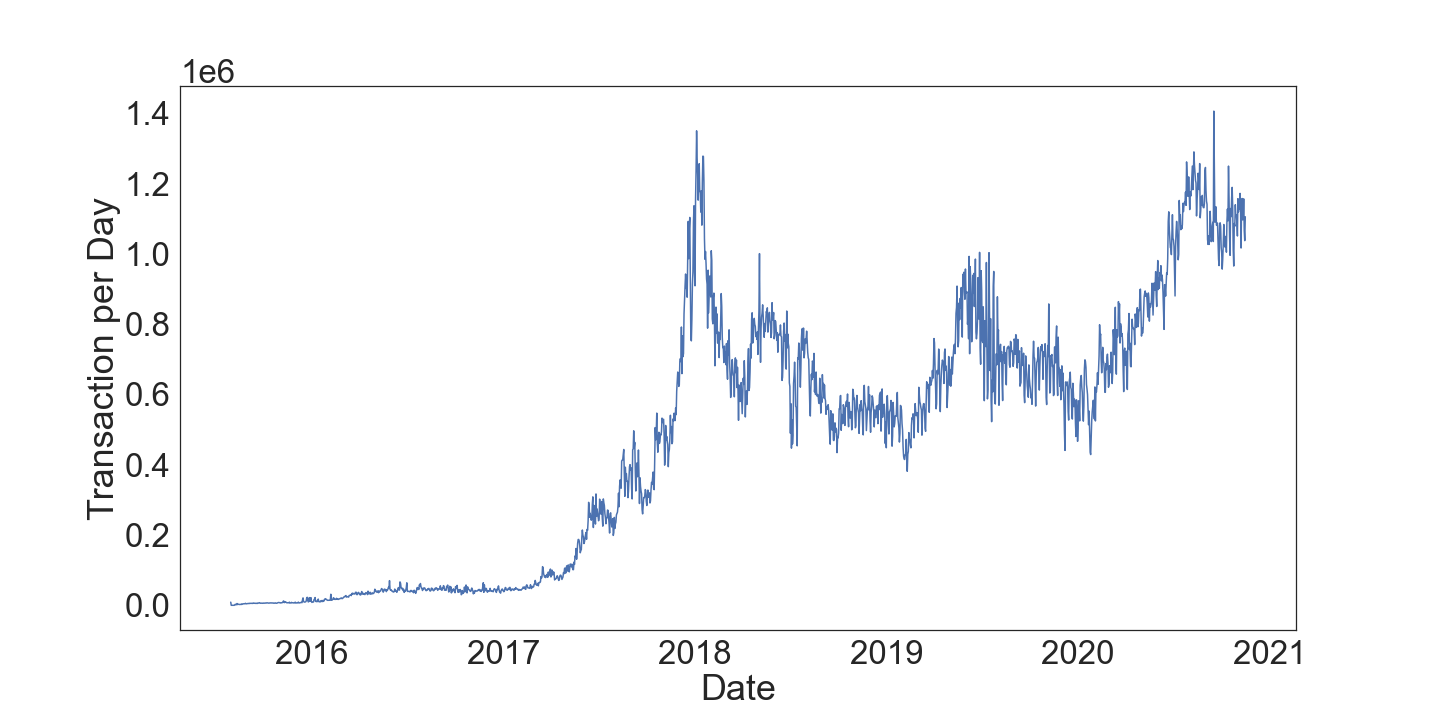}\par
    \includegraphics[width=\linewidth]{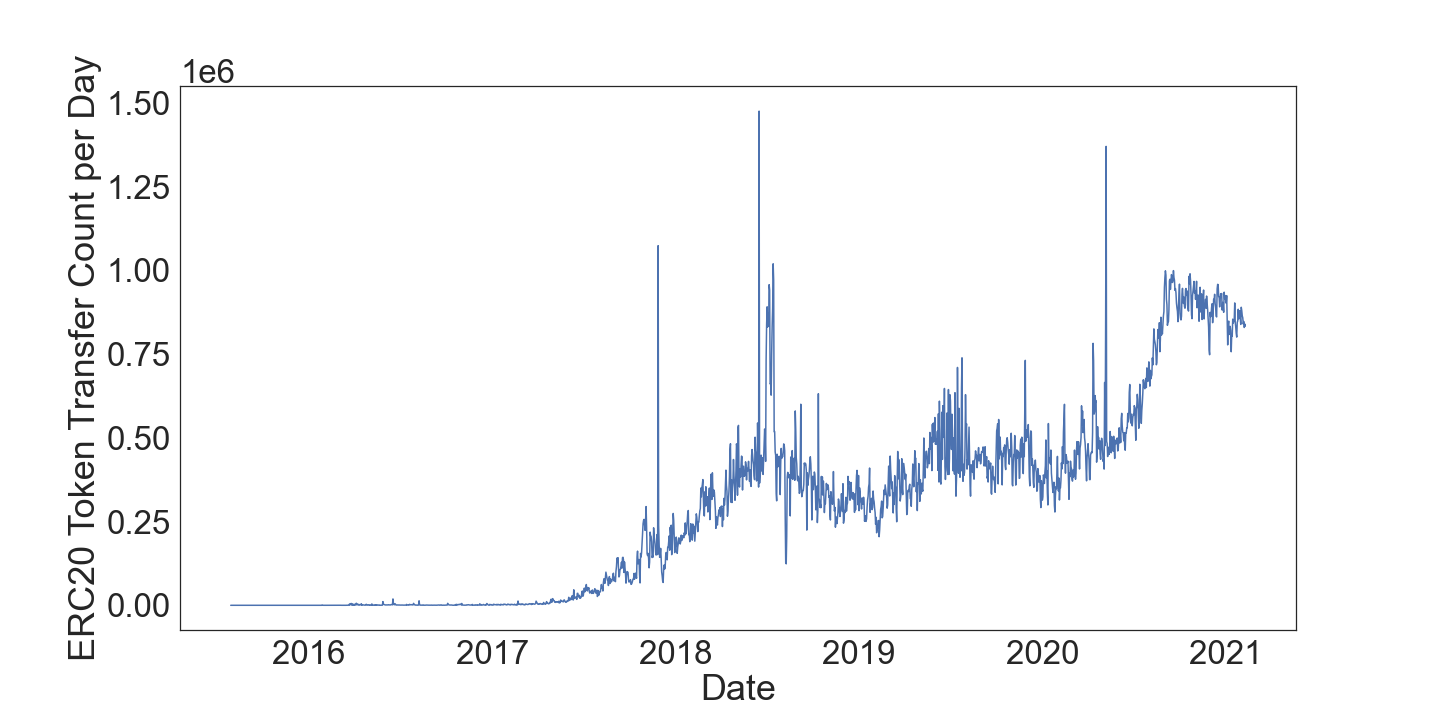}\par
\end{multicols}
\begin{multicols}{2}
    \includegraphics[width=\linewidth]{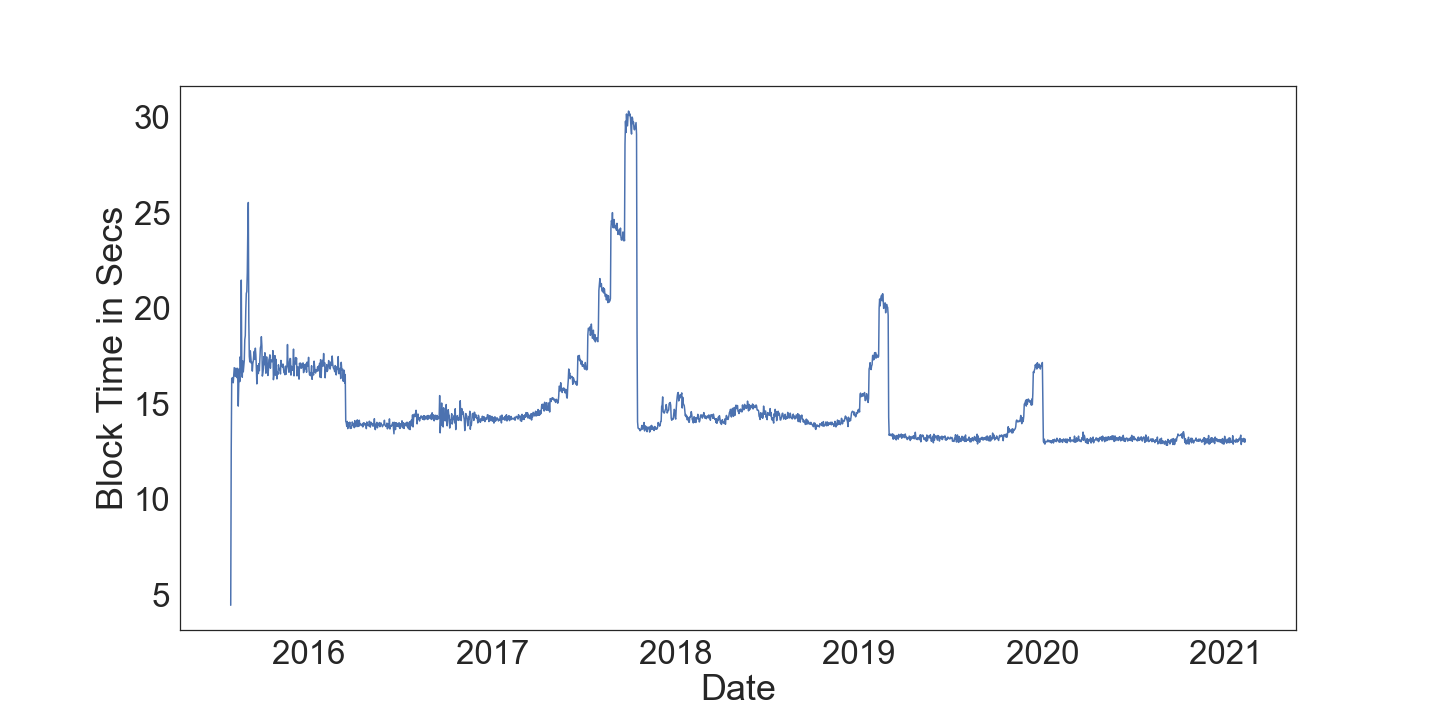}\par
    \includegraphics[width=\linewidth]{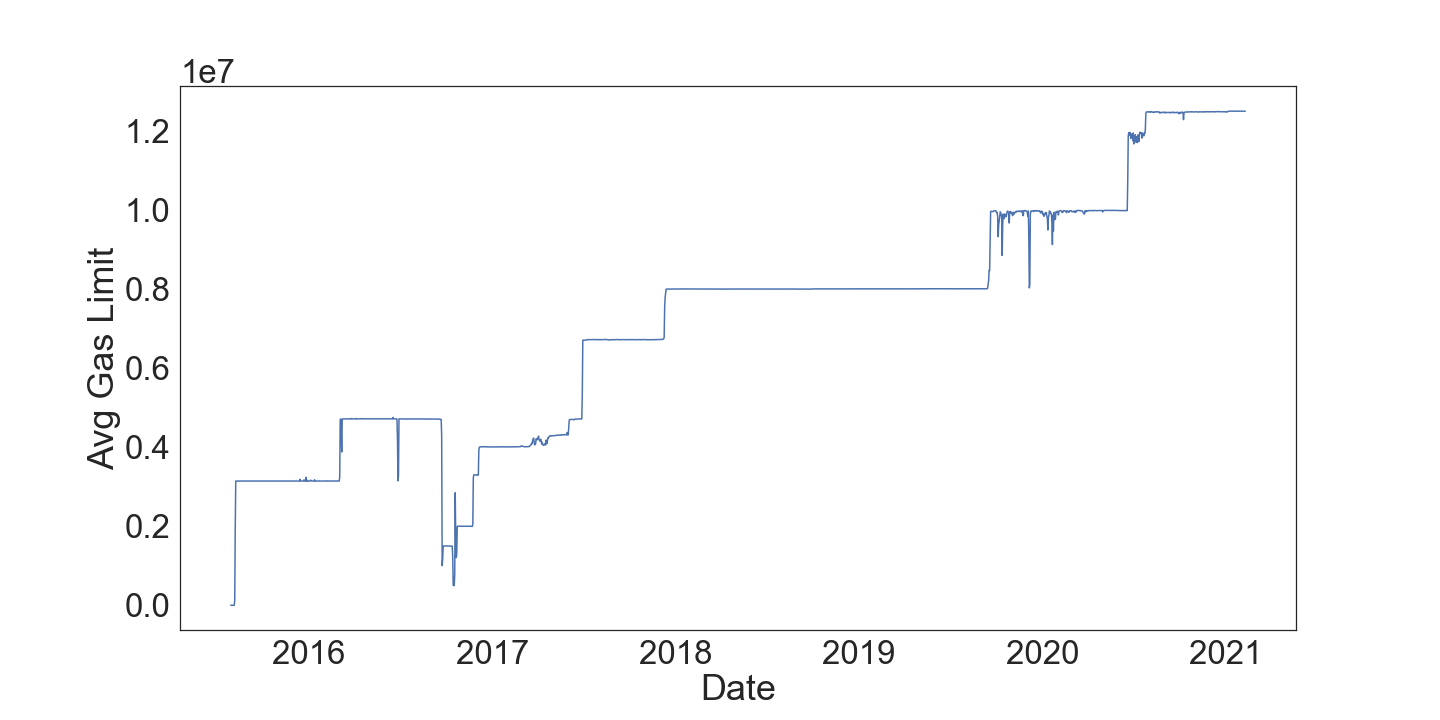}\par
\end{multicols}
\caption{Network Parameter time series, 20150630-20210211  \href{https://github.com/QuantLet/USC}{\includegraphics[keepaspectratio,width=0.4cm]{media/qletlogo_tr.png}}}
\label{Fig:Net_Param}
\end{figure*}

Contrary to the BTC realm -- where compensation is paid per financial transaction executed, hence per newly created/\textit{Proof-of-Work mined} block -- the EVM system calculates the costs per function/opcode called (gasPrice), i.e., per computational power consumed at that given time compared to the network utilization (average gas used capped by the gas limit) under the expected time of blocks to be added (\textit{block time}), and hence interaction can be cheaper given the amount of participating nodes and the amount of information sent to be validated at that time. This means, that the interactor can \href{https://ethgasstation.info/}{pay more Gas to have information validated more quickly}, as each block in the ETH system has a maximum amount of Gas that may be spent within a block -- i.e., a maximum of computational power and fee's that may be spend per block. Even if a block is created with no TX or no successful outcome at all, computational fees will be paid in Gas and there will be a block reward in Ether, which modifies the state trie and the state root -- otherwise two consecutive blocks would have to have identicial state roots. We will not touch at length on the so-called \textit{block gas limit} (gasLimit), as it can change over time and is a question heavily related to especially the \textit{scalability} of ETH's BC, see Figure \ref{Fig:Net_Param}.\\

%As the interpreter runs, it maintains a stack (where each element is 32 bytes) and a memory byte-array, and has access to the contract's storage tree. The stack and memory byte-array are dropped when the execution completes. 
%But yes, every single node runs the transaction - at any point in the execution, the state of the stack, the memory-byte array, the program counter, and the storage should be identical on each node.
%Every node runs the little script that executes when you try to spend from a previously unspent output (assuming you know about bitcoin).
%When the execution completes, the storage tree of the contract may be updated, so we recompute the merkle root hash of that tree (and any other contract which may have been called by 
%EVM byte code execution (which can cause account balances and storage values to change),that one!)
%The EVM/interpreter is typically just a for loop that increments a program counter and has a big switch statement telling it what to do for each operation in the byte code (pop/push the stack, load/store memory, load/store storage, call another contract, suicide, etc.)

%It is not only sandboxed but actually completely isolated, which means that code running inside the EVM has no access to network, filesystem or other processes.

The parameters relevant to mining and SC execution explained in a nutshell are: \textit{startGas} is the individual amount of Gas units assigned to individual opcodes and the maximum amount the creator assigns to be consumable. Partially synonymously, gasLimit sets the maximum price paid to execute an interaction in total and also denominates the total Gas that can be spent within a block. In the case of an interaction, gasLimit is paid upfront by an EOA. When an interaction is committed to the system, the value for the sum of individual startGas is reserved, leaving the value for \textit{remainingGas}. Each further individual computation uses up Gas, therefore lowers the value for remainingGas until reaching an ``Out of Gas" exception if the interaction is not finished successfully beforehand, causing the CA's state to be reverted back to its previous state -- there is no such thing like an SC running ad infinitum -- hence, even if the interaction is not successful, the participating nodes gets paid. The incentive for the participating nodes to validate and execute information -- i.e., to contribute computational power independent from its intentioned success -- is therefore, that Gas is paid in advance and miners receive a guaranteed compensation for participating in the system. Any interaction, i.e., anything that requires computation, can not consume and run in excess of the \textit{escrowed} amount of Gas, which prevents indefinite looping of executing commands. Reasons failed processing can be, for example, non-valid information and coding errors (i.e., non-compliance to standards, see further section \ref{Sec:FrameworkInfo}, or EVM exceptions with information processing), or just plainly missing sufficient funds on the interactors/senders end. Finally, the information processed by the miners is then stored in a new block on the BC and the miners receive their compensation, while unused Gas is refunded to the interactor to a maximum of half the Gas consumed. Therefore, if a code has a miniscule error that causes to EVM to run into an exception can waste a lot of resources, which is why ICO's (see further section \ref{Sec:FrameworkInfo}) may cost a lot of resources to begin with.\\

%To sum up: the amount a participating node receives fractionally as a reward is (startgas - gas_rem) * gasprice.
%paidEther = gasUsed * gasPrice
%the fee an originator pays a miner is the transaction’s (startGas — remainingGas) × gas price

\begin{figure}[H]
	\centering
	\includegraphics[keepaspectratio,width=15cm]{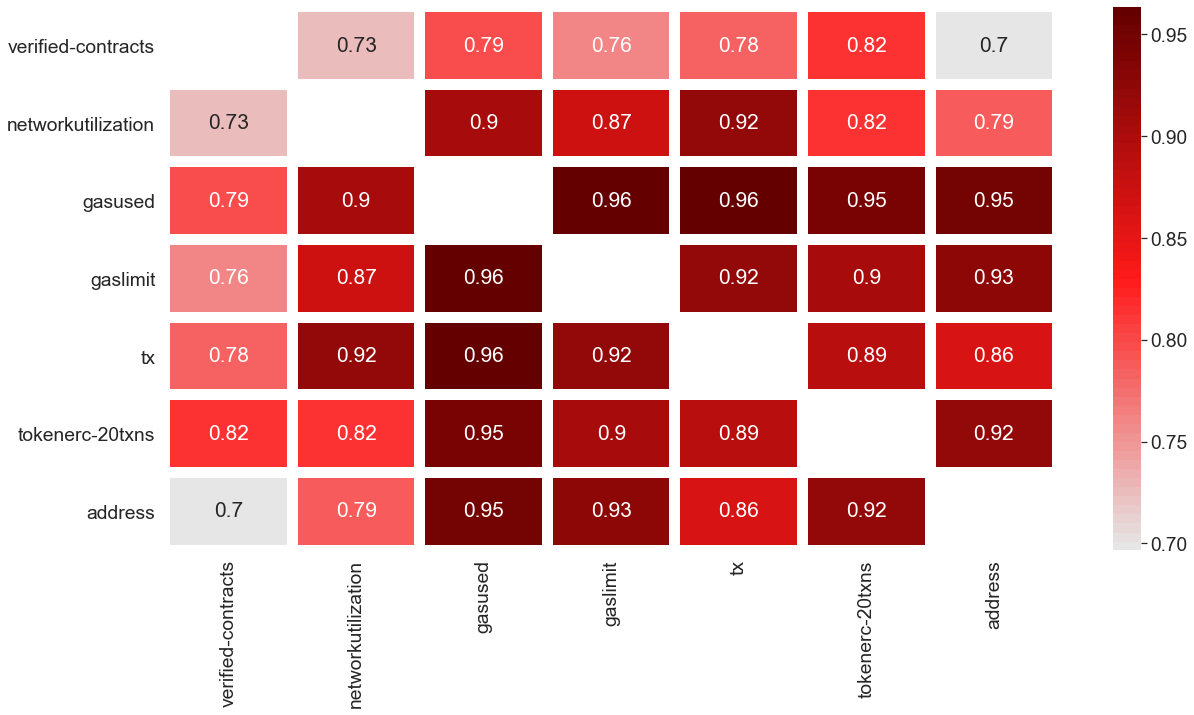}
        \caption{Network Parameter correlation to VSCs, 20150630-20210211
		\href{https://github.com/QuantLet/USC}{\includegraphics[keepaspectratio,width=0.4cm]{media/qletlogo_tr.png}}}
        \label{Fig:heat_corr}
\end{figure}

Figure \ref{Fig:heat_corr} presents the most important factors of SCs, like the obvious dependence of Network Utilization to used Gas. We explicitly underline the difference between mere TX and ERC20-Token TX. Interestingly, we observe a higher correlation of ERC20-Tokens to VSCs than to TX. This indicates the increased usage of transparent coding for such endeavours, certainly fostered through the sum of ICO scams that have been ongoing. However, VSCs do not have a high correlation to the Network Utilization, indicating a subordinate role compared to e.g. mere TX. Therefore we observe that whilst transparency issues have fostered the usage of VSCs, they are still not common ground for every application. With most applications being of a financial nature, the main point of interest is obviously Gas. Of course, the more complex an SC is, the more Gas it uses for its computational steps, yet Gas is more relevant to other proceedings on the ETH BC given the data analysis presented in Figure \ref{Fig:heat_corr}.

Importantly, the cost to run an SC can not be securely estimated upfront -- contrary to simple TX. This incentivises network participation, aside block rewards, and most importantly proper thought out coding. Clients like \href{https://geth.ethereum.org/}{Geth} inform interactors if not enough Gas is allocated for a given interaction to be executed. Entering a high Gas value can also end in spending more than what was originally intended, which once again incentivises proper coding. \textit{gasPrice} denotes the value of each computational step, i.e., each individual opcode. gasPrice is the amount in Ether to be paid for one unit of Gas utilized. As hinted to beforehand, miners can set a gasPrice threshold to prioritize or only accept respective interactions leading to others taking longer to be finished depending on the network utilization -- remember, that the outcome is not important, as the miners are always paid out, hence \textit{time is Ether}.\\

\begin{figure}[H]
	\hspace{0.5cm}
	\includegraphics[keepaspectratio,width=16cm]{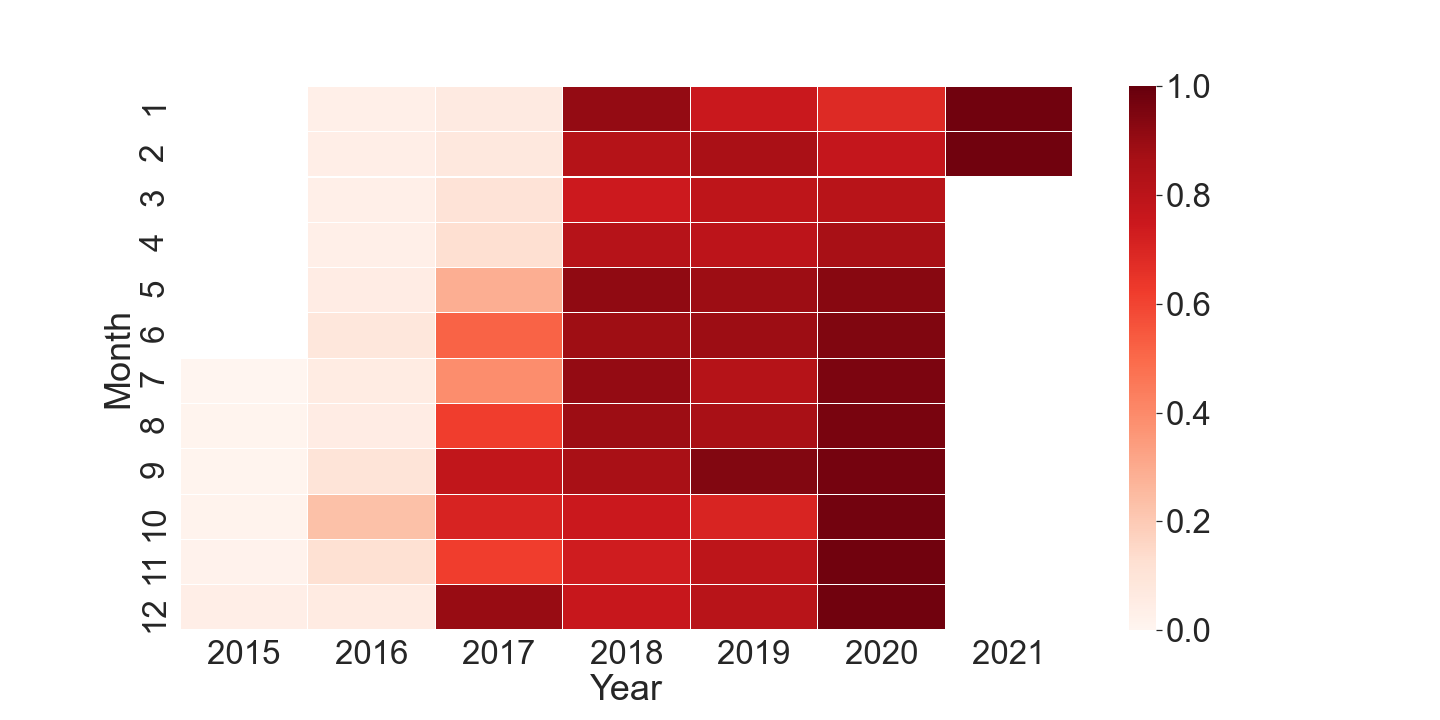}
        \caption{SC Value, 20150630-20210201
		\href{https://github.com/QuantLet/USC}{\includegraphics[keepaspectratio,width=0.4cm]{media/qletlogo_tr.png}}}
        \label{Fig:heat_value}
\end{figure}

SCs are employed to create tradeable digital tokens e.g. through employing the ERC20 standard. These may represent a ``coin", an asset, a virtual share, a proof of membership, amongst other codeable entities (see further \ref{Sec:FrameworkInfo}), and define how these tokens are meant to be distributed et cetera -- in other words, they can act like a central bank (see further section \ref{Sec:Implementation}). As SCs themselves can not access data from outside their coded environment, so-called \textit{oracles} are required -- wording related to off-chain information streams that can change SC values and trigger events if their input is respectively coded to be accepted. There are different kinds of \textit{Inbound} \textit{Oracles} that provide off-chain data, like Software Oracles (essentially API's that relay, for example, price data), Hardware Oracles (for example supply chain RFID sensors), or Consensus-based Oracles (like, for example, \href{https://gnosis.io/}{Gnosis}). As this is off-chain data, the idea of a decentralized environment is put ad absurdum with these -- potentially also nefarious -- centralised data sources. Being dependent on only one source can be risky, unreliable, and potentially lead to market manipulations. On the other hand, there are Outbound Oracles, which enable SCs to send data off-chain (like unlocking a flats' smart lock). Figure \ref{Fig:heat_value} represents the normalized value in Ether being committed and locked to SCs -- interestingly, this value has been on a high plateau since the COVID-19 crisis has started to really cripple real world based industries.\\

\vspace{-1.5cm}

% Figure \ref{Fig:radar} presents the activity parameters linked to SCs, as outlined above, which are actually not showing equal patterns over these months - yet in fact even completely diverging ones. When linked to figure \ref{Fig:SC_New_Total_Time}, we can presume, that a lot of value is kept within a minority of the total amount of SCs present in ETH. The figure therefore also presents very extreme patterns of activity, most obviously visible through the GasUsed variable for each month respectively. Furthermore, the variable verified-contracts, i.e. SCs with public source code (see further section \ref{sec:Classification}, underlines this argument regarding individual months being more active in the sense of SC propagation.

% \vspace{-0.5cm}

% \begin{figure}[H]
% 	\hspace{-5cm}
% 	\includegraphics[keepaspectratio,width=25cm]{images/radar_sc_parameters.png}
% 	\vspace{-1.5cm}
%         \caption{SC Network Parameters Differences, \textbf{\textcolor{cyan}{JAN}} \textbf{\textcolor{orange}{APR}} \textbf{\textcolor{green}{JUL}} \textbf{\textcolor{red}{OCT}}, 2020
% 		\href{https://github.com/QuantLet/USC}{\includegraphics[keepaspectratio,width=0.4cm]{media/qletlogo_tr.png}}}
%         \label{Fig:radar}
% \end{figure}

We observe that SCs are embedded in a very elaborate system with many functionalities. On the other hand, this means, that an entry into this realm and participation in it is bound to a quite demanding preliminary level of understanding how CCs work, or even how to access computer networks. The tale of ``banking the Unbanked" and creating financial inclusion, just to present the most obvious dissonance, is therefore debunked with either the possibility and/or infrastructure and/or knowledge of accessing this technology and its application are not given in most underdeveloped environment and/or does not present any visible benefits compared to traditional systems employed. Even with -- already classic -- Online-Banking, there still exist plenty of bank branches with employees to talk to customers, hence why should these -- on an scale that really matters beyond a proof-of-concept -- switch over to SC-banking? Furthermore, the strict structures imposed on SCs in order to function properly are in fact creating efficiencies, when only using them for simple tasks such as sending some value from A to B, but they therefore do not possess any kind of flexibility whatsoever -- i.e., any application beyond numeric transactions are therewith hardly implementable.

%There is enforced separation between “install time” and “run time”. No way to run the constructor twice.

%Smart contracts can use the same process to create other smart contracts.
%It is easy for a non-Solidity languages to implement.

%%%%%%%%%%%%NEXT%%%%%%%%%%%%%%

%%%%%%%%%%%%NEXT%%%%%%%%%%%%%%

%%%%%%%%%%%%NEXT%%%%%%%%%%%%%%

\section{Dataset}
\label{DatasetSection}

We cross-linked different sources. One of such sources is the \cite{stateDApps} \href{https://www.stateoftheDApps.com/}{(SDA)} -- a listing of decentralized Apps (DApps), where developers can list their DApps, additionally adding a category of the DApp \cite[inspired by][]{oliva2020exploratory}. We only included DApps listed as open source types into our datapool. One DApp can consist of multiple SCs, similar to a normal App where an SC can be seen as one coding script. As one App consists of multiple coding scripts, one DApp consists of multiple interacting SCs. Using SDA's API in December, 2020 we obtained a list of all ETH DApps and hash addresses of the SCs belonging to them. However, not all of them provide the information on the individual SCs belonging to them.\\

\href{https://etherscan.io/}{Etherscan}\nocite{etherscan1} -- a platform that allows to monitor transactions, blocks and SCs on the ETH BC -- enables the verification of SC source codes for informative purposes through their API. SCs with available source codes are called \textit{verified} SCs (VSC). However, not all DApps provide us with VSCs or even list their hashes / addresses. Therefore, the amount of possible codes is reduced. Yet, Etherscan provides a list of last 10.000 VSCs having open source licences. We hence got a labelled dataset with categories from SDA and a larger unlabelled dataset.\\
 
A list of all SC addresses on the ETH BC up to December 2020 was obtained using BigQuery on \href{https://www.kaggle.com/}{Kaggle}\nocite{ethereum_kaggle}. For that, we created an SQL query analogue to \cite{durieux2019empirical}, see \href{https://github.com/QuantLet/USC/tree/master/SC-over-time}{\includegraphics[keepaspectratio,width=0.4cm]{media/qletlogo_tr.png}}.\\

The first dataset that we are going to use for classification contains of 1428 observations belonging to 473 distinct DApps. 20\% of observations were put aside as the test set. As we have a very small dataset, and some categories have very few observation, it was decided to summarize the categories except the first most-frequent 5 to the ``other" category. The second dataset, which is unlabelled, consists of 16250 SCs.

\begin{table}[H]
\begin{center}
 \begin{tabular}{lcS[table-format=2.1]} 
 \hline
 \hline
 Topic & Amount of SCs &  \text{Relative Amount of SCs} \\ [0.5ex] 
 \hline
 Exchanges & 470 & 32.9\%\\ 
 Finance & 317 & 22.2\%\\
 Games & 272 & 19.0\% \\
 Gambling & 71 & 5.0\% \\
 High-risk & 60 & 4.2\% \\
 Marketplaces & 44 & 3.1\% \\
 Social & 41 & 2.9\%\\
 Media & 31 & 2.2\%\\
 Property & 29 & 2.0\%\\ 
 Development & 28  & 2.0\%\\ 
 Governance & 25 & 1.8\% \\
 Security & 13  & 0.9\%\\
 Identity & 12 & 0.8\%\\
 Wallet & 7 & 0.5\%\\ 
 Storage & 5 & 0.4\% \\
 Health & 3 & 0.2\% \\ [1ex] 
 \hline
 All & 1428 & \\
 \hline
 \hline
\end{tabular}
\end{center}
    \vspace{-0.5cm}
 \caption{Categories in the Dataset 1
 \href{https://github.com/QuantLet/USC/tree/master/SC-DApp-scraping}{\includegraphics[keepaspectratio,width=0.4cm]{media/qletlogo_tr.png}}}\label{Tab:Categories}
\end{table}

% \vspace{-0.5cm}

% \begin{figure}[H]
% 	\centering
% 	\includegraphics[keepaspectratio,width=16cm]{images/20210122 Sankey Plot SC.png}
%         \caption{Sankey diagramm  of the Dataset 1 categories  \href{https://github.com/QuantLet/USC/tree/master/SC-DApp-scraping}{\includegraphics[keepaspectratio,width=0.4cm]{media/qletlogo_tr.png}}}
%         \label{Fig:CRIX}
% \end{figure}

%%%%%%%%%%%%NEXT%%%%%%%%%%%%%%

\section{Clustering} 
\label{clustering}

\begin{figure}[H]
	\includegraphics[keepaspectratio,width=15cm]{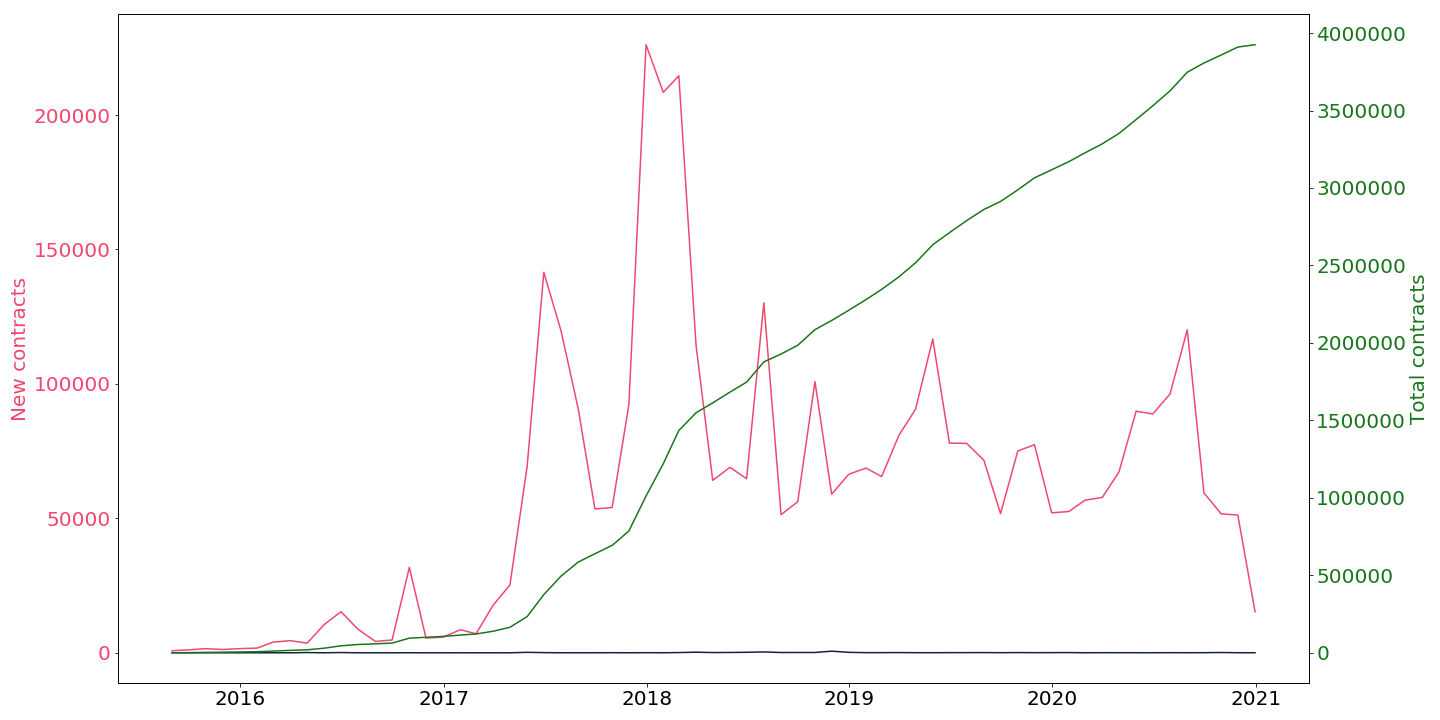}
        \caption{SCs created, 20150630-20201209
		\href{https://github.com/QuantLet/USC/tree/master/SC-over-time}{\includegraphics[keepaspectratio,width=0.4cm]{media/qletlogo_tr.png}}}
        \label{Fig:SC_New_Total_Time}
\end{figure}

In order to further understand the SC environment, it is hence not only necessary to understand the system they are deployed within, but moreover what kind of applications they are de facto be used for. Many media outlets, academic publications, and industry experts promise SCs that solve problems in digital notary, real-estate, managing transplantation organs, and many others, besides creating these processes to be more efficient and transparent. Can we actually find these categories in the existing applications on the ETH BC, or are these only hypothetical use-cases? The amount of TX that create individual SCs on the ETH BC count up to approximately 3.5B, see Figure \ref{Fig:SC_New_Total_Time}. We can observe the total amount of SDA-listed DApps and the amount of the SCs that these DApps are consisting of. There are currently around 3.5K DApps on the ETH BC and 6.49K SCs. So what are all these 99\% of the SCs saved on the BC used for?\\

To train a classifier that yields good performance for the data at hand can be very costly: researchers and industry experts would need to manually go through each observation and manually label it into pre-defined categories. In the case of source code, it is even more difficult, since only people having knowledge about this programming language may be able to properly categorize such data. Hence we first look into unsupervised approaches and other tools, like clustering. There exists academic work on the ETH data: grouping by the means of the bytecode \citep{norvill2017automated}, clustering the interfaces of SCs \citep{di2020assessing}, or the nodes on the ETH BC \citep{sun2019ethereum}. To the best of our knowledge though, there is no research on clustering the whole source codes of ETH SC, so far. \\

Every source code does usually contain comments of the developer who has written it. Some researchers even assess the quality of comments in the source code \citep{steidl2013quality} or develop methods to automatically create the comments \citep{liang2018automatic}. Therefore, one of our hypotheses is, that comments should describe what the code does and subsequently we can get a hint on what the whole program does. Comments are simply human-readable text, thus, based on the comments one may employ the same methodology used for natural language processing (NLP). One of the state-of-the-art techniques of NLP is the application of deep pre-trained transformers, such as BERT \citep{devlin2018bert} or DistilBert\citep{sanh2019distilbert}, and their sentence embeddings \citep{reimers-2019-sentence-bert}, which we are also going to use for our analysis.\\
 
As done in section \ref{Literature_Review}, we apply this strategy to our bigger unlabelled dataset to identify possible groups by using coder-comments present in SCs \citep[inspired by the BERT-approach in][]{topic2020}.
As we are dealing with a very high-dimensional dataset given by the comments in the source codes, our train dataset contains around 120K unique tokens and one SC can contain up to 14.6K of tokens (see Figure \ref{Fig:dist_length}) -- the first step is therefore to reduce dimensionality. For this purpose we use the uniform manifold approximation and projection for dimension reduction (UMAP) technique, recently proposed by \cite{mcinnes2018umap}. This advanced method combines fast computational results with the ability to preserve the global structures. First, we encoded our data via BERT and only load the embeddings, i.e., the lower dimensional representation of words. Even though these embeddings are of lower dimension than the original data, it still has too many dimensions to perform clustering. Clustering methods are based on distances and the more dimensions the given data has, the more options arise to separate such data. We reduced our dataset to 5 dimensions through UMAP and used $K$-means clustering methods to identify 14 topics in the case of the above mentioned literature research example, see section \ref{Literature_Review}. UMAP was employed once again to obtain a reduction to 2 dimensions to be able to plot our data points. \cite{topic2020} proposes a variant of TF-IDF -- the class-based variant of TF-IDF (c-TF-IDF), which the author formulates as the following: 

\begin{figure}[H]
	\includegraphics[keepaspectratio,width=15cm]{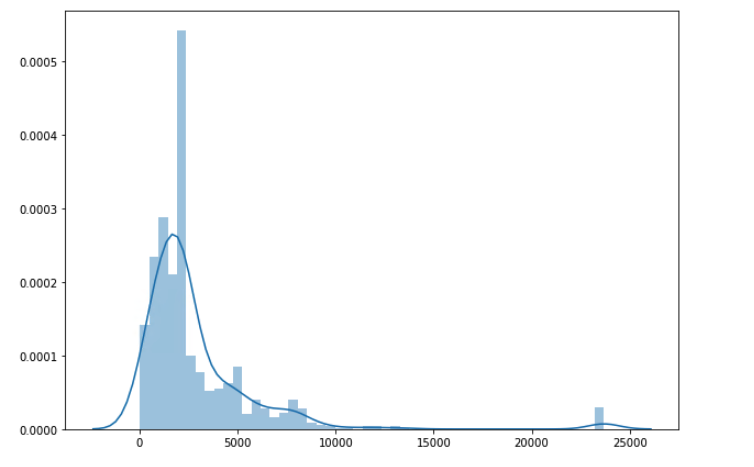}
	\centering
        \caption{Distribution of the amount of tokens per on SC source code
		\href{https://github.com/QuantLet/USC/tree/master/SC-classification-significance}{\includegraphics[keepaspectratio,width=0.4cm]{media/qletlogo_tr.png}}}
        \label{Fig:dist_length}
\end{figure}

\begin{equation*}
    c-TF-IDF_i = \frac{t_i}{w_i}  \times \log \frac{m}{\sum^n_jt_j},
\end{equation*}

where $t$ is the frequency of each word, $i$ (class), $w$ (total number of words for i-th class, $m$ (number of documents), and $\sum^n_jt_j$ (total frequency of word $t$ across all classes $n$). For each category, in our case topic, we identified the top ten of the msot important words. Similar approach was used to group unlabelled SCs using their comments.

%%%%%%%%%%%%NEXT%%%%%%%%%%%%%%

\subsection{Empirical results}\label{empirical_clustering}
As described in sections \ref{DatasetSection} and \ref{clustering}, our aim is not only to use labelled data but to try to gain knowledge from a by far larger unlabelled dataset. Here, unsupervised techniques become handy when no annotation is available.\\

\begin{figure}[H]
\hfill
\subfigure[Topic 1]{\includegraphics[width=5.2cm]{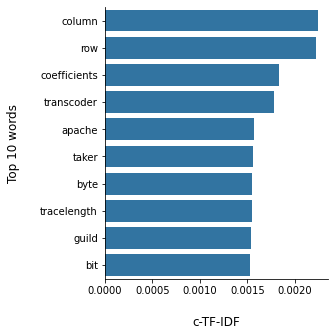}}
\hfill
\subfigure[Topic 2]{\includegraphics[width=4.8cm]{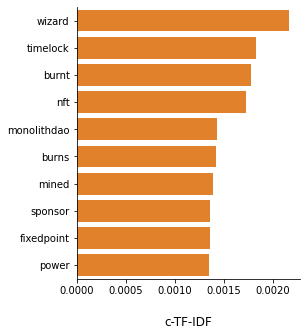}}
\hfill
\subfigure[Topic 3]{\includegraphics[width=4.8cm]{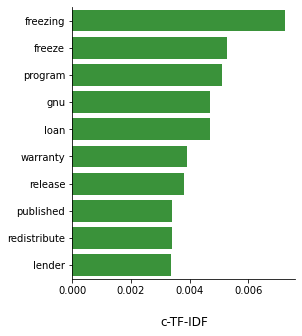}}
\hfill
\caption{Top 10 the most important words per cluster identified in the unlabelled SC dataset (Part 1) \href{https://github.com/QuantLet/USC/tree/master/SC-topics-unlabelled}{\includegraphics[keepaspectratio,width=0.4cm]{media/qletlogo_tr.png}}}
\label{fig:topics_unlab_2}
\end{figure}

\begin{figure}[H]
\hfill
\subfigure[Topic 4]{\includegraphics[width=5.5cm]{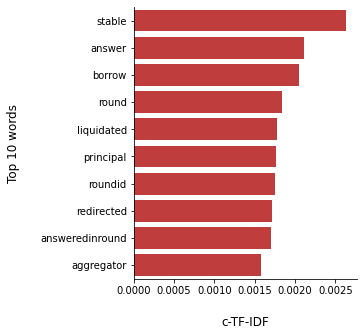}}
\hfill
\subfigure[Topic 5]{\includegraphics[width=4.6cm]{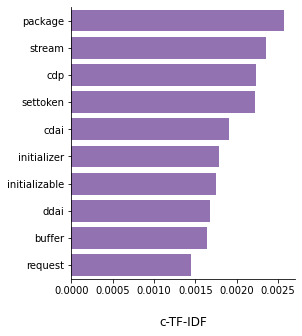}}
\hfill
\subfigure[Topic 6]{\includegraphics[width=4.9cm]{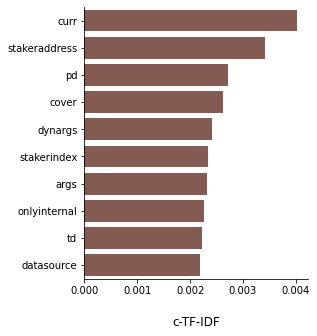}}
\hfill
\subfigure[Topic 7]{\includegraphics[width=5.4cm]{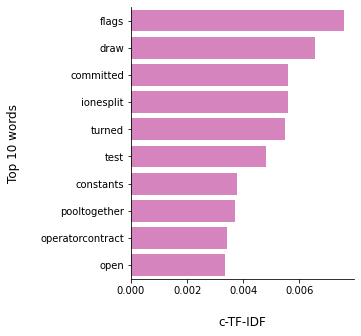}}
\hfill
\subfigure[Topic 8]{\includegraphics[width=4.6cm]{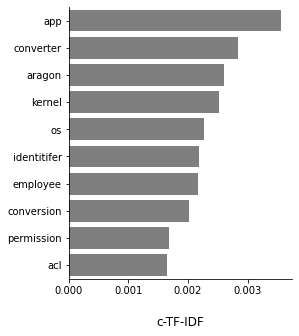}}
\hfill
\subfigure[Topic 9]{\includegraphics[width=4.7cm]{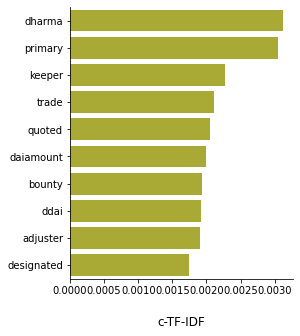}}
\hfill
\subfigure[Topic 10]{\includegraphics[width=5.7cm]{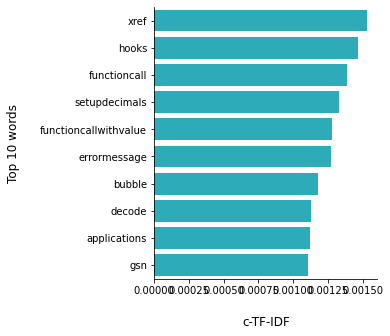}}
\hfill
\subfigure[Topic 11]{\includegraphics[width=4.5cm]{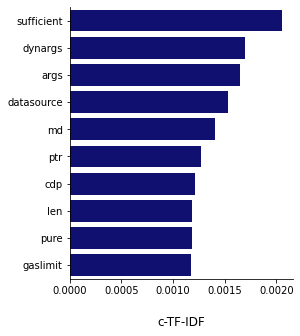}}
\hfill
\subfigure[Topic 12]{\includegraphics[width=4.6cm]{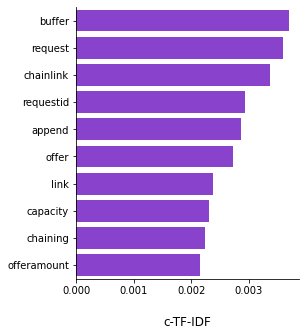}}
\hfill
\caption{Top 10 the most important words per cluster identified in the unlabelled SC dataset (Part 2) \href{https://github.com/QuantLet/USC/tree/master/SC-topics-unlabelled}{\includegraphics[keepaspectratio,width=0.4cm]{media/qletlogo_tr.png}}}
\label{fig:topics_unlab_1}
\end{figure}

\begin{figure}[H]
\centering
\includegraphics[width=13cm]{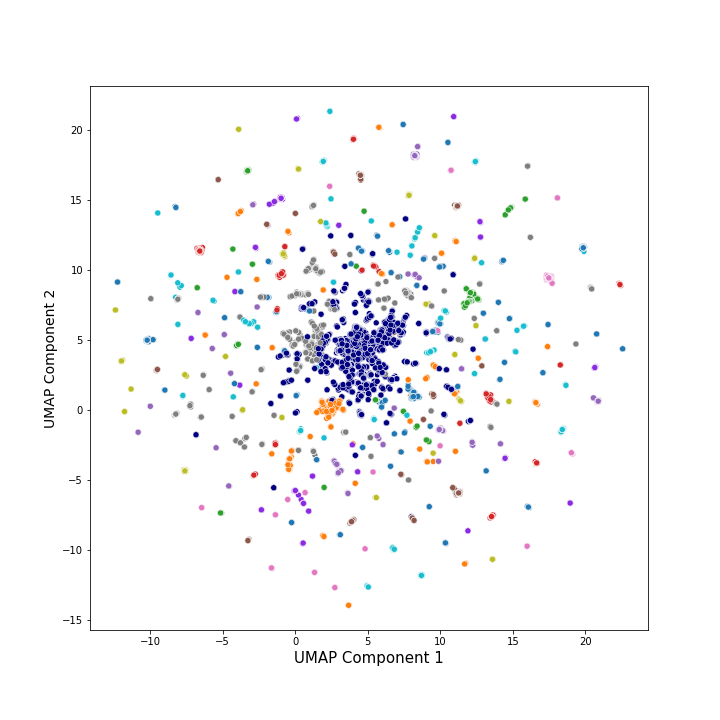}
\caption{Grouping the unlabelled SC dataset into 12 categories \href{https://github.com/QuantLet/USC/tree/master/SC-topics-unlabelled}{\includegraphics[keepaspectratio,width=0.4cm]{media/qletlogo_tr.png}}}\label{fig:Unlabelled_scatter}
\end{figure}

We apply the same technique of unsupervised clustering methods as in the beforehand section \ref{Literature_Review} on the unlabelled dataset, specifically on the comments of approximately 13k open source SCs each having on average 1.2K words in the comment. We hence test, if this delivers us similar results as in the beforehand classification, and also if it will picture whether comments can help to better understand what individual SCs do and possibly even understand in what domain the respective DApp belongs to. The first insight is, that of 16 250 open source SCs around 17\% do not contain any comments or their length is shorter than 10 characters or 5 words. These 17\% were excluded from the analysis due to lack of semantic meaning necessary for the analysis. On the procedural side, we first applied UMAP on the remaining source codes to reduce the dimensions to 5 and ran clustering multiple times with different parameter of number of clusters to decide on this parameter using the elbow curve method with Davies-Bouldin  score. Figures \ref{fig:topics_unlab_1} and \ref{fig:topics_unlab_2} present the 12 topics chosen as the optimal amount of clusters. Subsequently, the data was reduced further to 2 dimensions. Figure \ref{fig:Unlabelled_scatter} presents the scatter plot of our unlabelled dataset, where each color identifies a different cluster and axes correspond to the 1st and 2nd UMAP components. We can see that we have two main big clusters -- the navy- and gray-colored clusters, followed by the smaller orange cluster, and all the other are rather spread everywhere and in 2 dimensions do not look like building proper interconnected clusters.\\

Yet, how can we interpret these clusters in Figure \ref{fig:Unlabelled_scatter} and understand whether they are grouped by the category of their application (e.g., healthcare, notary, real-estate), their functionality, or rather just some other similarity measure? A closer look on the variables of the clusters reveals the 10 most important words per cluster in Figures \ref{fig:topics_unlab_1} and \ref{fig:topics_unlab_2}. On the first glance, we conclude that the majority of the keywords are very technical and reflect on the names of functions and/or variables commonly employed in SC codes. Apparently, the majority of comments do not really reveal any information apart technical implementations and describing what the respective function, interface or a library do. The only cluster having pretty much semantically-meaningful and understandable keywords from business application perspective is the \textcolor[HTML]{3a923a}{3rd cluster}, where some keywords are obviously related to loan business. Therefore, here we were able to show that the opportunities of directly trying to get the knowledge from the unstructured data without supervision are very limited and the supervised methods are needed.

\vspace{-1cm}

%%%%%%%%%%%%NEXT%%%%%%%%%%%%%%

\section{Classification} 
\label{sec:Classification}
 
Unlike traditional contracts, SCs are not written in natural language, which makes it hard to determine what each SC is about. We worked on a predictive model to classify SCs based solely on their coding. Previous approaches use  to annotate SCs manually, as done by \cite{bartoletti2017empirical}. We randomly picked the source codes of 800 verified SCs provided by \cite{etherscan}, using the functions of random sampling in Python, and tried to group the source codes into categories by looking at the code, reading the comments, searching for titles of the SC. The 9 categories that we came up with are:

\begin{itemize}[leftmargin=*]
	\item[$\boxdot$] \textbf{Token creation}, \textbf{Token sale} or \textbf{Token creation/Token sale}: SCs sole purpose is to define a token, or its crowdsale, or both.
\vspace{-0.3cm}
	\item[$\boxdot$] \textbf{Finance} or \textbf{Token creation/Finance}: whenever an SC has an implementation of a financial product or a derivative, a bidding mechanism, a lottery, a multisignature wallet, or in general any profit-creating mechanism. A token can be implemented in a separate SC and be evoked, or defined in a Finance SC directly. 
\vspace{-0.3cm}
	\item[$\boxdot$] \textbf{Airdrop} or \textbf{Token creation/Airdrop}: an implementation of a token airdrop, sometimes simultaneously with a token definition. An airdrop is a free distribution of a token, unlike a crowdsale.
\vspace{-0.3cm}
	\item[$\boxdot$] \textbf{DApp}: an implementation of an Ethereum DApp.
\vspace{-0.3cm}
	\item[$\boxdot$] \textbf{Scam}: pyramid schemes, or financial SCs that offer e.g. 5 percent daily interest.
\vspace{-0.3cm}
	\item[$\boxdot$] \textbf{Utility}: diverse helper functions, pieces of code that are necessary for running actual SCs, extensions, libraries, etc.
\end{itemize}

Obviously, these categories are slightly different then the ones described previously in Table \ref{Tab:Categories}. Only the category \textbf{Finance} came up in the manual selection as well in the dataset from SDA. Moreover, just by looking at the code ourselves we were not always able to identify the category of the code or what DApp's category it belongs to. Interestingly, some of the categories were identified through meaningful comments or naming the product. Sometimes however, it required additional search to specify what the code is about. Thus, it is not always possible to accurately identify the category of the SC without using additional check up by a human-being. It requires a lot of engineering and Solidity domain expertise, which makes manual labeling more difficult or at least its scaling and outsourcing. On the other side though, it might happen that automatic classifiers would pick up the features that we do not see as a human. Therefore, we would like to try to implement an automatic classifier that could improve this categorisation. Nonetheless, this difficulty to manually label the data speaks also for the difficulty to obtain a big dataset which is usually vital for automatic classification, therefore we need to rely on the very small dataset.

\subsection{Data Pre-processing and Feature Extraction} \label{classification_preprocessing}
We created three columns for obtaining regular expressions to extract comments from the codes: source code (full code), source code without comments (only code) and only comments (comments). Moreover, as additional feature engineering step, we extracted names of functions, SCs, events, interfaces and libraries and joined to one string. There exist different approaches to extract features from the source codes. Possible approaches are to extract features like names, design patterns or even abstract syntax trees. For our work, we decided to treat our individuall codes like text and used NLP techniques. In the case of traditional machine learning methods, since we do not need to preserve the sequential structure, we just use the TF-IDF weighting scheme. First, we separate our sequences into tokens -- words in the code or specific punctuation marks that are specific for the programming language like parentheses, then each token is encoded into a numeric representation. For the encoding we use TF-IDF scheme. For the deep learning methods, we just encode the tokens with the ID, which is then embedded into a matrix to obtain dense vector representations of the words depending on the surrounding context in the sentence.

\subsection{Machine Learning methods}
A classificator is trained on a variety of methods. Here, we employ a logistic regression, a Ridge regression (Ridge), and also an instance of Lasso regression (Lasso).  As we just mentioned in the section \ref{classification_preprocessing}, through encoding our source codes into TF-IDF vector representations, we obtain very sparse high-dimensional vectors. Since we have a such very high-dimensional data, we expect the Lasso regression to perform better due to inherent feature selection technique through L1 regularization. Moreover, we are going to use random forest (RF) and support vector machines (SVM).\\

Deep learning methods are seen to be standard for natural language processing, yet, they usually require huge datasets. However, we apply three classical deep learning architectures for testing on the above mentioned data columns: gated recurrent unit (GRU), bidirectional GRU, and convolutional neural network (CNN).\\

Pre-training became a standard for many natural language processing tasks. Bidirectional Encoder Representations from Transformers (BERT) architecture invented by \cite{devlin2018bert} allows the user to apply this deep learning architecture trained on a huge amount of data on a different task, e.g., text classification. These models achieve a state-of-the-art performance for many tasks. Moreover, they are able to work on relatively small datasets. Such architecture are usually pre-trained on huge text datasets. In our case, we have mostly the source codes, that differs from written and spoken human language, therefore it will not help us much. However, one of our tasks formulations includes machine learning on the comments. Therefore, we will be able to apply BERT on comments.\\

To summarize it, we are going to use traditional machine learning methods Ridge, Lasso, RF, and SVM and classical deep learning methods GRU, BGRU, and CNN on full code, only code, and comments. Moreover, we apply BERT to the comments. Therefore, we will try to answer the question, whether it is possible at all to train a classifier on the solidity SC source code, whether traditional machine learning methods can better work on such small and high-dimensional dataset or they will loose to deep learning. Moreover, we will see whether it is better to train a classifier on the code, code with comments, or just comments, or even that, as opposed to spoken and written text, the sequential structure of the source code is not relevant at all to identify the category of the SC. 

\subsection{Training and Evaluation}
For training we used cross-validation with 3 folds. Moreover, we performed a hybrid of manual and automatic grid hyper-parameter tuning. As we are dealing with a multiclass classification task  with one class containing examples from the rest of categories, we draw upon the well established one-vs-all classification approach for the first 5 classes. For the evaluation we used two metrics: Area under the ROC curve is equivalent to ``the probability that the classifier will rank a randomly chosen positive instance higher than a randomly chosen negative instance" \citep{fawcett2006introduction} and the average precision -- the metric which is seen as standard for unbalanced data sets.

\subsection{Empirical results}

Due to spatial reasons we present only the test performance, nonetheless, cross-validation performance can also be found in the linked GitHub repository. In Table \ref{Tab:results_FC}, we see the results of classificator trained on the full code -- without removal of comments. On the left part we see the performance in terms of AUC ROC, which is relatively high for all categories. Methods of traditional machine learning methods outperform on average the methods of deep learning. One of the possible and very probable reasons is the small amount of data points, whereas the big size of a dataset is essential for application of deep learning methods. On the other side, as opposed to our expectation Lasso regression showed the worst performance across all traditional machine learning methods. Random Forest (RF) shows the best performance in terms of AUC ROC in almost all categories, except high risk, where it is outperformed by the Ridge regression. Ridge shows the second best performance in all other categories. On the right part, we see the performance in terms of Av. Prec. Whereas the performance for classes with a higher amount of examples is still high, the classes that are outnumbered show much lower performance. Gambling and High-risk classes show the worst ability to be accurately classified.\\

Moreover, deep learning methods show a very poor performance for such unbalanced categories. In Table \ref{Tab:results_OC}, we see the performance of the classificators trained on the code cleaned from the comments. It shows very similar tendency: a good performance of rather simple traditional machine learning methods like RF and Ridge and very poor performance of neural networks. Interestingly, there is no obvious tendency of outperforming between results on the source code with and without comments. In Table \ref{Tab:results_comments}, we present results of predictors trained only on textual data -- the comments extracted from the code. Again, traditional methods outperform deep neural networks. Some categories, such as Exchanges, show even better performance than classificators  trained on code. A possible explanation could be the quality of the comments in the codes of this category. However, other groups show worse performance, except Gambling in terms of Av. Prec. Apparently, the comments allow the ridge regression to correctly pick the Gambling cases with higher precision and judging by lower AUC ROC, the False Positives rate increased. Interestingly, even BERT that is seen as a deep learning model that is able to deal with small datasets shows a very poor performance. \\

As we were able to show that deep learning networks, especially those based on recurrence and reflecting the sequential structure, show poor performance, it would be interesting to inspect, whether we are able to extract the most important features from the code. As explained in section \ref{classification_preprocessing}, we decided to extract the names of functions, events, libraries in the hope that they can be not only similar from the source code to source code, since as already mentioned many codes are cloned and copied, but also usually developers would name the variables and functions according to their behavior. In table \ref{Tab:results_features}, we show the results of classification training of traditional machine learning methods, which have shown the best performance in the previous tasks. The aim was to show that not only can we extract meaningful information and the sequential structure is not important for our task but also that we are able to achieve sufficiently high performance using lower amount of features. And as we see in this table, we still have high performance, even though the amount of features was reduced by more than 3k. \\

Even though the differences between each mode: \textit{Full Code} (FC), \textit{Only Code} (OC), \textit{Only Comments} (OCom), and \textit{Extracted Features} (EF) might not always seem significantly different from each other, or on the opposite seem obviously different for some classes, we decided to test whether their performance results are significantly different from each other. For that purpose, using 14 different seeds to split the data, we ran our Ridge regression classifier multiple times for each cross-validation fold. Thus, for each mode we obtained 42 predictions for each class. Since we want to have the same classifier for each class, we averaged predictions between classes. The non-parametric Wilcoxon test \citep{wilcoxon1992individual} was chosen to see whether the differences in AUC ROC and Av. Prec. of different modes are significant. Double-sided, left-sided and right-sided tests were conducted. Table \ref{Tab:results_wilcoxon} presents the pairs of modes and the null hypothesis $H_0$ which could not be rejected and the respective $p$-value. Using the test, we were able to show that even though when the classes are treated separately we do not see an obvious winning mode. If we want to use only one classifier, the best performance can be achieved while training it on the comments extracted from the source code (assuming the developers wrote accompanying comments).

\begin{table}[H]
    \begin{center}
    {\scriptsize
        \begin{tabular}{lcccccccc}
        \hline 
        \hline
            {Modes} &
            \multicolumn{4}{c}{AUC ROC} &
            \multicolumn{4}{c}{Av.Prec}\\ 
             & {FC} & {OC} & {OCom} &{EF}& {FC} & {OC} & {OCom} &  {EF}\\ 
             \hline
             FC & - & $FC<OC$ & $FC<OCom$ & $FC<EF$ & - & $FC<OC$  & $FC<OCom$ & $FC<EF$\\
              & & $p=0.99$ & $p=0.99$ & $p=0.99$    &   & $p=0.99$ & $p=0.99$  & $p=0.99$\\
             OC & - &  - & $OC<OCom$ & $OC<EF$      & - & - &   $OC<OCom$ & $OC<EF$\\
                &   &   & $p=0.99$ & $p=0.99$       &   &   &   $p=0.99$ & $p=0.99$\\
             OCom & - & - & - & $OCom>EF$ & - & - & - & $OCom>EF$\\
                  &   &   &   & $p=0.99$  &   &   &   & $p=0.99$\\
             EF   & - & - & - &     -     & - & - & - &  - \\ 
        \hline
        \hline
        \end{tabular}
        }
    \end{center}
    \vspace{-0.5cm}
        \caption{Results of the Wilcoxon Test
        \href{https://github.com/QuantLet/USC/tree/master/SC-classification-significance}{\includegraphics[keepaspectratio,width=0.4cm]{media/qletlogo_tr.png}}}  \label{Tab:results_wilcoxon}
\end{table}

Summarizing the results of these four tables: first, we can conclude that using supervised machine learning methods we are able to classify the source codes of SCs. Secondly, we highlighted the fact that no heavy computational architectures and algorithms such as neural networks are needed to classify the categories, and very simple Ridge logistic regression ran for each class independently will do the job sufficiently good not only for the over-represented classes like \textcolor[HTML]{050CF2}{Exchanges} and \textcolor[HTML]{2A702C}{Finance}, but will also show relatively good performance for rather smaller class of \textcolor[HTML]{F20C05}{High-risk} contracts. We were able to show that the common-sense telling us that the comments in the source code could give additional hint on the category of the SC is indeed correct. However, the presence of the comments in the code is required.\\

Therefore, we were able to show that it is possible to classify SCs by using their source codes quite accurately for the dominant categories: \textcolor[HTML]{050CF2}{Exchanges}, \textcolor[HTML]{2A702C}{Finance}, \textcolor[HTML]{9505F2}{Games}, \textcolor[HTML]{F20C05}{High-risk}. Such information can be useful not only for researcher who want to explore the remaining 99\% of SC on the ETH BC, but also for the developers and investors. Especially the high-risk category is interesting in identification of possible risks.

\begin{table}[H]
    \begin{center}
        {\scriptsize
        \scalebox{0.9}{
        \begin{tabular}{lcccccccccc}
        \hline 
        \hline
            Method &
            \multicolumn{5}{c}{AUC ROC} &
            \multicolumn{5}{c}{Av.Prec}\\ 
             & {Exchanges} & {Finance} & {Gambling} & {Games} & {High-risk} & {Exchanges} & {Finance} & {Gambling} & {Games} & {High-risk} \\ 
             \hline
             Ridge & 0.988 & 0.967 & 0.914 & 0.972 & 0.983 & 0.985 & 0.925 & 0.482 & 0.914 & 0.720 \\
             Lasso & 0.961 & 0.927 & 0.714 & 0.917 & 0.847 & 0.962 & 0.844 &	0.238 &	0.804 &	0.554\\
             RF & 0.988 & 0.973 & 0.944 & 0.971 & 0.974 & 0.984 & 0.942 & 0.572 &	0.912 &	0.700 \\
             SVM & 0.987 &	0.972 &	0.892 &	0.977 &	0.976 &	0.983 &	0.944 &	0.482 &	0.934 &	0.668  \\ 
             BGRU & 0.970 &	0.848 &	0.767 &	0.795 &	0.860 &	0.963 &	0.699 &	0.149 &	0.422 &	0.256	\\ 
             GRU & 0.966 &	0.712 &	0.749 &	0.818 &	0.897 &	0.956 &	0.443 &	0.102 &	0.442 &	0.486 \\ 
             CNN & 0.972 &	0.950 &	0.804 &	0.908 &	0.840 &	0.972 &	0.880 &	0.155 &	0.672 &	0.482\\
        \hline
        \hline
        \end{tabular}}
        }
    \end{center}
    \vspace{-0.5cm}
        \caption{Classification Results -- Full Code
        \href{https://github.com/QuantLet/USC/tree/master/SC-classification}{\includegraphics[keepaspectratio,width=0.4cm]{media/qletlogo_tr.png}}}  \label{Tab:results_FC}
\end{table}

\begin{table}[H]
    \begin{center}
        {\scriptsize
        \scalebox{0.9}{
        \begin{tabular}{lcccccccccc}
        \hline 
        \hline
            Method &
            \multicolumn{5}{c}{AUC ROC} &
            \multicolumn{5}{c}{Av.Prec}\\ 
             & {Exchanges} & {Finance} & {Gambling} & {Games} & {High-risk} & {Exchanges} & {Finance} & {Gambling} & {Games} & {High-risk} \\ 
             \hline
             Ridge & 0.983 & 0.965 & 0.916 &	0.957 &	0.988 &	0.981 &	0.914 & 0.483 & 0.868 & 0.795 \\
             Lasso & 0.960 & 0.915 & 0.832 &	0.922 &	0.934 &	0.959 &	0.813 & 0.363 & 0.769 & 0.656\\
             RF & 0.986 & 0.974 & 0.938 &	0.966 &	0.986 &	0.983 &	0.940 &	0.491 &	0.907 &	0.771 \\
             SVM & 0.980 & 0.973 & 0.905 &	0.970 &	0.976 &	0.979 &	0.943 &	0.574 &	0.919 &	0.717 \\ 
             BGRU & 0.946 &	0.282 &	0.461 &	0.744 &	0.763 &	0.952 &	0.157 &	0.049 &	0.448 &	0.508	\\ 
             GRU & 0.949 &	0.802 &	0.779 &	0.692 &	0.888 &	0.952 &	0.615 &	0.121 &	0.299 &	0.350 \\ 
             CNN & 0.952 & 0.895 &	0.707 &	0.868 &	0.781 &	0.953 &	0.794 &	0.100 & 0.663 &	0.168\\
        \hline  
        \hline
        \end{tabular}}
        }
    \end{center}
    \vspace{-0.5cm}
    \caption{Classification Results -- Only Code
    \href{https://github.com/QuantLet/USC/tree/master/SC-classification}{\includegraphics[keepaspectratio,width=0.4cm]{media/qletlogo_tr.png}}}\label{Tab:results_OC}
\end{table}

\begin{table}[H]
    \begin{center}
        {\scriptsize
        \scalebox{0.9}{
        \begin{tabular}{lcccccccccc}
        \hline 
        \hline
            Method &
            \multicolumn{5}{c}{AUC ROC} &
            \multicolumn{5}{c}{Av.Prec}\\ 
             & {Exchanges} & {Finance} & {Gambling} & {Games} & {High-risk} & {Exchanges} & {Finance} & {Gambling} & {Games} & {High-risk} \\ 
             \hline
             Ridge & 0.991 & 0.959 & 0.880 &	0.972 & 0.976 &	0.986 &	0.910 &	0.508 &	0.912 &	0.701 \\
             Lasso & 0.982 & 0.930 &	0.734 &	0.942 &	0.880 &	0.977 &	0.852 &	0.217 &	0.844 &	0.409\\
             RF & 0.984 & 0.972	& 0.890 & 	0.974 &	0.977 &	0.981 &	0.937 &	0.452 &	0.924 &	0.675 \\
             SVM & 0.988 &	0.963 &	0.879 &	0.978 &	0.976 &	0.983 &	0.921 &	0.511 &	0.933 &	0.626 \\ 
             BGRU & 0.971 &	0.639 &	0.656 &	0.853 &	0.839 &	0.968 &	0.276 &	0.086 &	0.520 &	0.361	\\ 
             GRU & 0.960 & 0.804 & 0.638 &	0.840 &	0.855 &	0.963 &	0.648 &	0.107 & 0.511 &	0.313 \\ 
             CNN & 0.937 &	0.907 &	0.829 &	0.865 &	0.703 &	0.937 &	0.799 &	0.155 &	0.600 &	0.070\\
             BERT & 0.968 &	0.777 &	0.353 &	0.789 &	0.602 &	0.965 &	0.637 &	0.038 &	0.628 &	0.056\\
        \hline  
        \hline
        \end{tabular}}
        }
    \end{center}
    \vspace{-0.5cm}
    \caption{Classification Results -- Only Comments
    \href{https://github.com/QuantLet/USC/tree/master/SC-classification}{\includegraphics[keepaspectratio,width=0.4cm]{media/qletlogo_tr.png}}}
    \label{Tab:results_comments}
\end{table} 

\begin{table}[H]
    \begin{center}
        {\scriptsize
        \scalebox{0.9}{
        \begin{tabular}{lcccccccccc}
        \hline 
        \hline
            Method &
            \multicolumn{5}{c}{AUC ROC} &
            \multicolumn{5}{c}{Av.Prec}\\ 
             & {Exchanges} & {Finance} & {Gambling} & {Games} & {High-risk} & {Exchanges} & {Finance} & {Gambling} & {Games} & {High-risk} \\ 
             \hline
             Ridge & 0.988 & 0.963 &	0.883 &	0.957 &	0.971 &	0.984 &	0.906 &	0.462 &	0.870 &	0.735 \\
             Lasso & 0.981 & 0.919 &	0.803 &	0.913 &	0.874 &	0.973 &	0.791 &	0.395 &	0.730 &	0.573 \\
             RF & 0.989	& 0.971 & 0.864 &	0.961 &	0.973 &	0.985 &	0.928 &	0.474 &	0.899 &	0.719\\
             SVM & 0.980 & 0.963 & 0.877 &	0.971 &	0.970 &	0.978 &	0.932 &	0.489 &	0.915 &	0.754 \\ 
        \hline   
        \hline
        \end{tabular}}
        }
    \end{center}
    \vspace{-0.5cm}
    \caption{Classification Results -- Extracted Features
    \href{https://github.com/QuantLet/USC/tree/master/SC-classification}{\includegraphics[keepaspectratio,width=0.4cm]{media/qletlogo_tr.png}}}
    \label{Tab:results_features}
\end{table}

After the algorithms were trained with satisfactory results regarding the achieved performance, we apply them onto new data. Here, we can apply our classifier on the unlabelled dataset to see what potential categories of open source SCs can be found on the ETH BC. We present the identifiable categories  discovered by the trained algorithms, without the implication that these datapoints were correctly identified. Since the most important category to categorise is high-risk, and moreover we cannot assume that all source codes contain comments, we decided to take the algorithm that predicts this category best to out final prediction on the unlabelled dataset, which is the Ridge logistic regression running on the source code cleaned from the comments. However, one could argue that we want to identify all classes correctly, then the mode of the interest would be to train the classifier solely on the comments.\\

Therefore, we identify the following observations in our unlabelled dataset:
\begin{table}[H]
    \begin{center}
        {\scriptsize
       \begin{tabular}{lcc}
        \hline 
        \hline
            Category & Amount of SCs & Relative Amount of SCs \\
            \hline
            \textcolor[HTML]{2A702C}{\textbf{Finance}}   & 10009 & 62\%  \\
            \textcolor[HTML]{A7B2B6}{\textbf{Other}}     & 4757  & 29\%  \\
            \textcolor[HTML]{050CF2}{\textbf{Exchanges}} & 756   & 5\%   \\
            \textcolor[HTML]{9505F2}{\textbf{Games}}     & 506   & 3\%   \\
            \textcolor[HTML]{F20C05}{\textbf{High-risk}} & 162   & 1\%   \\
            \textcolor[HTML]{05F2C0}{\textbf{Gambling}}  & 60    & $>$1\% \\
            \hline
            Sum & 16250 &  \\
        \hline   
        \hline
        \end{tabular}
        }
    \end{center}
    \vspace{-0.5cm}
    \caption{Frequency of categories in the predictions on unlabelled data
    \href{https://github.com/QuantLet/USC/tree/master/SC-classification}{\includegraphics[keepaspectratio,width=0.4cm]{media/qletlogo_tr.png}}}
    \label{Tab:unlabelled_freq}
\end{table} 

%Please note, that due to very high costs of manual annotation, we do not obtain them for the second dataset, which is everywhere called unlabelled. 

\vspace{-1.5cm}

\begin{figure}[H]
\centering
\includegraphics[width=13cm]{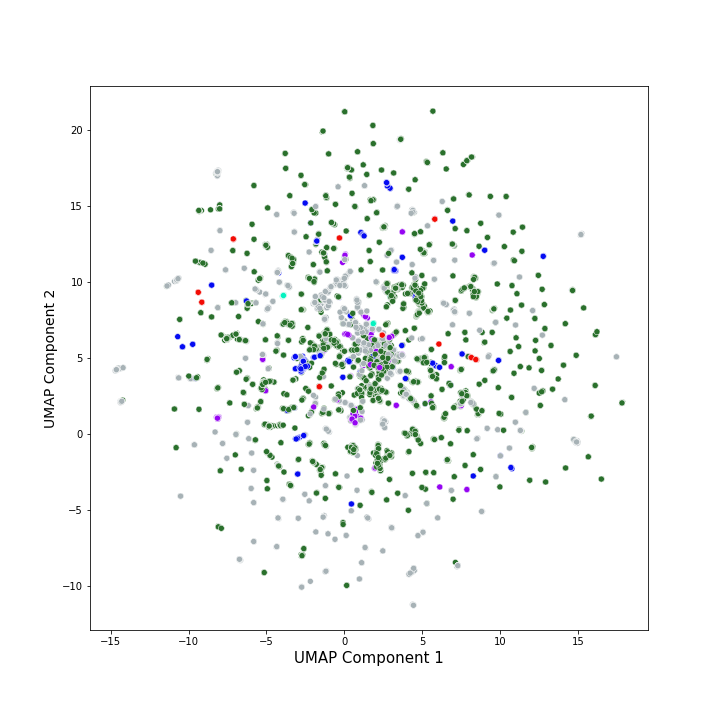}
\caption{Classification of the unlabelled dataset -- UMAP reduction to 2 dimensions: \textcolor[HTML]{2A702C}{\textbf{Finance}}, \textcolor[HTML]{F20C05}{\textbf{High-risk}}, \textcolor[HTML]{05F2C0}{\textbf{Gambling}}, \textcolor[HTML]{9505F2}{\textbf{Games}}, \textcolor[HTML]{050CF2}{\textbf{Exchanges}}, \textcolor[HTML]{A7B2B6}{\textbf{Other}}
\href{https://github.com/QuantLet/USC/tree/master/SC-classification}{\includegraphics[keepaspectratio,width=0.4cm]{media/qletlogo_tr.png}}}
\end{figure}\label{fig:unlabelled_scatter}

We note, that the \textcolor[HTML]{2A702C}{\textbf{Finance}} class outnumbers everything else, followed by the \textcolor[HTML]{A7B2B6}{Other} -- class having all observations that have too low probability to fit one of the specified classes. The other classes, such as \textcolor[HTML]{050CF2}{Exchanges}, \textcolor[HTML]{9505F2}{Games}, \textcolor[HTML]{F20C05}{High-risk}, and \textcolor[HTML]{05F2C0}{Gambling} represent only a very small fraction of the unlabelled dataset. This reveals, assuming our classification was similarly accurate as within cross validation and with test data, that the majority of VSC of this opportunity sample of SCs is dominated by the codes dedicated to financial services -- 68\% (\textcolor[HTML]{2A702C}{\textbf{Finance}} + \textcolor[HTML]{050CF2}{\textbf{Exchanges}} + \textcolor[HTML]{F20C05}{\textbf{High-risk}}) and around 10\% to other ``entertainment" purposes. And only around 30\% of them are dedicated to \textcolor[HTML]{A7B2B6}{\textbf{Other}} DApps -- potentially including the use-cases promised by media like smart rental services, voting through SCs, or even transplants management -- all the use-cases which made the SC technology such a hyped phenomenon.

%%%%%%%%%%%%NEXT%%%%%%%%%%%%%%

\section{Possibilities and Realities}
\label{Field}

New ideas are generally only accepted, if they seem to improve on a given state. This, of course, with respect to the technological and legal frameworks, as well as administrative possibilities. The abstract question to be kept in mind is: Does it make sense for a given enterprise to invest into creating BC/SC frameworks?

\subsection{Defining Smart Contracts}
\label{DefiningSC}

Whereas $Traditional$ $Contracts$ (TCs) are defined as a promise or a set of promises for the breach of which the law gives a remedy, or the performance of which the law in some way recognizes as a duty \citep{Bellia:2002}, SCs consequently are software that, for example, facilitates the generation and transfer of BC-based crypto-assets \citep[see further e.g.][]{Mik:2019}, and runs without human intervention once commanded to execute (see above PETs, section \ref{Sec:UnderstandSC}). Again, we must stress here: SCs do not execute without input, which can of course be a timed-coded threshold. SCs are $not$ self-executing and respective information might not be stored on the BC directly, but has to be accessed via different sources. SCs do not possess a $will$ to execute themselves, they just do what is being input. Legal laymen commonly define these as $contracts$, however, these terms are creating confusion: the terms of $account$ and $transaction$ can have very different meanings and so does obviously the term $contract$. Both fathers of this invention, \citet{Szabo:1997a} ("a computerized transaction protocol that executes the terms of a contract") and \citet{ETH:WP} ("systems which automatically move digital assets according to arbitrary pre-specified rules") define and underline, that SCs are $not$ contracts in the traditional terminus but means of contract execution -- a definition, which is conform with \citet{BoC18}.\\

The isolated full legal status of SCs, as pursued partially in the United States, does not change this and has no influence on completely different legal systems. However, for example, legislation in the state of Tennessee is of interest: ``Smart contracts may exist in commerce. No contract relating to a transaction shall be denied legal effect, validity, or enforceability solely because that contract is executed through a smart contract" \citep{TennCode:2021}. Yet, any further inclined discussion on this issue will open Pandora's box, as it misunderstands the technology way beyond its existence of being PETs, especially when mere supply chain reporting SCs are accepted as legal means of being ``contracts" instead of ``tools". Of concern would be the SC application when being used to spawn then forcibly legally accepted loophole-``currencies"  -- just think about ISO 4217 with its three different US dollars 840/USD ($cash$), 998/USS ($same day$), and 997/USN ($next day$) being simple compared to the chaos created by the number of existing and already abandoned CCs (see further e.g. \href{https://deadcoins.com/}{deadcoins.com}). The resulting issues of classification and handling these entities legally will result in an increase in insecurities. Hence, the discussion is a priori not even remotely leading to a result, as SCs were not even intended to $replace$ the traditional instruments.\\

%We will touch on definitions and code examples to eventually form a composite structure combining the complex arrangements of law and information technology, that is most essential for any digital advancement with real life impact.

We propose the following proposed definition to create the necessary -- yet missing -- connection between the factually more and more intertwined fields of information technology, statistics, and law:

\vspace{0.2cm}

\blockquote{$Smart$ $Contract$:\\
Software algorithm that is a blockchain application, digitally signed, computable and potentially dynamic in state, given the input potentially immutable, non-physical agreement between two or more parties, that relies on a consensus system and conditional constructions which can dependent on potentially off-chain third party information input, that is creating or based on mutual relations and/or obligations which it seeks to technically execute in the event of undisturbed predetermined code procedures.}

\vspace{0.2cm}

%Hence we will start with a short discussion, if the term $contract$ may be universally applied to smart contracts as well.

%Complex processes, overcrowded with intermediaries such as derivatives trading can certainly be restructured in a more efficient way, but at the given moment in time there exist many reasons why these have been implemented, as their value and knowledge can not be replaced by smart contracts today.\\

With the creation and emission of tokens being included in the above definition and whilst the legality of crypto-$currencies$ is not undisputed in many legal systems, SCs can be independent from these and are therefore not necessarily prima facie a negotium non grata for any given legal framework \citep[see further][]{XN:2020}. A discussion, if one does or does not accept $cryptocurrency$ as a $currency$ and consequently a $statutory$ $means$ $of$ $payment$ per se, and thus of which type the agreement at hand is in a legal sense (exchange or purchase contract in that meaning) is redundant as it is at least an intangible asset.\\

Yet, as with any given advancement, CCs and SCs first need to stand proof to the legal system before they may be accepted as valid modus of exchange. Improving efficiencies and facilitate dependable results, the promised key advantages of using SCs, are conclusive repetitions of the industrial automation. Technological advances have fostered the human thrive of rather trusting a machine, than human information, to a level that non-physical processes can now reliably be mirrored by machines. Nonetheless, even with $deep$ $reinforced$ $machine$ $learning$ and other consorts of technical marketing language, the basic question of $smartness$ is fundamental. Are SCs $smart$ enough for taking the $responsibility$ for decision making, thus can we speak of $contracts$ that can even replace legal entities? An argument against that would be of course, that the $smart$ human input is needed to kick-off SC processes -- but where is then the $smart$ $machine$? A code or document can only be as smart as the human initially writing it implying a plethora of possible sources for errors (see further section \ref{Sec:UnderstandSC} and appendix \ref{Appendix:ExampleSC}). We will not touch on the controversial ideas of $code$ $is$ $law$ \citep[$Note$: a replacement of law by code was literally not what was envisioned by][]{Lessig:99} as complex natural language can not be completely replaced by code at this given level of technological advance. There exists a wide variety of fields where SCs could nevertheless already be applied in a trade-off consideration. With many advances already accomplished towards the direction of mirroring traditional ways of interaction to digitized counterparts (recall IPO and ICO), a general comparison between SCs and TCs lead to a framework for understanding their potential:

 \begin{itemize}[leftmargin=*]
   \item[$\boxdot$] \textbf{TC}'s are created by legal professionals using their $lingua$ $franca$, handed out to each party as $physical$ copies, and enforced by institutions such as courts using an agreed upon $legal$ $codex$. These agreements are considered to be effective, if they stand the test of the law and according ruling. Their main problems can hence be broken down to their creation and enforcement being $time$ $consuming$ and $costly$ due to the controlling intermediaries involved, as well as their outcome being potentially $ambiguous$ given changing rulings or unclear wording.
   
   \vspace{-0.3cm}
   
   \item[$\boxdot$] \textbf{SC}'s, in contrast, promise to cut on these shortcomings by being $quickly$ createable through the help of $easy$ coding and $seamlessly$ understandable user interfaces. They purportedly define iura et obligationes the $same$ $way$ as TCs, hence are completely $deterministic$, while being $entirely$ $digital$ and therefore easy to mathematically proof and enforce, through algorithms and a given $consensus$ $codex$, adding to $efficiency$ and $verifiability$. These agreements are considered to be effective, if the agreed upon action, for example a TX, was performed without technological error.
 \end{itemize}

If an SC can really be seen as a universal replacement to traditional agreements in a legal sense may be clearly answered at this given state: $No$. It's a technical possibility to automatize given relationships according to pre-defined rules that depend on pre-defined coded events, not necessarily replicating legal standards or legal relationships. In other words: they facilitate traditional agreements technically without being classifiable as such themselves. However, their functions and proven fields of application can already replicate certain traditional functions of agreements/contracts.\\

We may mark the most basic understanding of what a traditional agreements should fulfill as we are pointing to the question of $replaceability$ and the likely $replacable$ parts of our contractual system:

 \begin{itemize}[leftmargin=*]
   \item[$\boxdot$]  Legality of the agreement itself
   \vspace{-0.3cm}
   \item[$\boxdot$]  Definition of the parties
   \vspace{-0.3cm}
   \item[$\boxdot$]  Definition of the subject et cetra
   \vspace{-0.3cm}   
   \item[$\boxdot$]  Offer and Acceptance 
   \vspace{-0.3cm}
   \item[$\boxdot$]  Ability and/or intention of the parties to be legally bound
   \vspace{-0.3cm}
   \item[$\boxdot$]  Capacity to satisfy contractual elements
   \vspace{-0.3cm}
   \item[$\boxdot$]  Signatures
 \end{itemize}

Taking this into account, the question of the $replaceability$ of TCs with SCs is more tangible. We can see that certain contracts can, for a fact, be reproduced as coded Doppelgänger's. From the standpoint of automation, SCs present an analogous form of agreement/contract, where only details such as the contractual parties, date, subject, et cetera have to be added -- the easiest to understand example would be an ICO. Potential can also be seen regarding $adhesion$ $contracts$, $shrink wrap$ $contracts$, or $boilerplate$ $contracts$, which refer to predefined standard form contract clauses and are commonly used for matters involving leases, purchases, insurance, mortgages, amongst many other use cases (with special regards to certain national limits these need to be differentiated from general terms and conditions). Such application fields could increase efficiency in contract law decrease processing time and hence also costs. Adhesion contracts are usually engaged as digitally signed $clickwrap$ agreements that offer individuals the opportunity to accept or decline digitally-mediated contents. These electronic agreements are requested to appear as identical to physical contracts as possible. Although not accepted in all legal frameworks, documents which are signed electronically possess equivalent legal validity to physically signed agreements/contracts.\\

The United Nations have worked on respective measures to replicate traditional frameworks for digital societies since at last 1996, resulting in a Model Law on Electronic Signatures that is of importance due to the beforehand mentioned algorithmic uniqueness of private keys/signatures \citep{UNICITRAL:2001}. The United States of America's  Uniform Electronic Transactions Act (UETA) from 1999, and subsequently the United States Electronic Signatures in Global and National Commerce (ESIGN) Act from the year 2000, present a national legal solution to the issues raised by a digital society. They grant legal recognition to electronic signatures and records, given that the contractual parties choose to use electronic means for their agreements. Besides special requirements, such as having the intent to sign such an agreement and common consent to do business electronically, the requirements of associating the signature with continuously updated records that can be fully accessed, plays into the hand of BC-based constructions such as SCs. Under this legislation, in order to be accepted as a prima facie effective agreement, it is mandatory to have an associated record reflecting how the signature was generated or an attached proof that this agreement was reached by using electronic signatures. Moreover, the electronic signature records need to be capable of retention and accurate reproduction -- once again, basic functions provided by BC-based systems as outlined in section \ref{Sec:UnderstandETH}. We can conclude that certain legal frameworks have been established that can be used directly or in an analogue way for SCs. BC users can: 

 \begin{itemize}[leftmargin=*]
   \item[$\boxdot$]  be uniquely identified and linked to a signature
   \vspace{-0.3cm}
   \item[$\boxdot$]  have sole control over the genesis of their signatures
   \vspace{-0.3cm}
   \item[$\boxdot$]  identify accompanying signed data stored on the BC
   \vspace{-0.3cm}
   \item[$\boxdot$]  easily prove the event that the accompanying data has been changed and revoke their consent
 \end{itemize}

This leads to the possibility to also accept digital signatures of uniquely identifiable BC participants as legally valid signatures for contracts -- in that sense SCs. Therefore, while certain steps towards the digitization of TCs have been accomplized, SCs can only be defined as technological means to facilitate TCs and not as contracts $sui$ $generis$, or $authenticated$ $agreements$.

%%%%%%%%%%%%NEXT%%%%%%%%%%%%%%

\subsection{Implementation}
\label{Sec:Implementation}

An ever increasing number of potential use cases are envisioned that try to mimic TCs in their function: insurance, mortgage, financial derivatives, human resources, supply chain management, to name a few. Typically each individual field is bounded by specific procedural legal standards and SCs are best applied where maximizing efficiency and optimizing cost-minimization, or the removal of traditional processes that require vast amounts of manpower to keep the checks-and-balances-system running, is a possibility. SCs presumably fit in these automation-gaps naturally, as many of the intermediaries become redundant with monitoring and information propagation being established via trustworthy standardized coding. Especially financial applications have regulatory requirements, like $Know$-$Your$-$Customer$ (KYC) to check if a participant is eligible for a given service according to regulatory watch-lists such as the Office of Foreign Assets Control (OFAC) Specially Designated Nationals (SDN) list. Keep in mind that, while one does not need a BC construction for these tasks of course, one does need it for fostering the idea of transparency and immutability (see further sections \ref{Sec:UnderstandBC} and \ref{DefiningSC}). SC applications are predestined for systems where currency is not a necessary medium of exchange and centralized allocations are not feasible or preferred. SCs do not necessarily depend on CCs to work (in ETH they do, see section \ref{Sec:UnderstandSC}), which imposes their adaptation also within governmental systems where CCs are deemed to be illegal. For example, organ donation and procurement could be improved through coded most favorable matching structures, where the inherent BC-based recording of input would be beneficiary especially regarding the legal patient's provisions that could be respectively stored data for and in an SC. We therefore note, that a crucial point to introduce an SC-based system is $risk$ $control$. Risk pooling can be achieved in both SCs and TCs as a risk-sharing structure, yet modern technologies are obviously superior in these fields than paper-based auditing systems could ever be. Another critical point is thus $monitoring$, as a measure to continuously audit the contractual terms and to detect potential problems or breaches.\\

To make the legal handling of $particular$ SCs replacing $particular$ TCs more tangible, consider a motel doing a series of standardized and proven rental SCs with digital access codes to the apartments. Instead of relying on an intermediary, for example a receptionist, to administratively keep watch on the payments, the landlord creates an SC factory to spawn individual SCs (see further \ref{Sec:UnderstandSC}) with all respective compulsory parameters. Additional actions like for example accommodation reporting to governmental institutions, may also be encoded through an beforehand outlined oracle to provide legit and trustworthy information on the individuals identity. If the conditions of this SC are met by the parties, i.e., that the exact rent was transferred to a given account and client information was forwarded successfully et cetera, then access to the rental object is granted. Yet, if the conditions required to be fulfilled are not met, then access is denied. This is arguably not a legal contract itself, but just a mean to process the corresponding TC itself. However, one could argue that this example could serve as proof for the replaceability of at least basic and often repeated TC cases. Adding to this is, that more than three quarters of the SCs we analyzed are serving the purpose of defining a value token. A selected few of them implement interesting distribution schemes like a \href{https://github.com/maurelian/dutch-auction}{Dutch auction} or a \href{https://godsunchained.com/}{collectible cards game} mechanism, but most of them are made of repeating code patterns for a distinct variety of use cases. As a consequence, since not that many SCs consist of more than 600 lines of code, these SCs are much easier to read and understand once one is familiar with a small subset of code examples. Although this shows that the developers want to employ best practices in their code this does not necessarily mean that the reused code is secure and bug-free  \citep[see further][]{pinna2019,ContractStacks:2020}. Furthermore, nefarious SCs are harder to discern from regular SCs since their codes and the codes of legitimate SCs are so similar. This is underlining the fact, that previously isolated fields of IT and LAW are required to, and will be, increasingly intertwined.\\

% Besides a handful of specific variables, the main indicative feature of scam contracts are comments. Sometimes the authors of the contract would explicitly announce in the comments, that a given contract is a pyramid scheme or promises unreal Ponzi scheme interest rates.\\

%. It means that this is a big promise for training a predictive model for a classification task
% Furthermore, when it comes to the minority of unique and interesting SCs, it will be another challenge for building a predictive model, since it will have to deal with a big imbalance in the training dataset.

An endless field of further imaginary applications in Sharing Economy, LegalTech, InsurTech, Dispute Resolvement, Royalty Management, voting, insurance contracts, electricity prosumer management, trade finance, escrow services, monitoring of regulatory compliance, copyright services and rights of use, traceability of products through supply chains and markets, or the famous field of Internet-Of-Things (IoT) applications, and many more buzzword-flooded fields can be envisioned.

\subsection{Decentralized App vs. ``traditional" App}
\label{Sec:DApp}
One of the key reasons for many startups to fail is the limited understanding of the hyped technology and most importantly its differences to already existing technologies. Some companies are even trying to follow the hype and force to implement such buzz-concepts as BC- and SC-based without properly understanding whether it is feasible and profitable on the stack they are using -- bigger ones silently abandon such endeavors eventually.\\

SC -- is one of such technologies, where the word  ``smart" promises futuristic digital tools, whereas what SC really are -- just a coding script running on the BC, a part of a decentralized App, similar to how interconnected multiple coding scripts written in Python, Java or other languages make up an App. Basically, DApps can do almost the same things (or even less, as it will be described further) as a respectively oriented normal Apps, while having the computational processing ran on the BC. Therefore, in this subsection, we are describing the key differences between Apps and DApps. Moreover, inspired by the flowchart proposed by \cite{wust2018you} ``Where does blockchain make sense", we are going to propose a flowchart for deciding whether one should choose DApp over a traditional App or vice versa. As one DApp consists of one or multiple SCs, our flowchart also aims to answer the question whether one should use SCs or not or in other words whether you should execute your code on the BC or not.

\vspace{-0.5cm}

\begin{figure}[H]
\hfill
\subfigure[Centralized]{\includegraphics[width=5cm]{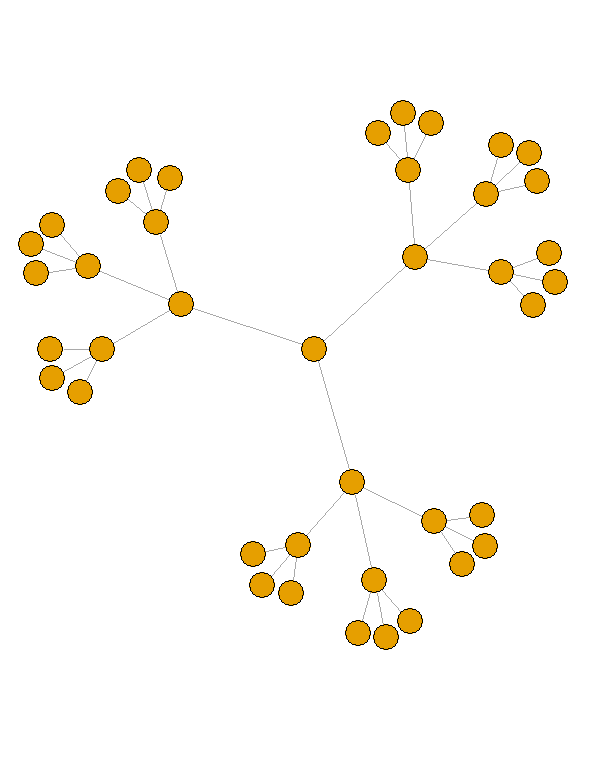}}
\hfill
\subfigure[Decentralized]{\includegraphics[width=5cm]{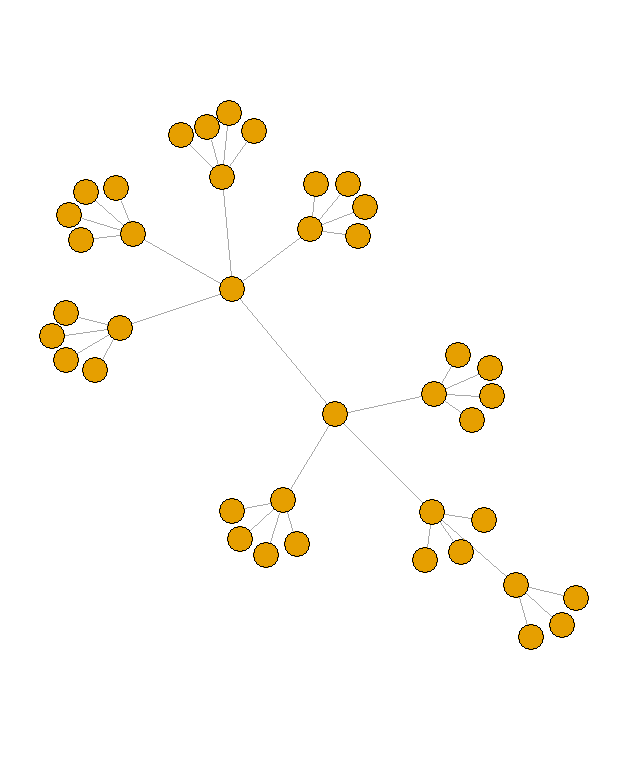}}
\hfill
\subfigure[Distributed]{\includegraphics[width=5cm]{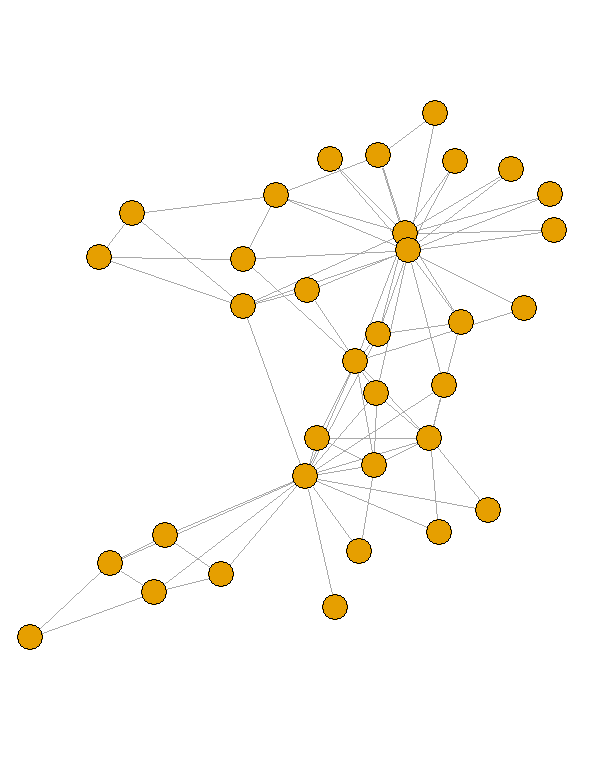}}
\hfill
\caption{Types of software applications Apps \href{https://github.com/QuantLet/CrixToDate}{\includegraphics[keepaspectratio,width=0.4cm]{media/qletlogo_tr.png}}}
\label{fig:apps}
\end{figure}

According to \cite{laplante2017dictionary} an application or app, is a program or a bundle of programs that are designed to solve a specific problem. Usually, one refers to apps or to application software -- to programs that run outside of the operating system. Usually apps are ran on the desktop, on mobile devices or in the webbrowser. Subsequently, we can identify webapps, mobile Apps or desktop apps. However, in this work, we will refer to all of these three types as to Apps. The majority of Apps, at least of webapps, follow a centralized server-client model. Moreover, some are distributed and the new type uses the decentralization \citep{raval2016decentralized}. In Figure \ref{fig:apps}, these three types of architecture are depicted. As we can see, in the centralized system, the most frequently found type of Apps, a central node controls the operation, flow of the information and interconnection between child nodes. All client-nodes are dependent of the central node. Distributed systems allocate computation to different nodes, many big companies make use of distributed computations. Therefore, a system can be centralized and distributed at the same time: e.g., a company provides computation services, the computations are distributed among many servers \citep{raval2016decentralized}, however the company is itself the central node holding the responsibility of the services. So if one server fails, nothing will happen to your data because of the distributed nature, however, if the whole company fails -- all your data could get lost. Decentralized systems, on the other hand, implies that no node in the hierarchy is higher than the other and can give instructions as to what to do. And all nodes hold similar amount of responsibility for the functionality of the system.\\

Therefore, looking at the definitions of the DApps and ``traditional" Apps, we cannot say that these are dichotomous concepts, but rather DApp is a subset of the Apps. Subsequently, we will still speak of DApps vs. Apps where Apps here can be web, desktop, and mobile apps relying on a centralized or centralized distributed systems (having one main responsible authority if something will fail). According to \cite{DApp_app}, for an App to be classified as DApp, it must comply to the following criteria:
\begin{enumerate}
    \item the app should not have a controlling entity 
    \vspace{-0.3cm}
    \item be completely open-source 
    \vspace{-0.3cm}
    \item operate autonomously 
    \vspace{-0.3cm}
    \item work on BC 
    \vspace{-0.3cm}
    \item use cryptography (for tokens that act as proof of the value of the node and are distributed through a rewarding system)
\end{enumerate}

Even though we mostly agree with this list, the criteria that everything should be open-source might be correct, however, while going through multiple DApps and SCs on \cite{stateDApps} and \cite{etherscan}, we found many SCs and DApps that do not have an open-source license. 
In the following Table \ref{Tab:pros_cons}, we present the advantages and disadvantages of using DApps compared to traditional Apps:

\begin{table}[H]
\begin{center}
 \begin{tabular}{ll} 
 \hline
 \hline
 DApps & Apps \\ [0.5ex] 
 \hline
 + decentralisation  &  + user-friedly,  \\
  \hspace{0.3cm} if one node fails, the data won't be lost & \hspace{0.3cm} no technical knowledge required \\
 + p2p validation & + nothing needs to be installed \\
 + mostly open-source & \hspace{0.3cm} (in case of webapps) \\
 + BC,  proof-of-work  & + can be designed to user's needs \\
 \hspace{0.3cm} make the systems trustworthy &  + can be adjusted if designed wrong \\
 \hspace{0.3cm} so it almost impossible to hack &  + new functionalities can be added \\
 + support CC & \\
 + transparrency & \\
 \hline
 - TXs are very slow & - can be hacked \\
 - the amount of TXs is bounded & - if data is stored centrally, one electricity \\
 \hspace{0.1cm} due to technical limitations of the system & \hspace{0.1cm} outage can lead to the data loss\\
 - TX costs: &  - TX costs: \\
  \hspace{0.1cm} creation of a contract or sending a  & \hspace{0.1cm} information exchange in a presence \\
   \hspace{0.1cm}  TX requires a fee to pay & \hspace{0.1cm} of a third party is costly \\
   
 - cannot be taken down from the network & - data is not always encrypted \\
 - cannot be adjusted or changed & \\
 - redundant storage of the data & \\
  \hspace{0.1cm} and redundant computations  & \\
  \hspace{0.1cm} thus very high energy costs \citep{eth_index} & \\
 - therefore - not sustainable & \\
 - complexity limitations of the code & \\
 \hline
 \hline
\end{tabular}
\end{center}
    \vspace{-0.5cm}
 \caption{Advantages and disadvantages of using DApps}\label{Tab:pros_cons}
\end{table}

While the majority of the criteria are quite clear without additional explanations or have been already mentioned in the previous sections, we would like to go in the details of complexity limitations. In this work, we address the complexity from two perspectives: first, the straightforward limitations of the code complexity, which is also related to the TX costs: to compile the code on the BC, payment of a fee is required. Code of different complexity requires a different fee. And this fee is bounded to a certain value in order not to exceed it and not to consume all the balance for the code, which could mistakenly run too long. The second perspective of the complexity, which we are working with is the ability to make complex math computations. And even though, the DApp consists of ``Smart" contracts, their ability to do mathematical computation is very limited: only basic operations like Addition, Subtraction, Multiplication, Division, Modulus and Exponential \citep{math_sol} can be done on the ETH BC.\\

\begin{figure}[H]
\centering
\includegraphics[width=13cm]{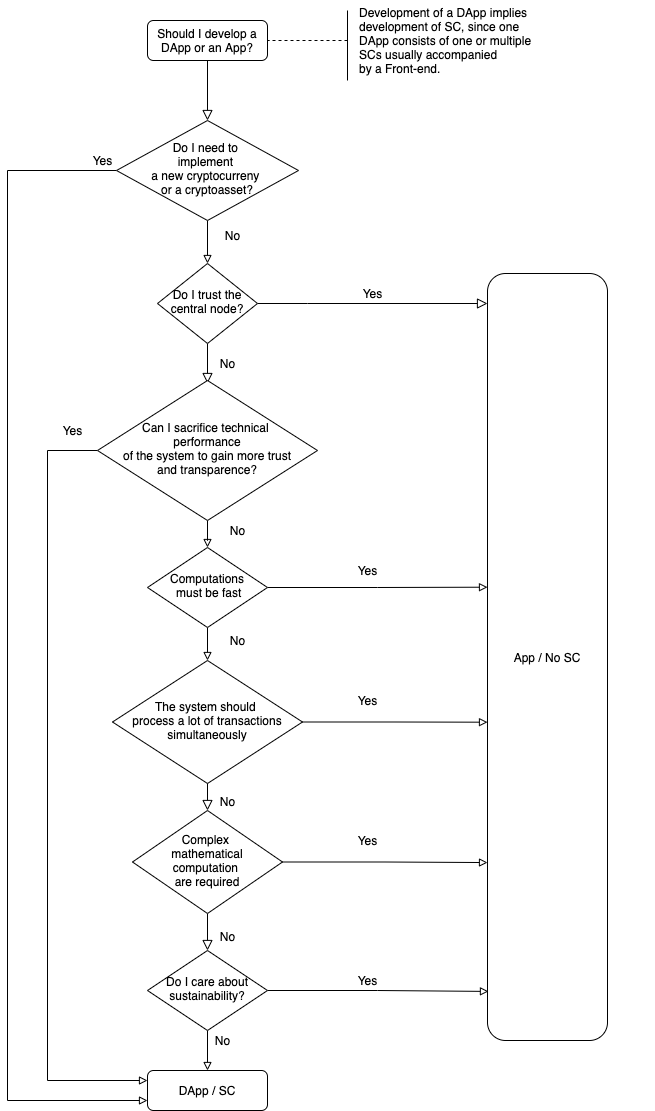}
\caption{A flowchart to identify whether a DApp or an App is a proper solution.}\label{fig:flowchart}
\end{figure}

% Talk about BC applications, revolve most frequently around finance. And sometimes media promises simultaneously security of the BC while giving the ability to invest money in different financial instruments. \\

We propose a decision flow in Figure \ref{fig:flowchart}, which can help to decide whether it is useful to develop a DApp or if one should stick to a traditional Apps. Further, we refer to trust from three perspectives: first, as to resilience of the system in case of failure or outage -- whether the data will not get lost, the system saves the last state and the usage of the App can be immediately continued, secondly -- that the node provides sufficient security that no third-party can alternate the data or retrieve personal data they do not have access to, and thirdly -- probably the most important perspective -- whether we trust the central authority that they will not manipulate data or the state of the system themselves. Therefore, we see that if we are not aiming to implement a CC or a cryptoasset and trust our central node, there are almost no incentives to use BC for developing Apps -- DApps. Thus, speaking of DApps, we have one main question which needs to be answered -- \textit{do we trust our central node}?

\subsection{Shortcomings and Legal Handling }
\label{Sec:Shortcomings}

Immutability is an obvious benefit with BC system constructions being coined as \textit{trust machines}, yet, absolute immutability can also be insurmountable hurdle. To provide a brief framework one may keep the following points on BCs \& SCs in mind:

 \begin{itemize}[leftmargin=*]
   \item[$\boxdot$]  \textbf{Interactability}: software and operational errors hardly removable or reversible, if at all. 
   \vspace{-0.3cm}
   \item[$\boxdot$]  \textbf{Scalability}: data storage and availability with permanent fixation of information to BCs and adjacent structures like the EVM, require increasingly sophisticated cloud- and physical-based management structures.
   \vspace{-0.3cm}
   \item[$\boxdot$]  \textbf{Sensitive content}: illegal, nefarious, or confidential information hardly removable
   \vspace{-0.3cm}
   \item[$\boxdot$]  \textbf{Regulatory issues}: conflicting national frameworks on consumer rights and data privacy, or personal legal register information like the EU General Data Protection Regulation or Rome I Regulation, the US Gramm-Leach-Bliley Act and Fair Credit Reporting Act, or the SEC Regulation S-P.
 \end{itemize}

\begin{figure}[H]
\hspace{-1cm}
	\includegraphics[keepaspectratio,width=17cm]{images/density.png}
        \caption{Network Usage Comparison, \textbf{\textcolor{cyan}{2018}}  \textbf{\textcolor{orange}{2019}} \textbf{\textcolor{green}{2020}} \href{https://github.com/QuantLet/USC}{\includegraphics[keepaspectratio,width=0.4cm]{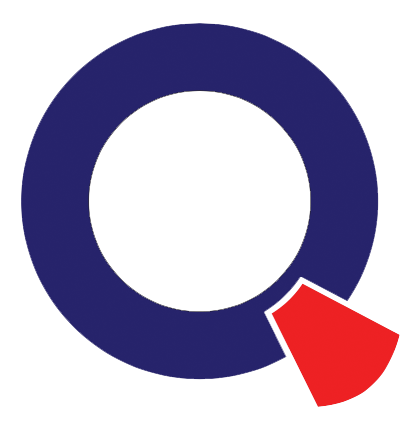}}}
        \label{Fig:NetUtilDensity}
\end{figure}

Despite SCs offering interesting new opportunities in various use cases (see further section \ref{Sec:Implementation}), not many new solutions for existing problems are actually employing them, particularly in the legal industry, although they were intended as a replacement for TCs. Of course, there exist decent examples of companies putting the technology to use, such as the likes of \cite{tracr} and \cite{inmusik}, yet the scale of the technology adoption is to this day rather small. While novelty and unfamiliarity surely are factors, another reason for that may be the beforehand mentioned scalability issues. An outstanding example is the very well-known \href{https://www.cryptokitties.co/}{CryptoKitties} game SC \citep{cryptokitties}. At its peak accounted for roughly 12\% of all TX on the ETH BC \citep{quartz_kitties} leading to the amount of pending TX increasing by a sixfold \citep{bbc_kitties} with execution time and costs of TX considerably high as the system got clogged with ``cats" (see further section \ref{Sec:UnderstandSC}). Vitalik Buterin expressed more than once that scalability remains a bottleneck and hinders widespread adoption of BC platforms \citep{star_buterin}. The ETH network utilization remains quite consistently over the 80\% mark moving towards a constantly higher than the 90\% region, as visualized in Figure \ref{Fig:NetUtilDensity}. While some big players like \cite{ubs_smart_bonds} experimented with SCs, not much was heard about it afterward with disintermediation and decentralization through SCs, and BCs are rarely able to replace common structures effectively \citep{Greenspan_2016}.

\vspace{-0.5cm}

As outlined in section \ref{Sec:UnderstandSC} and appendix \ref{Appendix:ExampleSC}, it is important to note that first and foremost, the code of SCs must be absolutely bug-free and secure from nefarious inputs as much as possible. Otherwise, exploitation of bugs or security flaws can be an insurance business and puts the saying of ``BC-based systems mean security" ad absurdum. An important factor affecting the quality of SC source codes and their legal handling is that only 1\% of all SCs have a publicly available source code \citep[amongst other sources, so-called \textit{verified} SCs on Etherscan. See further][]{nikolic2018}. The remaining 99\% can only be read via the compiled source code. This does not necessarily mean that those other 99\% of SCs have qualitatively better or worse code, but it would reinforce good practices and enrich the knowledge base of SC developers to have access to all source codes, preventing possible further DAO-esque incidents.\\

\textit{The DAO} \citep{dao} -- a construction typus, which can be seen legally as a cooperative, partnership, or unincorporated association with participants having usufructuary or usage rights -- lead to a split/\textit{forking} of the initial ETH BC into an altered BC, now known as ETH, and a BC that kept the exploited status quo, now known as Ethereum Classic (ETC). In fact, the ETH community regulated itself in the absence of jurisdiction, as the value lost during this happening due to bugged coding and nefarious exploitation of that error could legally not be seen as value back then -- hence ``nothing" was stolen. This has since changed and respective assets are commonly seen as non-tangible assets, which can also be subject to taxation. Similarly, the software hosting project Edgeware -- which accumulated investments of 300 000 000 USD in July 2019, a month after going live -- held an exploitable coding, which would make it unable to unlock and emit funds from their SC construction \citep{ContractStacks:2020} -- an issue , which is also present nowadays, if the ETH SC coder simply forgot to include the function or to set the contract correctly as \textit{payable} (see further appendix \ref{Appendix:ExampleSC}). Finding bugs and imperfections inside the code just by looking at it is not easy, therefore it is hard to assess how good or bad the code of more complex contracts is and throughout testing is essential. This induces costs as repeated auditing and testing is required to be performed by specialists. Additionally, as it could be expected from Solidity being an object-oriented programming language (see section \ref{Sec:UnderstandSC}), the \textit{open-closed principle} of the object-oriented software design \citep[i.e., open for extension, but closed for modification. See further][]{meyer1988} is especially important in the case of SCs, since it cannot be altered once an SC is deployed with a CA on the ETH BC unless the BC is forked at a point prior to the deployment. We did not encounter extension properties and outstanding modular development in the subset of the SCs we analyzed, yet, one could imagine that with time more and more examples of such missing elements could become available. \\

 %\textcolor{red}{DAO Hacker = CC illegal, Real legal -> erwerb von Genuss-/Nutzungsrechten}\\
 
 %\textcolor{red}{DAO = Energy e.g., Genossenschaft, oHG, nichtrechtsfähiger Verein}\\

\begin{figure}[H]
	\centering
	\includegraphics[keepaspectratio,width=13cm]{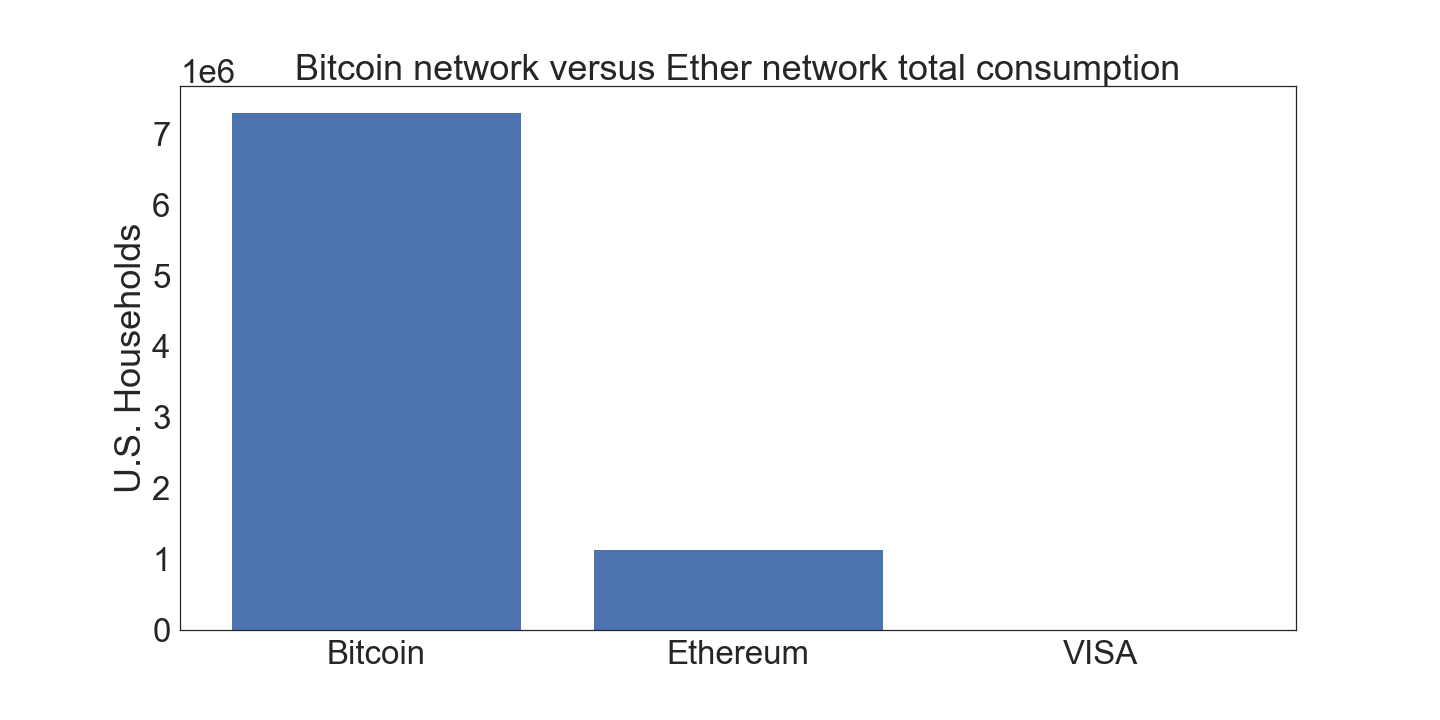}
        \caption{Energy Consumption of BTC and ETH compared to VISA, 20170527 - 20201210 based on \cite{eth_index}}
        \label{Fig:EnergyVISA}
\end{figure}

\begin{figure*}[!h]
\begin{multicols}{2}
    \includegraphics[width=\linewidth]{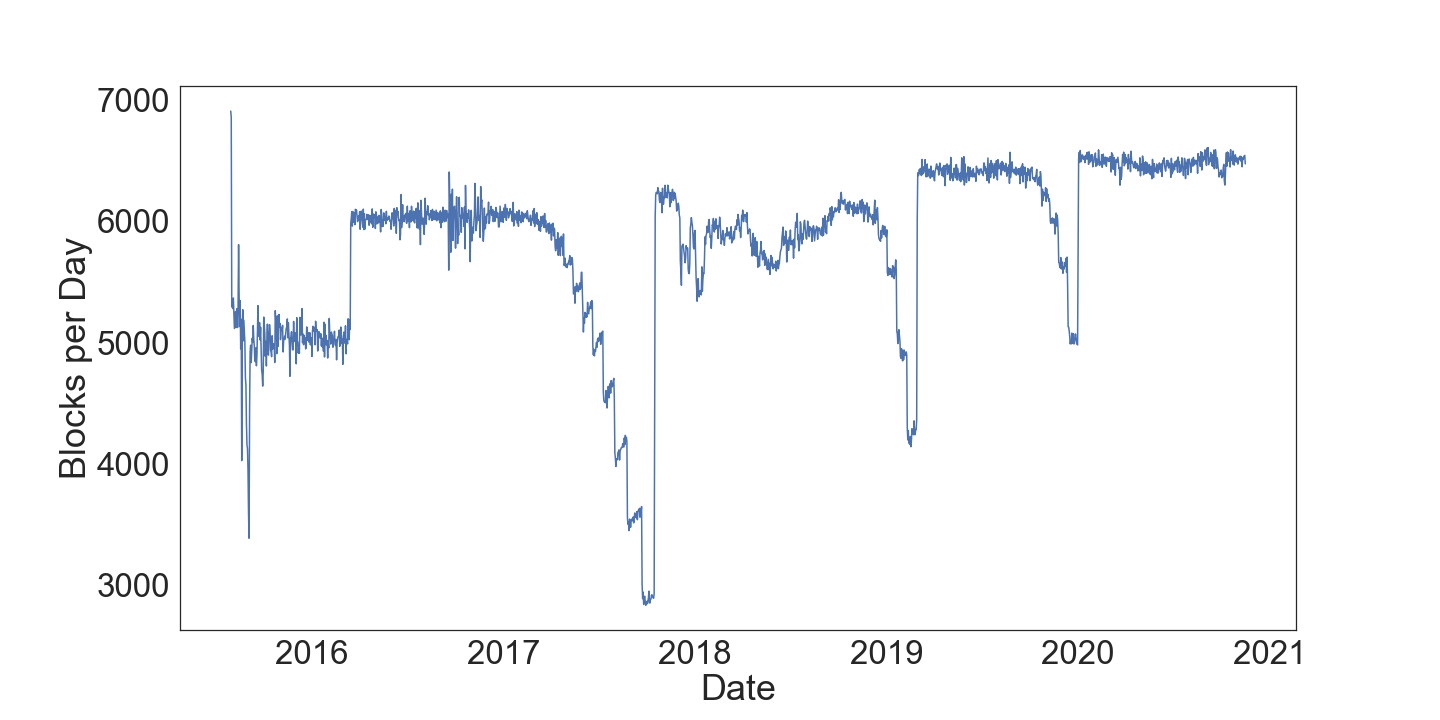}\par 
    \includegraphics[width=\linewidth]{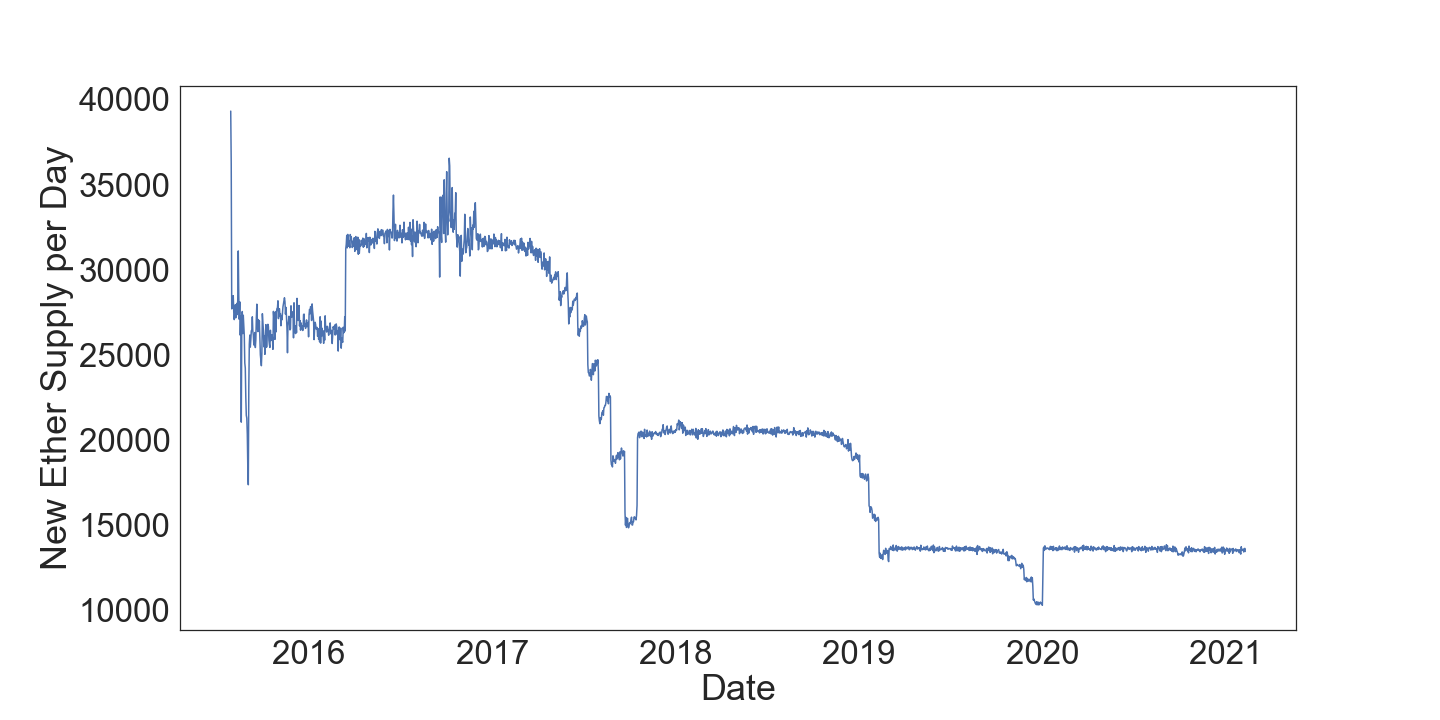}\par 
\end{multicols}
\begin{multicols}{2}
    \includegraphics[width=\linewidth]{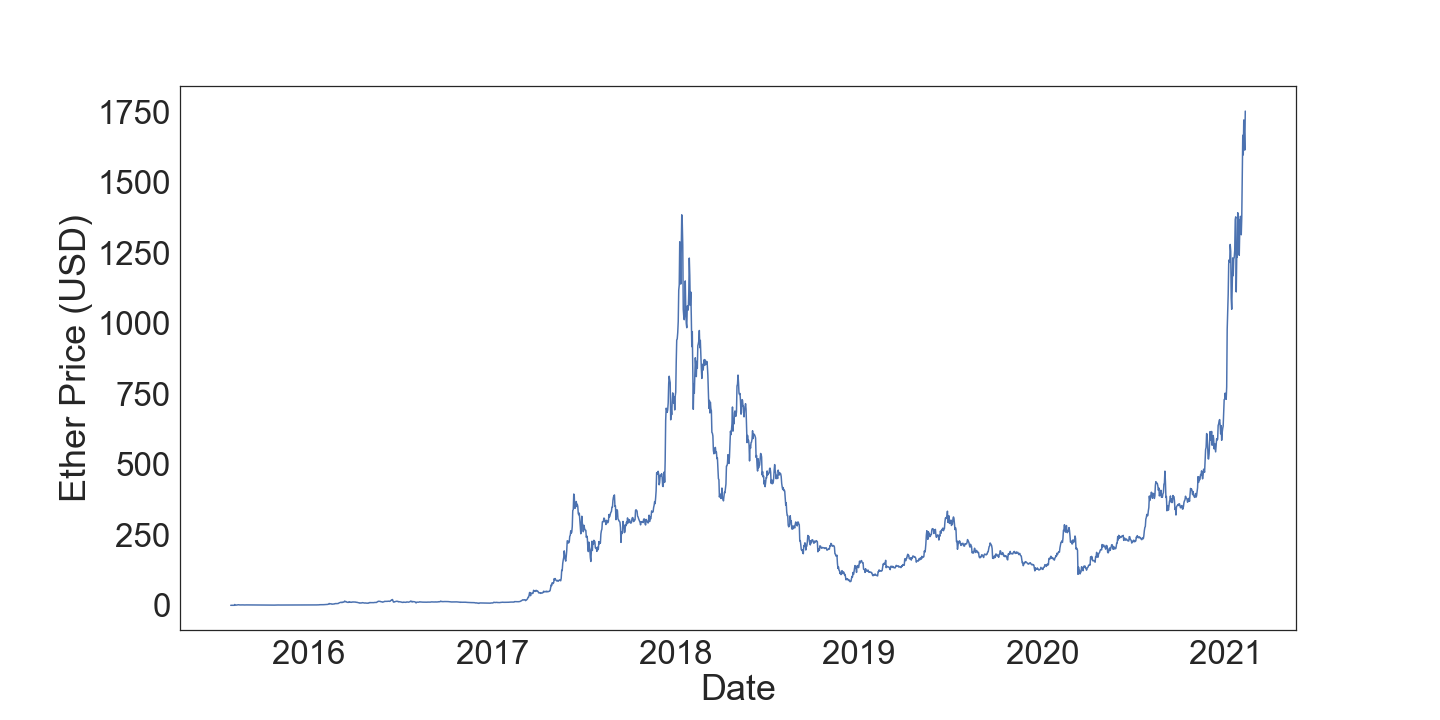}\par
    \includegraphics[width=\linewidth]{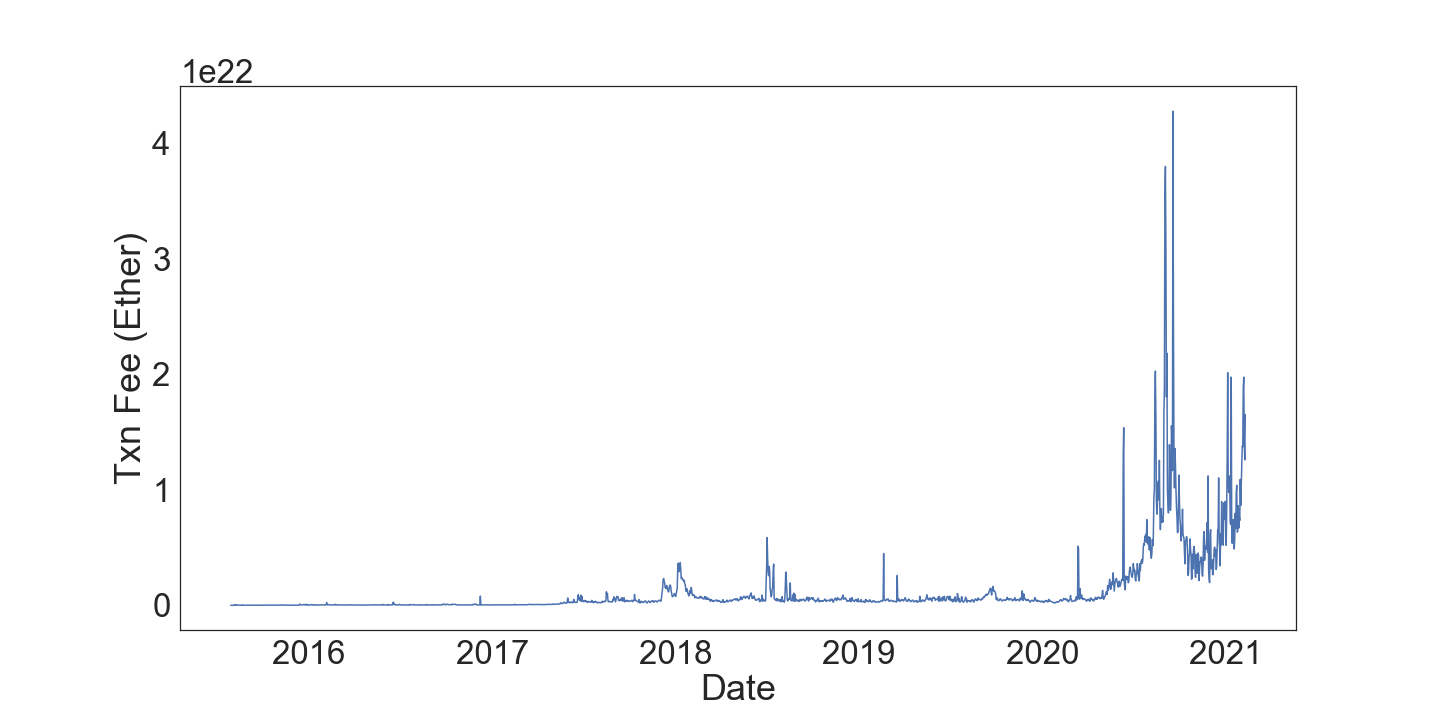}\par
\end{multicols}
\caption{Network Parameter time series, 20150630-20210211  \href{https://github.com/QuantLet/USC}{\includegraphics[keepaspectratio,width=0.4cm]{media/qletlogo_tr.png}}}
\label{Fig:Mining}
\end{figure*}

 If SCs are really cheaper than their traditional counterparts is also often up to the field of employment. First off is the question, if it makes sense for a small or medium enterprise to actually invest in such an endeavor. This consumes a lot of development time and human resources, as well as a considerable amount of funds to make sure the codes are (possibly) free of bugs and exploits. On a second look the question is, if such an SC-based system makes sense in the long run for a plethora of cases. Here it is interesting to look at the price of Gas related to certain opcodes (introduced in section \ref{Sec:UnderstandSC} and denoted in Appendix \ref{Appendix:Wei}). The following table \ref{Tab:OpcodeCosts} presents the Gas cost for certain opcodes \citep[see further: Appendix G: Fee Schedule;][]{YellowPaper} and how much that would cost if done a million times with an arbitrary stable Ether exchange price of 250 USD. Executing the ADD opcode once will therefore cost $3*10^{9}$ Gwei which gives a cost of 0.00000009 Ether and 0.0000225 USD -- respectively done a million times, this gives 22.5 USD \citep[see further][]{Ryan:2017}. When comparing this data storage and propagation technique -- without taking the energy intensive mining process into consideration -- to the pricing of \href{https://aws.amazon.com/s3/pricing/}{Amazon S3 AWS Services}, the benevolent reader may turn pale looking at the \textit{SSTORE} opcode and ideas like: ``Just store a natural language version with the SC code" given volatile price structures like the previous maximum of 1 427.05 USD for 1 Ether on the 13. January 2018 (see further Figure \ref{Fig:Mining} -- not mentioning the costs to set up and understand such a system). Data received through the \href{https://digiconomist.net/ethereum-energy-consumption}{Ethereum Energy Consumption Index} \citep{eth_index} is providing a self explanatory overview regarding the cost in energy compared to traditional proceedings in the Figure \ref{Fig:EnergyVISA}. \cite{eth_index} and \cite{de2018bitcoin} propose the Bitcoin Energy Consumption Index, where they estimate energy costs of TXs lying on the BC, the same approach they propose using to estimate the energy costs of ETH. \cite{eth_index} show that the footprints of a single ETH transaction is equivalent to the power consumption of an average U.S. household over 2.05 days and the annual footprint could be compared to the power consumption even of a whole country (e.g., compared to Lybia as of 15 March 2021\cite{eth_index}). For a deeper look into the statistics of the EVM we recommend, for example, \cite{Swende:2019}.

\begin{table}[H]
\centering
\begin{tabular}{llrrr}
\hline
\hline
opcode  & description                       & Gas & Ether & USD*1M \\
\hline
ADD     & Addition operation.               & 3           & 0.00000009    & 22.5           \\
MUL     & Multiplication operation.         & 5           & 0.00000015    & 37.5           \\
SLOAD   & Load word from storage.           & 200         & 0.000006      & 1.5            \\
BALANCE & Get balance of the given account. & 400         & 0.000012      & 3              \\
SSTORE  & Save word to storage.             & 20000       & 0.0006        & 150   \\
\hline
\hline
\end{tabular}
\caption{opcode costs examples in Ether and in USD when done one million times \linebreak (1 Ether = 250 USD, mining fees and energy costs not taken into consideration) Network Energy Consumption Comparison with ETH, 20170527 - 20201210 \href{https://github.com/QuantLet/USC}{\includegraphics[keepaspectratio,width=0.4cm]{media/qletlogo_tr.png}}}
\label{Tab:OpcodeCosts}
\end{table}
 
 \vspace{-0.5cm}
 
As we have already touched on the issue of mining in section \ref{Sec:UnderstandSC}, we just briefly want to add on the issues related to the costs of mounting and running SCs contrasted to the grand scale of things happening in ETH. We saw, that the price of Gas, the fuel needed for every action of SCs, is directly bound to the price of Ether. Therefore, not only the setup of some SC may be very expensive to ensure bug-free runtime, but on the other hand bug-free yet laboriously written SCs may consume more Gas than actually needed for the individual application at hand. One can compare this to the energy consumption of badly coded simple action applications on a handheld device: having, for example, a badly written step counter on your mobile phone is a nuisance and needlessly consumes energy through requiring more computing power. Figure \ref{Fig:Mining} visualizes, that hence the price of running the same SC can be very expensive to begin with, but depending on the market price of Ether, it can be even more expensive.

A discussion of the term \textit{smart} is hence needed for a basic understanding of the potential decoupling between technological advances (code) and legal frameworks (natural language) which require insight into the source codes analogue to raw natural language. At the moment, SCs can not adapt to situations dynamically, but instead only return based on the fulfillment or non-fulfillment of the predefined conditions. Indefinite legal terms commonly used, like ``to strive for" or similar, are not yet codeable and hence the technological aspects lack considerably behind flexibility offered by TCs. Is this state, however, a consequentially bad outcome? By recalling the quote by Nick Szabo from the introduction, this is not necessarily a bad result given the human nature \citep{Szabo:1997}. Without the intention to create an Osborne effect, extended usage of SCs is momentarily quite limited due to the lack of \textit{smartness}. For example: any given SC, on a more advanced level than just financial TXs, cannot identify any violation as these can come in infinite forms and SCs are only coded to react according to respectively written triggers. As most issues for SCs are of a time-varying, stochastic, and humanoid nature -- especially identifying problems \textit{ante factum} -- they require an agile technological and legal framework that is hard to implement at the given technological state.\\

A failsafe ``bugout" solution in the hands of a trustful third party would be a solution, which however would make it a circle argument for some governmental backing of such systems to induce trust into the \textit{intermediary-less BC trust machine}. Plenty of legal literature and decisions are available on non-/wrong-fulfillment of any kind of agreements, ranging from, for example, an automobile being delivered in the (partially) wrong color, to not having the expected quality of lacquer et cetera perge perge. These fine nuances are not representable by SCs unless the contractual parties agree \textit{post id factum} to alter the SC -- which is unfortunately only safely possible by deactivating the initial SC and by redeploying an altered one as a different instance. The fundamental problem of \textit{legal} \textit{effectiveness} of a \textit{particular} contract also has to be taken into consideration, like for example, the legal age or a mental illness of a party, or if digital signatures can be accepted au lieu de physical signatures in a given case. However, these discussions revolve more about the regulation of oracles and, therefore, data service providers -- a field that is not linked to SCs, but to classic service contract law for which a plethora of literature exists. It is \textit{not} necessary to create yet another legal framework, as SCs  are trying to be technological portrayals of their legal counterparts -- a rental contract will still be a rental contract, just in a different presentation.\\

This serves to underline the redundancy of an elaborated exegesis of fulfillment problems in the context of SCs, as there are more legal problems that may hinder SCs to be classified as actual and effective contracts in their legal meaning. While an aficionado may certainly be able to write a technically effective smart ``contract", it does not mean that this ``contract" is legally effective, leave alone legally enforceable. Depending on respective legal constructions, a contract may also only be partially effective, i.e., the obligation to fulfill the contract may be ineffective, yet the transfer of ownership/property part of \textit{the} contract may be effective. Also, the contract needs to follow a certain phenotype, like being in a certain language -- where it can be up to discussion, if a programming language can be chosen given the envisioned type of agreement. In the end, this is defined by the already existing respective national limits of contractual freedom. Once again we recurse on the fact that a third party -- possibly governmentally backed system is needed to solve problems if an agreement should \textit{ex tunc} be deemed as \textit{ineffective} as BC-based actions are considered to be eternal. This is a question primarily referring to technical solutions to make such constructions post factum editable (hence against the initial idea of BC), as having, for example, illegal content on a BC could potentially invalidate it completely for further usage.\\

The solution to fork the BC before this event and to abandon the ``illegal BC" is highly problematic given governmental power and activity structures on such a system. The idea of ETH's SCs to grant different \textit{states} (stages of contractual fulfillment so to say, see section \ref{Sec:UnderstandETH}) proposes a solution if a \textit{fallback function} to a given previous state or \textit{reverse TX} can be coded. Courts often use the ``doctrine of reasonable expectations" as a fallback function and justification for invalidating parts or all of a given agreement: the weaker party will not be held to adhere to contract terms that are beyond what the weaker party would have reasonably expected from the contract, even if what the party reasonably expected was outside the strict letter of agreement. The necessary ambiguity in the TC and SC situation alike arises when there are plausible and competing interpretations of a certain term. Given that the contractual parties can choose the language of the agreement in certain legal frameworks -- may it be French, Chinese, or Solidity -- it is important to note that this exemplary doctrine is not a rule granting substantive rights to any party if there is no doubt about the meaning of the used terms. Respectively, \textit{clickwrap/-through agreements} can pose considerable sources for problems, especially when thinking of \textit{adhesion clickwrap}. In a traditional context, these can be invalidated, which is not as easy to do in SC/BC software code cases. As the parties agreed to use a particular language, both parties are seen as being able to understand what is written in the agreement. The limit of freedom of contract, however, is obviously reached when the counterparty does not understand the given agreement at all and is respectively exploited. Unlike natural language on physical paper, it is harder to hide nefarious proceedings in standardized code forms (see further section \ref{Sec:UnderstandBC}), as both parties can de facto read the source code in any font they want and can check for all functions present to not be of disadvantage for them. Unless the disadvantageous party can prove to have been in a state of mental distress or similar while agreeing to this code, the only way out would be through \textit{bona fides} rulings -- which in itself are not helping the idea of legal security and may infringe on the freedom of contract.\\

As long as this level of \textit{smartness} is not present with SCs, the solution to most of these problems is not to employ an armada of legal scholars and practitioners to forcibly create completely new legal frameworks but to employ the already existing legal knowledge analogously rewritten as these have proven their efficiency to eventually serve as \textit{ius} \textit{consuetudinis} with specialized international courts. Given the scope of applicability and the inherent nature of these agreements, we can obviously observe that unfair and surprising outcomes, lack of notice or understanding of clauses can cause problems especially, when considering SCs as replacements for TCs. There exists a potential for unconscionability if the TC and SC do not represent equal information. However, such an event of discrepancy and ambiguity will be resolved contra proferentem against the party drafting the contract and its mean of execution -- again, there is no need to specialised codices. The European Union most notably works around these classic issues, for example, in the Rome I and Rome II regulations \citep[see further discussion e.g.,][]{Ruehl:2021}.\\

%\textcolor{red}{Adhersion Contract / Boilerplate Contract vs. Terms & Conditions vs. AGB}\\

%\textcolor{red}{305 ii bgb - programmcode vs. öffentlicher aushang}\\
 
%\textcolor{red}{Meeting of the Minds}\\

%This basal example of an automated agreement on the purchase of some value shows, that once set up, the landlord should not worry about chasing after the tenant for a rent payment given that if the rent is not paid, the conditions of the contract are not met, and hence access the apartment is denied. No further enforcement, for example, by law is purportedly needed.\\

The question of \textit{natural} or \textit{legal} persons is also interesting to be looked at with advancing digitization. Can \textit{Artificial Intelligence} (AI) can be considered as contractual partner in the sense of a legal person? Basically, if a given SC is coded to send or receive specific MSGs, it can be considered to be a smart post box or receptionist for inbound interaction and a messenger for outbound messages. Yet, if AI can hypothetically be considered a persona itself, or if the derelictio of software is possible and forces such entities to exist, can also be up to discussion. Any action of an AI should not pose a problem as long as the expressed will of the AI can be attributed to a natural or legal person. Considering the AI to be a legal person itself would only overly complicate the matter at hand (unless one would want to introduce an \textit{intelligent} person, besides a natural and legal one), as there will always be a wetware point-of-origin. The person who initially wanted to express a will via a code, may it be an SC or an AI, needs to reassure that unwanted results are prevented, otherwise has to accept the unwanted consequences by implication or estoppel (attributable to a natural or human person in the sense of \textit{falsus procurator} or similar to  a blanquette signature). Eventually this becomes relevant, when looking at BC-based IoT applications of say autonomous vehicles, which are designed to transmit sensor data for insurance reasons in case of an accident, or also might address repair/supply stations to keep it operational.\\

Another grave aspect of SCs versus the legal reality is their incompatibility of eventual execution according to the coding and potential legal expectations or pleads of, say, a consumer as the counterparty. Whereas it has been often discussed how to employ SCs, these applications can be plainly nonsense with serious consequences for the applicator in certain legal frameworks. Whereas the example of locking a rentee out of a flat has been often used to outline this issue, it is important to also think of possible scenarios of the landlord selling a lien. This could be an autonomous driving car with an SC-coupled electronic lock which the rentee needs to go to earn a living. The SC would act like a vigilante given that this scenario can be coded and executed after being triggered. ``Fiat iustitia, et pereat mundus" does not work in the state that SC-applying civilizations have reached.\\

%\textcolor{red}{unvereinbarkeit automatische durchführung ansprüche vs einreden / einwendungen - 309 nr 2,  307 i, ii ;; 310 III !! B2B, B2C}\\

% \textcolor{red}{Algorithmic vigilante: Possessorisch vs. Petitorisch}\\

Whereas the idea of using digital signatures is already widely accepted in legal frameworks, advances such as the EIP712 (\href{https://eips.ethereum.org/EIPS/eip-712}{Ethereum typed structured data hashing and signing}) should be observed critically. A major point for friction between legal systems and BC-based systems is present in the European Union (EU), most notably regarding the \href{https://gdpr-info.eu/}{GDPR}. Given this regulation, SCs as a replacement for TCs de facto \textit{not} possible in the EU (at least in B2C realms). Notably, the SC standard ERC1812 (\href{https://eips.ethereum.org/EIPS/eip-1812}{Ethereum Verifiable Claims}) records Identity Claims containing Personal Identifying Information (PII) on the ETH BC(remember: immutable public database) is conflicting with the GDPR. However, the \textit{legal legibility} of SCs should be seen as a \textit{time decaying} \textit{problem}. Knowledge and incorporation of codes as part of traditional constructions, just like the eventual acknowledgement of electronically saved parts of legal documents, will happen as the level of safety and standardization of SCs takes place. If SCs will become \textit{smart}, in that if a piece of code is able to become a real competition to paper, is yet to be seen. Therefore, we are still waiting for the realm envisioned by \citet{Susskind:13}. While it is certainly true that members of the tech or law guild enjoy to create dense and impenetrable \oe uvres, one always has to take the elementary self-defining aspect of many BC enthusiasts of being \textit{anti-establishment} into account -- or as \cite{Sklaroff:2017} put it: ``How to Lease a Car from an Anarchist". Most of what we commonly encounter as ``Smart Contract" is running against the self-definition of what the SC inventors themselves envisioned for their creation. Therefore many overused terms such as \textit{self-enforcing} and \textit{unbreakable} need to be understood with a pinch of realism -- from both a legal, as well as a technical point of view. As it stands today: Any smart contract will not protect from a smart lawyer in court or an enthusiastic hacker -- not even a smart SC will.

\subsection{Legal Example}
\label{Sec:Legal}

As a hands-on-example, let us look at a SC as a mean to execute a long term flat rental TC under German law: Landlord (L) and Tenant (T) sign a TC on renting an unrenovated ground floor road-view flat at BC-wonderland Berlin ``X-Berg"'s Tourist and Club area, i.e., EUR 2.000,-- gross warm rent for 45 square meters, on the 01. August 2021. The flat sports a well maintained and shiny smart-lock belonging and connected to a dedicated and extremely hyped Berlin-Startup (S) BC system that runs exclusively for such businesses -- obviously L and T agree in the TC to have an SC to manage respective value flows, as this is portrayed as saving them money and time, instead of making a musty repeating standing order via online banking.\\

After they have forwarded all their contractual, professional, and private information, besides creditworthiness et cetera, to the service provider to set up the SC, the complete heating system collapses on the 01. November 2021 and is not fixed for the whole month. This is the first time that T realizes that the SC has no function coded for any kind of such rent reduction due to heating shortages -- the smart meters on the heating devices are not linked to the SC via an Oracle to cut on respective costs, as it is expected that lodgers in this area will always be hot. T declares to L a rental reduction of 100\% for the month of November, as he was not able to use the flat accordingly -- a shortcoming in the sense of § 536 Par. 1 S. 1 Bürgerliches Gesetzbuch (BGB; German Civil Code). As the L does not react and the service provider S is not reachable, T addresses his credit institute -- from where the SC automatically deducted the full rent as advised repeatedly -- to cancel the permission to have any funds going off towards this address. Meanwhile, T tries to reach L and S, the SC is unable to automatically deduct the rent from the banking account of T on the 01. December and locks T out of the flat via a coded command to the smart-lock. After T -- who suspected a typical Berlin-esque ``Digital Advancement" technical error -- had the door broken open by a locksmith and having the smart-lock replaced by a conventional lock, the S tries to call T why the data-stream from the device has broken off -- just seconds before the L also tries to call T after having received an automatic alert message, that T has not paid the rent.\\

The days pass -- S has gone bankrupt in the meantime due to an overflow of unpaid marketing bills -- and L prompts T -- who has now access to the flat via the installed traditional lock-key-system and transfers the rent via a classic standing order -- to pay the missing rent for October 2021 on the firsts of December 2021, January and February 2022. Eventually, L files a lawsuit against T in March 2022 to pay the rent for October 2021 stating that the SC is an exhaustive and final provision regarding the rights and obligations of this rental contract. T eventually terminates the TC properly and moves back to his Uckermark home village in April 2022 after not becoming a successful Avantgarde fashion designer.\\

In the following, we provide a rough sketch on how this case of interaction between SC and TC would be handled legally in Germany and programmatically respectively in adjacent legal system constructions:\\

\begin{lstlisting}[frame=single, escapeinside={(*}{*)}]
I. Admissibility of the lawsuit of L against T to pay the rent (*for*) November 2021

1. Juristiction
    Given the TC (*and*) missing any contradicting information - the SC is completely irrelevant - 
    the Amtsgericht Tempelhof-Kreuzberg is factually (*and*) locally responsible (*for this case*),
    (*§*) 23 Nr. 2a Gerichtsverfassungsgesetz (GVG; Courts Constitution Act) in conjunction with 
    (*§*) 29a Par. 1 Zivilprozessordnung (ZPO; German Code of Civil Procedure).

2. Defensible Interest
    In (*case*) one should see an SC as a fully viable replacement (*for*) a TC, then one could argue,
    that there could be an easier way of justice than going to the courts (*and*) have a lengthy 
    process. As we (*do not*) see SCs as equal to TC, but merely as a mean of execution, we (*do not*) 
    need to discuss "self executing and enforcing justice".
    

II. Reasonableness of the lawsuit of L against T to pay the rent (*for*) November 2021
    Conditioned that L's claim to pay the rent (*for*) October 2021 based on the rental TC holds.

1. Contractual Base
    Again, we (*do not*) consider an SC to be a contract in the legal sense, hence only the TC is 
    important here with no contradicting information regarding issues with the TC. The SC is 
    just a mean of execution of the TC. Therefore the plead of L, that the SC is an exhaustive 
    (*and*) final provision is meaningless. 
    
    Thinking of an SC of being just a mere mean on execution, problems regarding a potentially 
    required written form of an agreement are of redundant nature. Importantly, it is commonly 
    agreed, that eMails, Telefaxes, (*or*) Computerfaxes (*do*) (*not*) hold up to fulfill the required 
    written form.
    
    On the other hand, (*if*) one would think of an SC being a replacement of a TC, one would pay 
    attention to respective regulations on the way the agreement is presented, i.e., the written 
    form requirement of (*§*) 126 Par. 1 BGB (*for*) a contract in the sense of (*§*) 550 S. 1 BGB can 
    be replaced according to (*§*) 126 Par. 3 BGB given that respective signatures in the sense of 
    (*§*) 126a BGB are given (recall our multiple hints towards the importance of signatures in a 
    technical (digital signature) (*and*) legal (electronic signature) sense - nevertheless a missing 
    written form according to (*§*) 550 S. 1 BGB does (*not*) provoke a nullity according to (*§*) 125 S. 1 
    BGB, but just an indefinite rental length in (*this*) particular (*case*)). 
    
    The full legal reference of (*§§*) 126a, 126 Par. 3, 127 Par. 1 (*and*) 3 BGB however hints towards 
    the border of the freedom of contract in the sense of (*§§*) 13, 125 BGB, as a given law may 
    define (*this*) otherwise. Examples would be the (*§§*) 484 Par. 1 S. 2, 492 Par. 1 S. 2, 
    623 Half-S. 2, 766 S. 2 BGB, (*or*) the (*§*) 780 S. 2, 781 S. 2 BGB.
    

2. Rights based on Irregularity In Contractual Performance
    Rights resulting from an irregularity in contractual performance can be called given that a 
    contractual performance is (*not*) (*or*) (*not*) as owed given.
    
    In (*this case*), the T ows the L to rent (*for*) November 2021. We could consider that T exercised 
    his right to set-off (*this*) obligation according to (*§*) 389 BGB with the rent of October 2021 
    given that he has a right to reduction.
    
    The SC itself does (*not*) have any other function besides checking (*for*) cash flows (*and*) 
    smart-lock data stream - it is obvisouly (*not*) (*$that$*) smart as it will never be able to represent 
    the status of the rental flat, unless that flat (*if*) full of sensors (*and*) these are connected 
    flawlessly to a respectively coded SC. (*This*), however, does (*not*) change anything in the rights a 
    tenant has by law. Here, as T was unable to live in the flat (*for*) the month of November 2021, 
    he has the right to reduce the rent of November 2021 by 100(*\%*) according to 
    (*§*) 536 Par. 1 BGB (*and*) to set-off (*this*) position according to (*§*) 389 BGB. L can (*not*) demand 
    the rent (*for*) November 2021.
    
    L's statement, that the SC is an exhaustive (*and*) final provision is on one hand irrelevant, as 
    they also (*signed*) a TC (*for*) which the SC is only the mean of execution, (*and*) the rights given by 
    law to T are (*not*) of a  dispositive nature. In general, any contractual agreement needs to stand 
    up to the questions of, (*for*) example, (*if*) they stand up against control towards their legality.
    
    
3. Cancellation of Contract
    The right to exercise the termination of a contract is a right to influence a legal relationship 
    just like a contestation, a revocation, (*or*) others besides (*§*) 134 BGB. As outlined beforehand, (*if*) 
    an SC should be seen as a complete surrogate (*for*) TCs, there can be issies (*if*) a respective 
    cancellation is (*not*) coded (*or*) (*not*) coded properly.
    
    Given that an SC is (*not*) coded properly, the respective legal rights are (*not*) superseded by them 
    missing in an ``exhaustive (*and*) final provision" as outlined above regarding the reduction of 
    the rent or the extraordinary termination of the agreement. In our example case, the T could 
    have terminated the TC immediately in December when the smart-lock did not let T in the 
    flat which represents an important issue according to (*§*) 543 Par. 1 (*\&*) Par. 2 S. 1 BGB, as the 
    respective requirements for locking a tenant out of a flat via a forced eviction or clearance 
    order according to (*§*) 940a ZPO were missing.
    
    
III. Conclusion
    The abovehand outlined case would be admissible to the courts, but it can not be seen as 
    being reasonable.
    
    Should one see an SC as a replacement for a TC, without a given legal foundation that 
    identifies this equally, then one could argue, that in the case of problems arising the people 
    utilising this framework willingly let go of their otherwise existing frameworks of rights - 
    caveat emptor. One can compare that to illicit employment or moonlighting to save on taxes 
    and issues resulting from a normally given warranty for defects or in the case of the customers 
    insolvency. Using surrogate systems to circumvent the existing regulated system does not 
    deserve to be protected.
    
    At least the German legal system is not in need of a specialised BC-law or similar lex 
    cryptographia creatures, as it is an effective tool given its abstractness to adapt to a given 
    case, as well as given its neutrality towards any kind of technological advances, may they have 
    been electronic signatures in the past or BC-systems at the given time of this writing.
    
         

\end{lstlisting}

\vspace{0.5cm}

With this rudimentary example -- we are not going deeper into any claims of T against L (e.g. § 231 BGB due to being locked out -- beyond §§ 858 ff. BGB), or of L against S (e.g. consequential damage in the sense of § 280 Par. 1 BGB due to a primitively coded SC), or of S against L and/or T (e.g. due to the willingly damaged smart-lock in the sense of the civil and penal law), et cetera -- we can quite easily present the frictions between the ``vision" and the reality of SC in other realms than being a financial vehicle like peer-to-peer lending or data transmitter in areas such as say renewable energy Prosumer situations.\\

We want to underline, that law enforcement and legal security do not work in the SC-framework we have at hand at the time of writing this -- especially when thinking about, for example, §§ 273, 320, 229, 230, 539 Par. 2, 997 BGB, or §§ 765a, 811 ZPO. Hence (nearly) every thinkable contractual construction will run into problems when being pushed in an SC application. Gravely, the exclusive right of courts to rule on what is eventually right or wrong is circumvented and can lead to unjust overenforcement, like locking the tenant out of the flat in the above shown example. Legal systems all over the world, and to our knowledge especially such coming from a Franco-Germanic genesis, are clear enough on what individual rights are and on what these individual rights borders are -- to this point, a \textit{embedded legal knowledge} can not be represented by SCs.

%%%%%%%%%%%%NEXT%%%%%%%%%%%%%%

%%%%%%%%%%%%NEXT%%%%%%%%%%%%%%

%%%%%%%%%%%% NEXT %%%%%%%%%%%%

\section{Closing remarks}
\label{Sec:Closing}

\blockquote{There is no reason anyone would want a computer in their home.}
\vspace{-0.4cm}
\begin{displayquote}
\citet{Olsen:1977}\\
\end{displayquote} 

\vspace{-0.6cm}

Wetware, Hardware, and Software have an inherent need for evolvability in response to changing requirements, and SCs are no different. Though interest in SCs has taken off in the past few years, they have created -- at the time of writing this -- a bulk of ``hyped hopes" with only a fraction of the said miracles having been delivered. Yet, is there really no reason anyone would want to buy their home via an SC?\\

We have shown how the first system that enabled SCs works, and what SCs can and can not accomplish in contrast to the common narratives. Moreover, we proposed a methodological solution for the decision support on the appropriateness of using DApp and SCs for solving specific problems. While we critically assessed the narrative of SCs, we see the chance and hope that SCs provide the incentive to move onwards to more efficient and stable proceedings. Work towards this on the intersection with Ricardian Contrachts is already underway \citep{Grigg2015,GriggRC}. However, the fields involved in the matter at hand -- explicitly Statistics, Information Technology, and Law -- must cooperate and not claim the sole right to exist coupled with papal-esque infallabilitas.\\

We have proven, that the application of SCs is very restricted to this point, but every technology requires some patching before being able to cater the promised deeds to the target group. Hence, SCs have the potential to be Door Kickers to evolve the idea dramatically with most legal frameworks being abstract or adaptive enough to get it accustomed to it. Otherwise, at the given state, the question what SCs can and can not deliver will stay to be a question of Realism and Idealism.

%%%%%%%%%%%% literature  %%%%%%%%%%%%

%% \section{References}
%% \label{sec7}

\begin{footnotesize}

\bibliography{literature}

\raggedright

\end{footnotesize}

\newpage

\section{Appendix}
\label{Appendix}

\vspace{1cm}

\subsection{List of cryptocurrencies in this research}
\label{Appendix:ListCC}

\begin{table}[H]
\centering
\label{my-label}
\begin{tabular}{lll}
\hline
\hline
Abbrev.   & CC               & Website                    \\ \hline
BTC (XBT) & Bitcoin          & \url{bitcoin.com}, \url{bitcoin.org}   \\
ETC       & Ethereum Classic & \url{ethereumclassic.github.io}  \\
ETH       & Ethereum         & \url{ethereum.org}               \\
LEO       & UNUS SED LEO         & \url{bitfinex.com} (iFinex ecosystem) \\
USDC       & USD Coin            & \url{centre.io/usdc}              \\
\hline
\hline
\end{tabular}
\end{table}

\subsection{List of abbreviations}
\label{Appendix:Abbrev}

\begin{table}[H]
\centering
\label{my-label}
\begin{tabular}{ll}
\hline
\hline
Terminus   & Abbrev.                \\ \hline
Blockchain  &     BC         \\
Cryptocurrency       & CC \\
Smart Contract (general terminus)      & SC \\
Verfified Smart Contract (Source Code public)     & VSC            \\
decentralized Apps     & DApps            \\
State of the DApps     & SDA            \\
Externally Owned Account       & EOA          \\
Contract Account       & CA      \\
Transaction (BC recorded)      & TX            \\
Call/Message  (``internal TX")     & MSG            \\
\hline
\hline
\end{tabular}
\end{table}

\subsection{Topics in the literature research}
\label{Appendix:Topics_literature}

\begin{figure}[H]
\hfill
\subfigure[Topic 1]{\includegraphics[width=4.2cm]{images/topics/Topic_0_t.png}}
\hfill
\subfigure[Topic 2]{\includegraphics[width=4.2cm]{images/topics/Topic_1_t.png}}
\hfill
\subfigure[Topic 3]{\includegraphics[width=4.2cm]{images/topics/Topic_2_t.png}}
\hfill
\subfigure[Topic 4]{\includegraphics[width=4.7cm]{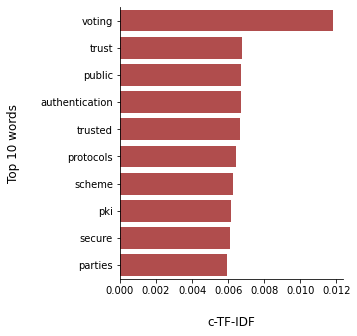}}
\hfill
\subfigure[Topic 5]{\includegraphics[width=4.8cm]{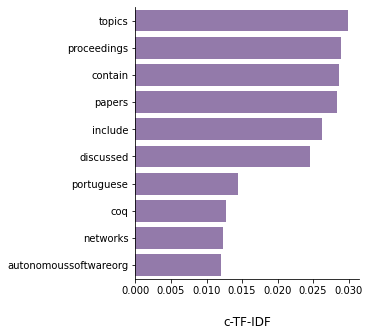}}
\hfill
\subfigure[Topic 6]{\includegraphics[width=4.1cm]{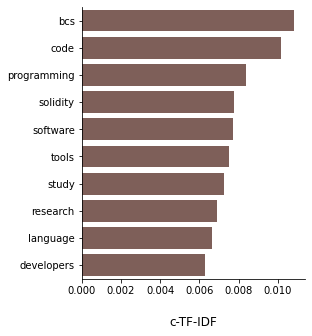}}
\hfill
\subfigure[Topic 7]{\includegraphics[width=4.4cm]{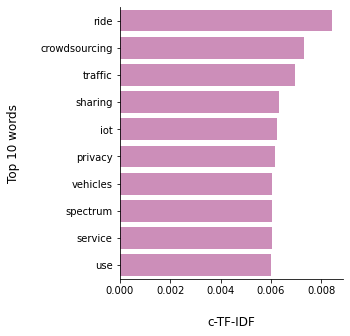}}
\hfill
\subfigure[Topic 8]{\includegraphics[width=3.8cm]{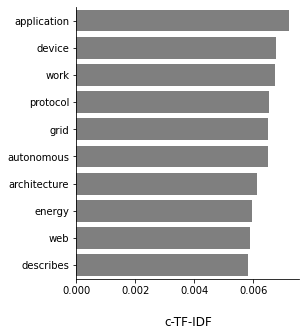}}
\hfill
\subfigure[Topic 9]{\includegraphics[width=4.0cm]{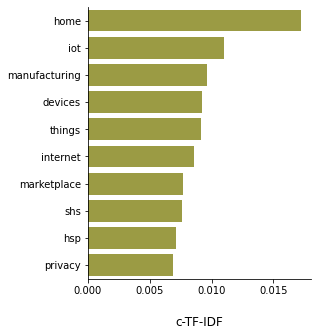}}
\hfill
\subfigure[Topic 10]{\includegraphics[width=4.4cm]{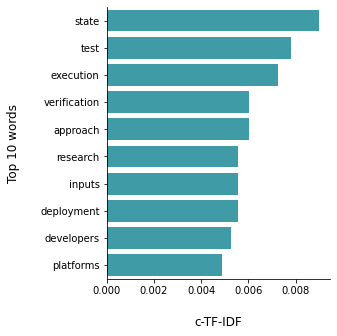}}
\hfill
\subfigure[Topic 11]{\includegraphics[width=3.9cm]{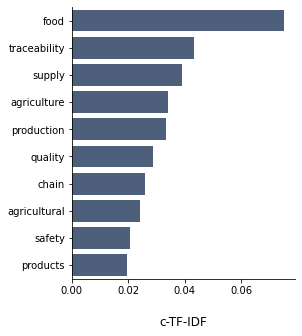}}
\hfill
\subfigure[Topic 12]{\includegraphics[width=4.4cm]{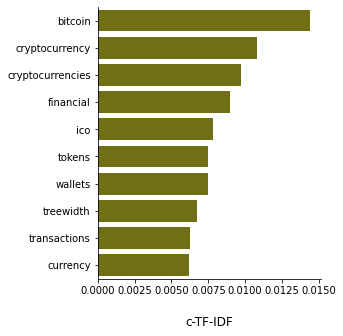}}
\hfill
\caption{Top 10 the most important words per topic identified in the existing SC research (Part 1) \href{https://github.com/QuantLet/USC/tree/master/SC-literature-research}{\includegraphics[keepaspectratio,width=0.4cm]{media/qletlogo_tr.png}}}
\label{fig:topics_research_1}
\end{figure}

\begin{figure}[H]
\hfill
\subfigure[Topic 13]{\includegraphics[width=4.85cm]{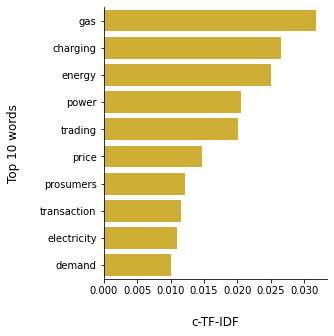}}
\hfill
\subfigure[Topic 14]{\includegraphics[width=4.2cm]{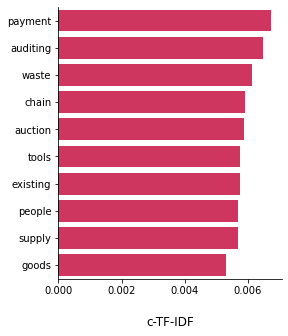}}
\hfill
\includegraphics[width=4.2cm]{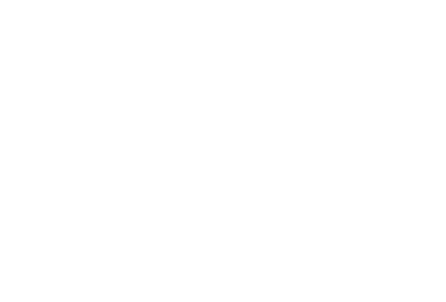}
\hfill
\caption{Top 10 the most important words per topic identified in the existing SC research (Part 2) \href{https://github.com/QuantLet/USC/tree/master/SC-literature-research}{\includegraphics[keepaspectratio,width=0.4cm]{media/qletlogo_tr.png}}}
\label{fig:topics_research_2}
\end{figure}

\subsection{ETH value denominations}
\label{Appendix:Wei}

Further Units and Globally Available Variables used in Solidity can be accessed through \href{https://solidity.readthedocs.io/en/v0.4.21/units-and-global-variables.html}{Ethereum - Read the Docs}.

\begin{table}[H]
\centering
\begin{tabular}{lll}
\hline
\hline
\textbf{Unit}                & \textbf{Wei Value} & \textbf{Wei}              \\ \hline
\textbf{Wei}                 & 1 wei              & 1                         \\
\textbf{Kwei (babbage)}      & 1e3 wei            & 1,000                     \\
\textbf{Mwei (lovelace)}     & 1e6 wei            & 1,000,000                 \\
\textbf{Gwei (shannon)}      & 1e9 wei            & 1,000,000,000             \\
\textbf{Microether (szabo)}  & 1e12 wei           & 1,000,000,000,000         \\
\textbf{Milliether (finney)} & 1e15 wei           & 1,000,000,000,000,000     \\
\textbf{Ether}               & 1e18 wei           & 1,000,000,000,000,000,000 \\ 
\hline
\hline
\end{tabular}
\end{table}

\subsection{Elliptic Curve Digital Signature Algorithm}
\label{ECDSA}

Asymmetric cryptography is used to create accounts in ETH in three steps: Firstly, the private key associated with an EOA is randomly generated as SHA256 output and could look like this: b032ac4de581a6f65c41889f2c90b3a629dc80667bf7167611f8b5575744f818 - a random 256 bit/32 bytes big and 64 hex character long output. Secondly, the public key is derived from the private key via the Elliptic Curve Digital Signature Algorithm (secp256k1 ECDSA) and could then look like this:  fc2921c35715210aeb0f2fffeaad94 \linebreak d906aaf2feda8a71e52c5d0a0da8ada4a44099e8ea65e3ec214b6686189255bba2373bd2ee6b05\linebreak 520dfd4dc571b682ccc3 - a 512 bits/64 bytes big and 128 hex character long output. The public key is therefore subsequently calculated from the private key, but as we are dealing with a $trapdoor$ $function$, this is not possible vice versa being an irreversible calculation. A private key is therefore kept non-public, whereas the public key is used to derive, in a third step  -  via Keccak-256 hashing of that public key, which results in a bytestring of length 32, from which the first 12 bytes are removed and result in a bytestring of length 20 - the individual ETH $address$, that could look like \href{https://rinkeby.etherscan.io/address/0x9255bba2373bd2ee6b05520dfd4dc571b682b349}{0x9255bba2373BD2Ee6B05520DfD4dC571B682b349} - a 160 bits/20 bytes big and 40 hex character long output (leaving out the 0x prefix). A public key can hence be used to determine, that the given signature of the information at hand is genuine, that means created with the respective public key and address of the interactor, without requiring the private key to be divulged \citep[see further appendix \ref{Appendix:ZeroKnowledgeProof} and][]{Buchanan}.

\subsection{Zero-Knowledge Proofs}
\label{Appendix:ZeroKnowledgeProof}

The standard notion of a mathematical proof can be related to the definition of an nondeterministic polynomial (NP) time complexity class of decision problems. That is, to prove that a statement is true one provides a sequence of symbols on a piece of paper, and a verifier checks that they represent a valid proof. Usually, though, some additional knowledge, other than the sole fact that a statement is true, is gained as a byproduct of the proof. Zero-knowledge proofs were introduced as a way to circumvent that, i.e., to convey no additional knowledge beyond proving the validity of an assertion. \cite{GoMiRa:89} first described zero-knowledge proofs as interactive proof systems.\\

As the term itself suggests, interactive zero-knowledge proofs require some interaction between a prover and a verifier. Intuitively, a proof system is considered zero-knowledge if whatever the verifier can compute, while interacting with the prover, it can compute by itself without going through the protocol \citep{GoOr:94}. Formally they can be defined as follows, as per \cite{BoSa:07}.\\

Given an interactive proof system, or an interactive protocol $(P, V)$, where $P$ and $V$ can be seen as interactive probabilistic polynomial-time (PPT) Turing machines symbolizing a Prover and a Verifier, for a formal NP-language $L \subset \{0,1\}^*$ and an input $x$, the $output$ of $V$ on $x$ at the end of interaction between $P$ and $V$ can be written as 
\[out_V[P(x), V(x)].\] 
$(P, V)$ is called a zero-knowledge protocol for $L$ if the following three conditions hold:

\textbf{Completeness:} 
\[\forall x \in L, u \in \{0,1\}^*, \qquad Pr[out_V[P(x,u),V(x)]] \geq 2/3,\]
where $u$ is a certificate for the fact that $x \in L$. In other words the prover can convince the verifier of $x \in L$ if both follow the protocol properly.

\textbf{Soundness:} If $x \notin L$, then 
\[\forall P^*,u \in \{0,1\}^*, \qquad Pr[out_V[P^*(x,u), V(x)]] \leq 1/3.\]
I.e., the prover cannot fool the verifier, except with small probability.\\

\textbf{Perfect Zero-Knowledge:} For every strategy $V^*$ there exists an expected PPT simulator $S^*$ such that 
\[\forall x \in L, u \in \{0,1\}^*, \qquad out_{V^*}[P(x,u), V*(x)] \equiv S^*(x).\]
The last condition prevents the verifier from learning anything new from the interaction, even if she does not follow the protocol but rather uses some other strategy $V^*$. Otherwise she could have learned the same thing by just running the simulator $S^*$ on the publicly known input $x$. $S^*$ is called the simulator for $V^*$, as it simulates the outcome of $V^*$’s interaction with the prover without any access to such an interaction.

\subsection{\textit{t}-SNE (\textit{t}-Distributed Stochastic Neighbor Embedding)}
\label{Appendix:tsne}

\textit{t}-SNE is a non-linear technique for dimension reduction and data visualisation. it allows to preserve the local structure and is proposed by \cite{maaten2008visualizing}. It aims to design an embedding of high-dimensional input to low-dimensional map while preserving much of a significant structure. On the algorithm \ref{algo} below, the reader can find the pseudocode of the \textit{t}-SNE computation. \\
\begin{equation*}
    X = \{x_1, x_2,\dots,x_n \} \xrightarrow[]{} Y = \{y_1, y_2,\dots,y_n \} 
\end{equation*}
$x_i$ is the $i^{th}$ object in high-dimensional space. \\
$y_i$ is the $i^{th}$ object in high-dimensional space. \\

\begin{algorithm}[H]
\caption{\textit{t}-SNE Pseudocode}\label{algo}
\SetAlgoLined  
\KwData{data set $\chi=\{x_1, x_2, \dots, x_n\}$ \\
cost function parameters: perplexity $Perp$, \\
optimization parameters: number of iterations $T$, learning rate $\eta$, momentum $\alpha(t)$} 
\KwResult{low-dimensional data representation $Y^{(T)}=\{y_1, y_2, \dotsm  y_n\}$ }
\Begin{
compute pairwise affinities $p_{j|i}$ with perplexity $Perp$ (using Equation \ref{eq:e1})\;
set $p_{ij}=\frac{p_{j|i}+p_{i|j}}{2n}$\;
sample initial solution $Y^{(0)}=\{y_1,y_2,\dots,y_n\}$ from $N(0,10^{-4}I)$\; \\
\For{$t=1$ to $T$}{
 computer low-dimensional affinities $q_{ij}$ (using Equation \ref{eq:e2}))\; \\
 compute gradient $\frac{\delta C}{\delta Y}$ (using Equation \ref{eq:e3})) \; \\
 set $Y^{(t)}=Y^{(t-1)}+\eta \frac{\delta C}{\delta Y} + \alpha(t)(Y^{(t-1)}-Y^{(t-2)})\;$
}
}
%\hline 
\end{algorithm}

\begin{equation} \label{eq:e1}
    p_{j\vert i}=\frac{\exp(-\| x_i-x_j\|^2/2\sigma_i^2)}{\sum_{k\neq i}\exp(-\| x_i-x_k\|^2/2\sigma_i^2)}
\end{equation}

\begin{equation}  \label{eq:e2}
    q_{ij}=\frac{(1+\| y_i-y_j\|^2)^{-1}}{\sum_{k\neq l}(1+\| y_i-y_l\|^2)^{-1}}
\end{equation}

\begin{equation}   \label{eq:e3}
    \frac{\delta C}{\delta y_i}=4\sum_j(p_{ij}-q_{ij})(y_i-y_j)(1+\| y_i-y_j\|^2)^{-1}
\end{equation}

%%%

\subsection{UMAP vs. \textit{t}-SNE}
\label{Appendix:umap}
The dimensionality reduction algorithm that we are using here is the  Uniform Manifold Approximation and Projection for Dimension Reduction
(UMAP) \citep{mcinnes2018umap} and is like \textit{t}-SNE -- a neighbor graph algorithm. The mathematical foundations of the UMAP rely on Laplacian Eigenmaps and are very extensive. The important differences between \textit{t}-SNE and the UMAP are the following.The UMAP aims to better preserve more of the global structure while requiring less computational time. As compared to the \textit{t}-SNE, the UMAP relies not only on the the Kullback-Leibler divergence measure, but on the cross-entropy.

\begin{equation} \label{eq:4}
    C_{UMAP}=\sum_{i\neq j}\bigg\{ v_{ij}\log\left(\frac{v_{ij}}{w_{ij}}\right)+\left(1-v_{ij}\right)\log\left(\frac{1-v_{ij}}{1-w_{ij}} \right) \bigg\}
\end{equation}
where $v_{ij}$ are the pair-wise similarities in the high dimensional space and $w_{ij}$ - in the low-dimensional. The optimization problem used in the UMAP is the stochastic gradient descent instead of gradient descent used in the \textit{t}-SNE, which speeds up the computations and decreases the required memory resources. Moreover, UMAP does not require the distance Euclidean.

\subsection{Example Code}
\label{Appendix:ExampleSC}

Adding to the source code review and possible use cases, we are presenting a simple $Hello$ $World$-esque SC as provided by the \href{https://github.com/ethereum/ethereum-org/blob/master/views/content/greeter.md}{Ethereum Github} with explanatory adaptions to outline some of the technical complexity that has a grave impact on every adjacent structure. We have observed many non-technical outlets -- especially Blogs and the such -- discussing a theme, that they have apparently never seen as code itself -- consequently, we will also keep this to a very brief overview to introduce the structures in a very simple example. Each comment starts with /** and ends with /*. \\

%&\Comment{This is the version control for the compiler, see above section \ref{Sec:UnderstandSC}, as each Solidity version may have different commands that can be coded - higher versions will obviously improve efficiency of the coding and lead to better code controlling.}&

\begin{lstlisting}[frame=single, escapeinside={(*}{*)}]
    /** Version control for the compiler, see above section 6, as each Solidity version may have different commands that can be coded - higher versions will obviously improve efficiency of the coding and lead to better code controlling. /*
pragma solidity >=0.4.22 <0.6.0; 

    /** "Mortal" is the name of this SC. /*
contract Mortal {
        /** Defines the variable "owner" of the type "address". /*
    address owner;
    
        /** The "constructor" is executed at initialization and sets the owner of the SC, i.e., it is executed once when the CA/SC is first deployed. Similar to other class-based programming languages, it initializes state variables to specified values. "msg.sender" refers to the address where the CA is being created from, i.e., here in the constructor setting the "owner" to the address of the SC creator. SCs depend on external TX/MSG to trigger its functions, whereas "msg" is a global variable that includes relevant data on the given interaction, such as the address of the sender and the value included in the interaction. This is assured by the "public" function, which can be called from within the CA/SC or externally via MSG's, like here getting the address of the interactor. "private" functions are not callable and can only reached by the SC itself - a particular source for grave errors, as you can not change the SC once deployed. /*
    constructor() public { owner = msg.sender; }
    
        /** Another important function, and source for grave errors if missing, follows and represents a mean to recover funds stored on the CA. Alternatively, calling "selfdestruct(address)" sends all of the SCs current balance to address specified. Remember, that once deployed the SC can not be changed unlike non-BC-based software. The only way to modify an SC is to deploy a corrected one - best after deactivating and recovering all funds in the problematic one. Interestingly, "selfdestruct" consumed "negative Gas", as it frees up BC/EVM space by clearing all of the CA/SCs's data./*
    function kill() public { if (msg.sender == owner) selfdestruct(msg.sender); }
}


    /** After "Mortal", "Greeter" is another SC presented to visualize, that CA/SCs can "inherit" characteristics of CA/SCs enabling SCs to be written shorter and clearer. By declaring that "Greeter is Mortal", "Greeter" inherits all characteristics of "Mortal" and keeps the "Greeter" code herewith crisp and clear to to point, where is has individual functions to be executed. In this example, the inherited characteristic of "Mortal" gives, as defined beforehand in "Mortal", that "Greeter" can be deactivated with all locked funds being recovered. /*
contract Greeter is Mortal {

        /** Defines the variable "greeting" of the type "string", i.e., a sequence of characters. /*
    string greeting;

        /** This is defined as beforehand in "Mortal", whereas in this case the underscore in "\_greeting" is a style used to differentiate between function arguments and global variables. There is no semantic difference between "greeting" and "_greeting", whereas the latter one is defined as such not to shadow the first one. Here, the underscore differentiates between the global variable "greeting" and the corresponding function parameter. Strings can be stored in both "storage" and "memory" depending on the type of variable and usage. "memory" lifetime is limited to a function MSG and is meant to be used to temporarily store variables and respective values. Values stored in "memory" do not persist on the network (EVM (*\&*) BC) after the interaction has been completed. /*
        
    constructor(string memory  (* \_*)greeting) public {
        greeting = (* \_*)greeting;
    }

        /** Main function of the SC that returns the greeting once "greet" function is MSG'ed /*
    function greet() public view returns (string memory) {
        return greeting;
    }
}
\end{lstlisting}

\end{document}